\def\IEEEsubmission{0}
\def\reviewColor{black}

\def\complexNumbers{\mathbb{C}}
\def\aDomain{\mathbb{A}}
\def\compactInterval{\mathbb{E}}
\def\aCompactMetricSpace{\mathbb{S}}
\def\aMetricSpace{{\Gamma}}
\def\functionSpace[#1]{\mathcal{F}(#1)}
\def\realNumbers{\mathbb{R}}
\def\integers{\mathbb{Z}}
\def\constante{{\rm e}}
\def\constantj{{\rm j}}
\def\expectationOperator[#1][#2]{{\text{E}_{#2}}\left\{#1\right\}}
\def\uniformDistribution[#1][#2]{{\mathcal{U}_{[#1,#2]}}}
\def\traceOperator[#1]{{\textrm{tr}}\{#1\}}
\def\identityMatrix[#1]{\textbf{\textrm{I}}_{#1}}
\def\zeroVector[#1]{\textbf{\textrm{0}}_{#1}}
\def\exponentialIntegral[#1]{\textrm{Ei}(#1)}
\def\channelAtSubcarrier[#1]{h_{#1}}
\def\affineEnc{g}
\def\affineDec{h}
\def\weight[#1]{w_{#1}}
\def\indicatorFunction[#1]{\mathbb{I}\left[{#1}\right]}
\def\thresholdTCI{t}

\def\normalizationCoef[#1]{\eta_{#1}}
\def\base{\beta}
\def\functionArbitrary[#1]{f_{#1}}
\def\functionArbitraryEstimate[#1]{\hat{f}_{#1}}

\def\signNormal[#1]{\text{sign}\left(#1\right)}
\def\diagOperation[#1]{\text{diag}\left\{#1\right\}}
\def\idftSize{N}

\def\probability[#1]{\textrm{Pr}\left({#1}\right)}
\def\numberOfActiveSubcarriers{M}
\def\complexGaussian[#1][#2]{\mathcal{CN}({#1,#2})}
\def\gaussian[#1][#2]{\mathcal{N}({#1,#2})}
\def\meanG{\mu}
\def\sigmaG{\sigma}
\def\normalPDF[#1]{\phi\left(#1\right)}
\def\normalCDF[#1]{\Phi\left(#1\right)}

\def\Preference{P_{\rm ref}}
\def\referenceDistance{r_{\rm ref}}

\def\distanceED[#1]{r_{#1}}
\def\powerED[#1]{P_{{#1}}}
\def\powerEDdistance[#1]{P_{\text{rx}}({#1})}
\def\pathlossExponent{\alpha}
\def\powerControl{\beta}

\def\symbolVector[#1]{\textrm{\textbf{s}}{[#1]}}
\def\symbolVectorED[#1]{\textrm{\textbf{s}}_{#1}}
\def\preProcessedVector[#1]{\textrm{\textbf{p}}_{#1}}
\def\encodedVector[#1]{\textrm{\textbf{c}}_{#1}}
\def\modulationVector[#1]{\textrm{\textbf{m}}_{#1}}
\def\transmittedVector[#1]{\textrm{\textbf{x}}_{#1}}
\def\superposedVector{\textrm{\textbf{y}}}
\def\modulationVectorSuperposed{\acute{\textrm{\textbf{{m}}}}}
\def\demappedVector{\acute{\textrm{\textbf{{c}}}}}
\def\decodedVector{\acute{\textrm{\textbf{{p}}}}}
\def\postProcessedVector{\textrm{\textbf{z}}}
\def\noiseVector{\textrm{\textbf{n}}}

\def\permutation{\sigma}
\def\preProcessingFunction[#1][#2]{\psi_{#1}{\left(#2\right)}}
\def\postProcessingFunction[#1][#2]{\varphi_{#1}{\left(#2\right)}}
\def\preProcessingFunctionW[#1]{\psi_{#1}}
\def\postProcessingFunctionW[#1]{\varphi_{#1}}
\def\encoder[#1]{\epsilon_{#1}}
\def\decoder{\delta}
\def\projectionMatrix{\textbf{\textrm{G}}}
\def\precoder[#1]{\textbf{\textrm{B}}_{#1}}
\def\channelMatrix[#1]{\textbf{\textrm{H}}_{#1}}
\def\channelVector[#1]{\textbf{\textrm{h}}_{#1}}
\def\aggregator{\textbf{\textrm{A}}}
\def\mapperResource{\mathcal{M}}
\def\demapperResource{\mathcal{M}^{-1}}

\def\outageProbability[#1]{P_{\rm out}(#1)}
\def\errorProbability{P_{\rm cer}}
\def\errorProbabilityblock{P_{\rm bcer}}
\def\errorProbabilityCont[#1]{P_{\rm bout}(#1)}

\def\coefInField[#1]{q_{#1}}
\def\integerVector{\textbf{\textrm{a}}}
\def\integerVectorEle[#1]{{a}_{#1}}
\def\mappingFromFieldtoInteger{g^{-1}}
\def\finiteField[#1]{\mathbb{F}_{#1}}

\def\kroneckerDelta[#1][#2]{\delta_{#1#2}}
\def\indexED{k}

\def\indexSampleTime{n}
\def\indexEncodedVector{b}

\def\indexTXsubcarrier{l}
\def\indexRXsubcarrier{\ell}
\def\indexRXsample{n}

\def\indexCommunicationRound{n}
\def\indexKolmogorovSum{\ell}

\def\spaceContinuous[#1]{\mathcal{C}^0(#1)}
\def\spaceContinuousNomographic[#1]{\mathcal{N}^0(#1)}
\def\spaceContinuousAprxNomographic[#1]{{\mathcal{N}^0_\precision}(#1)}
\def\spaceNomographic[#1]{\mathcal{N}(#1)}
\def\inverseMap{g^{-1}}
\def\forwardMap{g}
\def\coefa{\alpha}
\def\coefb{\beta}

\def\symbolVectorEle[#1]{{{s}}_{#1}}
\def\preProcessedVectorEle[#1]{{{p}}_{#1}}
\def\encodedVectorEle[#1]{{{c}}_{#1}}
\def\modulationVectorEle[#1]{{{m}}_{#1}}
\def\transmittedVectorEle[#1]{{{x}}_{#1}}
\def\superposedVectorEle[#1]{{{y}}_{#1}}
\def\modulationVectorSuperposedEle[#1]{{{\acute{m}}}_{#1}}
\def\demappedVectorEle[#1]{{{\acute{c}}}_{#1}}
\def\decodedVectorEle[#1]{{{\acute{p}}}{[#1]}}
\def\postProcessedVectorEle[#1]{{{z}}{[#1]}}
\def\noiseVectorEle[#1]{{{n}}_{#1}}

\def\numberOfDigitsBits{{b_0}}
\def\numberOfEdgeDevices{K}
\def\numberOfDimensionAtRX{N_{\rm r}}
\def\numberOfDimensionAtTX{N_{\rm t}}
\def\numberOfTransforms{N_{\rm access}}
\def\numberOfFunctions{N_{\rm f}}
\def\numberOfChannelUse{D}
\def\blockLength{B}
\def\numberOfModulationSymbols{L}
\def\sequenceLength{N}
\def\sequenceLengthOther{M}
\def\numberOfActiveSubcarriers{M}

\def\numberOfGroups{S}
\def\integersGoldenbaum[#1]{w_{#1}}
\def\numberOfParametersGoldenbaum{\tau}
\def\computationRate{\mathcal{R}}
\def\computationThroughput{R}
\def\generatorMatrix{\textbf{\textrm{G}}}
\def\aprime{p}

\def\timeSymbol{t}
\def\transmittedSignal[#1][#2]{\mathfrak{s}_{#1}(#2)}
\def\receivedSignal[#1][#2]{\mathfrak{r}_{#1}(#2)}

\def\aUnitaryMatrix{\textbf{\textrm{A}}_{\rm n}}
\def\transmittedSignalPass[#1][#2]{\mathfrak{s}_{#1}'(#2)}
\def\receivedSignalPass[#1][#2]{\mathfrak{r}_{#1}'(#2)}
\def\aFixedPowerValue{P_0}
\def\outageThreshold{\epsilon}
\def\estimationError[#1]{e_{\text{FEE}}(#1)}
\def\meanSquareError[#1]{{\tt{MSE}}(#1)}
\def\meanSquareErrorOfMod[#1]{{{e}}(#1)}
\def\bayesianMeanSquareError{{\tt{BMSE}}}
\def\normalizedMeanSquareError[#1]{{\tt{NMSE}}(#1)}
\def\functionEstimationError[#1]{e_{\text{NFEE}}(#1)}
\def\meanFunctionSquareError{{\tt{MSFE}}}
\def\fmax{f_{\text{max}}}
\def\fmin{f_{\text{min}}}
\def\fcarrierES{f_{\text{ES}}}
\def\fcarrierED[#1]{f_{\text{ED},#1}}

\def\phaseES{\theta_{\text{ES}}}
\def\phaseED[#1]{\theta_{\text{ED},#1}}

\def\entropy[#1]{H\left({#1}\right)}
\def\errorProbabilityConst{\delta}
\def\error{\epsilon}
\def\computationRateAchievable{\mathcal{R}^\text{comp}(\functionArbitrary[],\error)}
\def\computationRateAchievableNoArg{\mathcal{R}^\text{comp}}

\def\coefPoly[#1]{c_{#1}}

\def\pForNorm{p}

\def\dirichletKernel[#1][#2]{D_{#1}\left(#2\right)}
\def\kernelInput{x}
\def\kernelScalar{N}

\def\syncPoint{t_{\text{0}}}

\def\precision{\epsilon}
\def\precisionF{p_0(\precision)}
\def\asmallNumber{\alpha}

\def\ncfo[#1]{\eta_{\text{FO,}#1}}
\def\nto[#1]{\eta_{\text{TO,}#1}}

\def\cfo[#1]{\Delta f_{#1}}
\def\to[#1]{\Delta t_{\text{ED,}#1}}
\def\toComplete[#1]{\Delta t_{#1}}
\def\po[#1]{\Delta \theta_{#1}}
\def\so{\Delta t_{\text{ES}}}

\def\delaySymbol{\tau}
\def\delay[#1]{\delaySymbol_{#1}}
\def\gain[#1]{a_{#1}}
\def\gainComplex[#1]{a_{#1}^{\rm b}}
\def\channelPass[#1]{\mathfrak{h}_{\indexED}'(#1)}
\def\channelBaseband[#1]{\mathfrak{h}_{\indexED}(#1)}
\def\pathIndex{p}
\def\numberOfPaths{P}
\def\deltaFunction[#1]{\delta(#1)}

\def\symbolDuration{T_{\text{sym}}}

\def\noiseVariance{\sigma_{\text{n}}^2}
\def\optimumGeometricMean{{\rm \bf z^*}}
\def\variableForGeoemtricMean{{\rm \bf z}}
\def\variableForGeoemtricMeanIteration[#1]{\variableForGeoemtricMean^{#1}}

\def\weightDistance[#1]{\alpha_{#1}}
\def\medianDistance[#1]{\beta_{#1}}
\def\smoothFactor{v}
\def\modelParametersAtIteration[#1][#2]{\textbf{p}^{}_{#2}}
\def\bandwidth{W}
\def\duration{T}
\def\numberOfSpatialStreams{S}

\if\IEEEsubmission1
\documentclass[journal,12pt,onecolumn,draftclsnofoot,]{IEEEtran}
\else
\documentclass[journal]{IEEEtran}
\fi

\usepackage{multicol}
\usepackage{lipsum}
\usepackage{mathtools}
\usepackage{amsthm}
\usepackage{acronym}
\usepackage[bookmarksopen=true]{hyperref}
\usepackage{stfloats}
\usepackage{amsfonts}
\usepackage{acronym}
\usepackage{cite}
\usepackage{multirow}
\usepackage{bm}
\usepackage{amsmath,amssymb}
\usepackage{graphicx}
\usepackage{epstopdf}
\usepackage[table,xcdraw]{xcolor}
\usepackage[caption=false,font=footnotesize]{subfig}
\usepackage[geometry]{ifsym}
\usepackage{array}
\usepackage[utf8]{inputenc}
\usepackage[T1]{fontenc}
\usepackage{pifont}

\let\norm\undefined 
\DeclarePairedDelimiter\norm{\lVert}{\rVert}
\newcommand\mydots{\hbox to 1em{.\hss.\hss.}}
\DeclarePairedDelimiter\ceil{\lceil}{\rceil}

\DeclarePairedDelimiter\abs{\lvert }{\rvert }

\newtheorem{theorem}{Theorem}
\newtheorem{definition}{Definition}

\newcommand{\cmark}{\ding{51}}%
\newcommand{\xmark}{\ding{55}}%

\makeatletter
\newif\ifAC@uppercase@first%
\def\Aclp#1{\AC@uppercase@firsttrue\aclp{#1}\AC@uppercase@firstfalse}%
\def\AC@aclp#1{%
	\ifcsname fn@#1@PL\endcsname%
	\ifAC@uppercase@first%
	\expandafter\expandafter\expandafter\MakeUppercase\csname fn@#1@PL\endcsname%
	\else%
	\csname fn@#1@PL\endcsname%
	\fi%
	\else%
	\AC@acl{#1}s%
	\fi%
}%
\def\Acp#1{\AC@uppercase@firsttrue\acp{#1}\AC@uppercase@firstfalse}%
\def\AC@acp#1{%
	\ifcsname fn@#1@PL\endcsname%
	\ifAC@uppercase@first%
	\expandafter\expandafter\expandafter\MakeUppercase\csname fn@#1@PL\endcsname%
	\else%
	\csname fn@#1@PL\endcsname%
	\fi%
	\else%
	\AC@ac{#1}s%
	\fi%
}%
\def\Acfp#1{\AC@uppercase@firsttrue\acfp{#1}\AC@uppercase@firstfalse}%
\def\AC@acfp#1{%
	\ifcsname fn@#1@PL\endcsname%
	\ifAC@uppercase@first%
	\expandafter\expandafter\expandafter\MakeUppercase\csname fn@#1@PL\endcsname%
	\else%
	\csname fn@#1@PL\endcsname%
	\fi%
	\else%
	\AC@acf{#1}s%
	\fi%
}%
\def\Acsp#1{\AC@uppercase@firsttrue\acsp{#1}\AC@uppercase@firstfalse}%
\def\AC@acsp#1{%
	\ifcsname fn@#1@PL\endcsname%
	\ifAC@uppercase@first%
	\expandafter\expandafter\expandafter\MakeUppercase\csname fn@#1@PL\endcsname%
	\else%
	\csname fn@#1@PL\endcsname%
	\fi%
	\else%
	\AC@acs{#1}s%
	\fi%
}%
\edef\AC@uppercase@write{\string\ifAC@uppercase@first\string\expandafter\string\MakeUppercase\string\fi\space}%
\def\AC@acrodef#1[#2]#3{%
	\@bsphack%
	\protected@write\@auxout{}{%
		\string\newacro{#1}[#2]{\AC@uppercase@write #3}%
	}\@esphack%
}%
\def\Acl#1{\AC@uppercase@firsttrue\acl{#1}\AC@uppercase@firstfalse}
\def\Acf#1{\AC@uppercase@firsttrue\acf{#1}\AC@uppercase@firstfalse}
\def\Ac#1{\AC@uppercase@firsttrue\ac{#1}\AC@uppercase@firstfalse}
\def\Acs#1{\AC@uppercase@firsttrue\acs{#1}\AC@uppercase@firstfalse}

\acrodef{WSN}{wireless sensor network}
\acrodef{USRP}{universal software radio peripheral}
\acrodef{SN}{sensor node}
\acrodef{FC}{fusion center}
\acrodef{MAC}{multiple-access channel}
\acrodef{FL}{federated learning}
\acrodef{ED}{edge device}
\acrodef{CS}{compressed sensing}
\acrodef{ES}{edge server}
\acrodef{DCN}{data center network}
\acrodef{RIS}{reconfigurable intelligent surfaces}
\acrodef{IMC}{in-memory computing}
\acrodef{FPGA}{field-programmable gate array}
\acrodef{SDR}{software-defined radio}
\acrodef{PS}{processing system}
\acrodef{SS}{soft synchronization}
\acrodef{IQ}{in-phase/quadrature}
\acrodef{IP}{intellectual property}
\acrodef{DMA}{direct-memory access}
\acrodef{RAM}{random access memory}
\acrodef{CC}{companion computer}
\acrodef{FEE}{function estimation error}
\acrodef{MSK}{minimum-shift keying}
\acrodef{TDMA}{time-domain multiple access}
\acrodef{PLNC}{physical-layer network coding}
\acrodef{UAV}{unmanned aerial vehicle}
\acrodef{LoRa}{Long-Range}
\acrodef{DC}{direct-current}
\acrodef{DAC}{digital-to-analog converter}
\acrodef{ADC}{analog-to-digital converter}

\acrodef{OBO}{output-power back-off}
\acrodef{ACLR}{adjacent-channel-leakage ratio}

\acrodef{LDPC}{low-density parity check}

\acrodef{TBMA}{type-based multiple access}

\acrodef{MSFE}{mean-squared function error}
\acrodef{FEE}{function-estimation error}
\acrodef{CER}{computation error rate}
\acrodef{BCER}{block-computation error rate}
\acrodef{CFO}{carrier frequency offset}
\acrodef{TO}{time offset}
\acrodef{PO}{phase offset}
\acrodef{RSSI}{received signal strength  information}

\acrodef{STLC}{space-time line code}
\acrodef{CCI}{co-channel interference}
\acrodef{CSIT}[CSIT]{\ac{CSI} at the transmitter}
\acrodef{CSIR}[CSIR]{\ac{CSI} at the receiver}
\acrodef{MIMO}{multiple-input-multiple-output}
\acrodef{PC}{phase correction}
\acrodef{ZF}{zero-forcing}
\acrodef{ANOVA}{analysis of variance}

\acrodef{PCA}{principal component analysis}
\acrodef{TIG}{Technical Interest Group}

\acrodef{FSK}{frequency-shift keying}
\acrodef{PPM}{pulse-position modulation}
\acrodef{PAM}{pulse-amplitude modulation}

\acrodef{MRC}{maximum-ratio combining}
\acrodef{HP}{hard-coded participation}
\acrodef{HPA}{hard-coded participation with absentees}
\acrodef{SP}{soft-coded participation}
\acrodef{FSK-MV}{\ac{FSK}-based \ac{MV}}
\acrodef{RF}{radio-frequency}
\acrodef{MF}{matched filter}
\acrodef{PPM}{pulse-position modulation}
\acrodef{CSK}{chirp-shift keying}
\acrodef{PPM-MV}[PPM-MV]{\ac{PPM}-based \ac{MV}}
\acrodef{DFT-s-OFDM}{\ac{DFT}-spread \ac{OFDM}}
\acrodef{SC}{single-carrier}
\acrodef{SGD}{stochastic gradient descent}
\acrodef{signSGD}{sign stochastic gradient descent}

\acrodef{SL}{split learning}
\acrodef{SNR}{signal-to-noise ratio}
\acrodef{RMSE}{root-mean-squared error}
\acrodef{OFDM}{orthogonal frequency division multiplexing}
\acrodef{DFT}{discrete Fourier transform}
\acrodef{PSK}{phase-shift keying}
\acrodef{QAM}{quadrature amplitude modulation}
\acrodef{QPSK}{quadrature phase-shift keying}
\acrodef{PMEPR}{peak-to-mean envelope power ratio}
\acrodef{BER}{bit-error ratio}
\acrodef{SNR}{signal-to-noise ratio}
\acrodef{PSD}{power spectral density}
\acrodef{SE}{spectral efficiency}
\acrodef{CP}{cyclic prefix}
\acrodef{AWGN}{additive white Gaussian noise}
\acrodef{CFR}{channel frequency response}
\acrodef{CIR}{channel impulse response}
\acrodef{MMSE}{minimum mean-squared error}
\acrodef{LMMSE}{linear minimum mean-squared error}
\acrodef{BPSK}{binary phase shift keying}
\acrodef{BLER}{block-error rate}
\acrodef{ML}{maximum likelihood}
\acrodef{PHY}{physical layer}
\acrodef{PA}{power amplifier}
\acrodef{IDFT}{inverse DFT}
\acrodef{DoF}{degrees-of-freedom}
\acrodef{IoT}{Internet-of-Things}
\acrodef{FDE}{frequency-domain equalization}
\acrodef{RF}{radio-frequency}
\acrodef{IM}{index modulation}
\acrodef{BS}{base station}
\acrodef{MF}{matched filter}
\acrodef{PPM}{pulse-position modulation}

\acrodef{MSE}{mean-squared error}
\acrodef{MRT}{maximum-ratio transmission}
\acrodef{ERC}{equal-ratio combining}
\acrodef{BAA}{broadband analog aggregation}
\acrodef{OBDA}{one-bit broadband digital aggregation}
\acrodef{FEEL}{federated edge learning}
\acrodef{FL}{federated learning}
\acrodef{ED}{edge device}
\acrodef{ES}{edge server}
\acrodef{UL}{uplink}
\acrodef{DL}{downlink}
\acrodef{OAC}{over-the-air computation}
\acrodef{TCI}{truncated-channel inversion}
\acrodef{MV}{majority vote}
\acrodef{CNN}{convolution neural network}
\acrodef{ReLU}{rectified-linear unit}
\acrodef{CSI}{channel state information}
\acrodef{PAPR}{peak-to-average power ratio}
\acrodef{SC}{single-carrier}
\acrodef{iid}[IID]{independent and identically distributed}
\acrodef{RMS}{root-mean-square}
\acrodef{4G}{Fourth Generation}
\acrodef{5G}{Fifth Generation}
\acrodef{NR}{New Radio}
\acrodef{LTE}{Long-Term Evolution}
\acrodef{DFT-s-OFDM}{\ac{DFT}-spread \ac{OFDM}}
\acrodef{OFDMA}{orthogonal frequency division multiple access}
\acrodef{HARQ}{hybrid automatic repeat request}
\acrodef{D2D}{Device-to-Device}
\acrodef{NOMA}{non-orthogonal multiple access}
\acrodef{OMA}{orthogonal multiple access}

\acrodef{IMT}{International Mobile Telecommunications}
\acrodef{ITU}{International Telecommunication Union}

\begin{document}

\title{A Survey on Over-the-Air Computation
	\\
	\thanks{
Alphan~\c{S}ahin and Rui~Yang are affiliated with the University of South Carolina, Columbia, SC and InterDigital, New York, NY, USA, respectively. E-mails: asahin@mailbox.sc.edu\IEEEauthorrefmark{1}, rui.yang@interdigital.com\IEEEauthorrefmark{2}.
}
	\author{Alphan~\c{S}ahin\IEEEauthorrefmark{1},~\IEEEmembership{Member,~IEEE} and Rui Yang\IEEEauthorrefmark{2},~\IEEEmembership{Member,~IEEE}} 
}


\maketitle

\begin{abstract}
Communication and computation are often viewed as separate tasks. This approach is very effective from the perspective of engineering as isolated optimizations can be performed. 
However, for many computation-oriented applications, the main interest is a function of the local information at the devices, rather than the local information itself.
 In such scenarios, information theoretical results show that harnessing the interference in a multiple access channel for computation, i.e., \ac{OAC}, can provide a significantly higher achievable computation rate than separating communication and computation tasks. Moreover,
 the gap between OAC and separation in terms of computation rate increases with more participating nodes. Given this motivation,  in this study, we provide a comprehensive survey on practical \ac{OAC} methods. After outlining fundamentals related to \ac{OAC}, we discuss the available \ac{OAC} schemes with their pros and cons. We provide an overview of the enabling mechanisms for achieving reliable computation in the wireless channel. Finally, we summarize the potential applications of \ac{OAC} and point out some future directions.
\end{abstract}


\acresetall

\section{Introduction}
\begin{figure}
	\centering
	\subfloat[Separation of communication and computation. {\color{\reviewColor}Computation occurs after receiving the arguments of the function at the ES.}]{\includegraphics[width =3.5in]{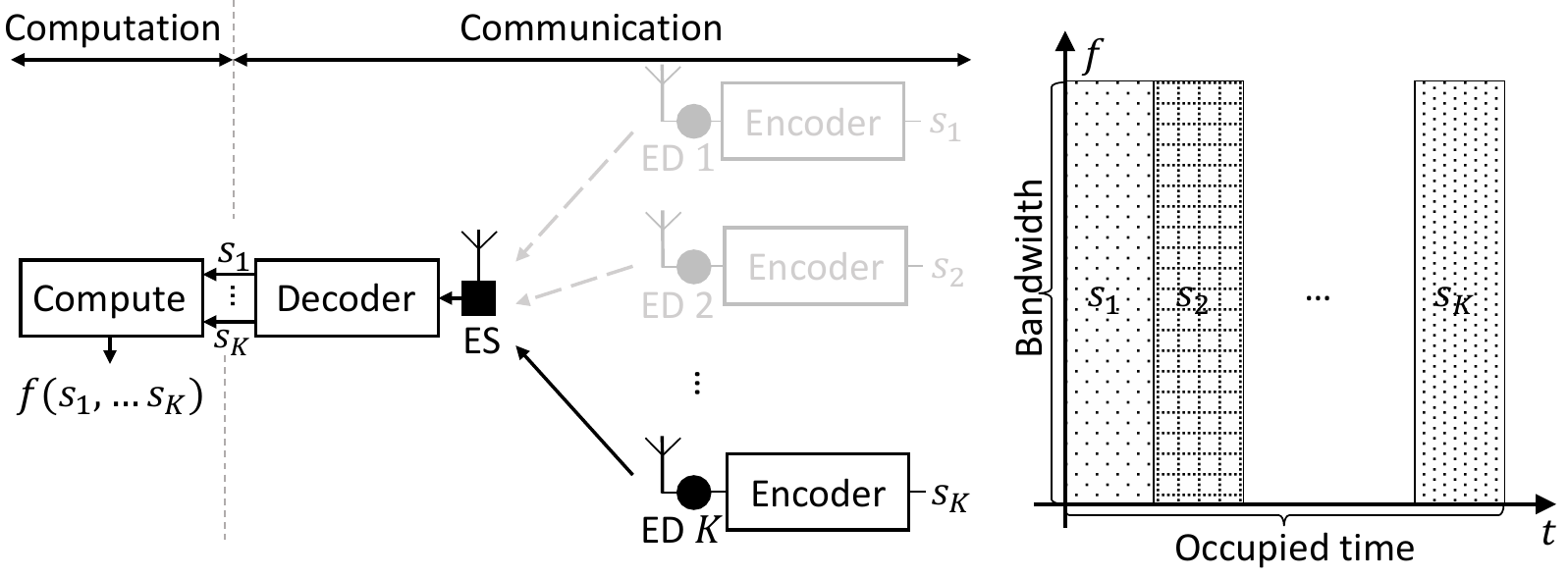}
		\label{subfig:sep}}
	\\	
	\subfloat[Computing function by using the signal superposition property of multiple access channels via simultaneous transmissions.]{\includegraphics[width =3.5in]{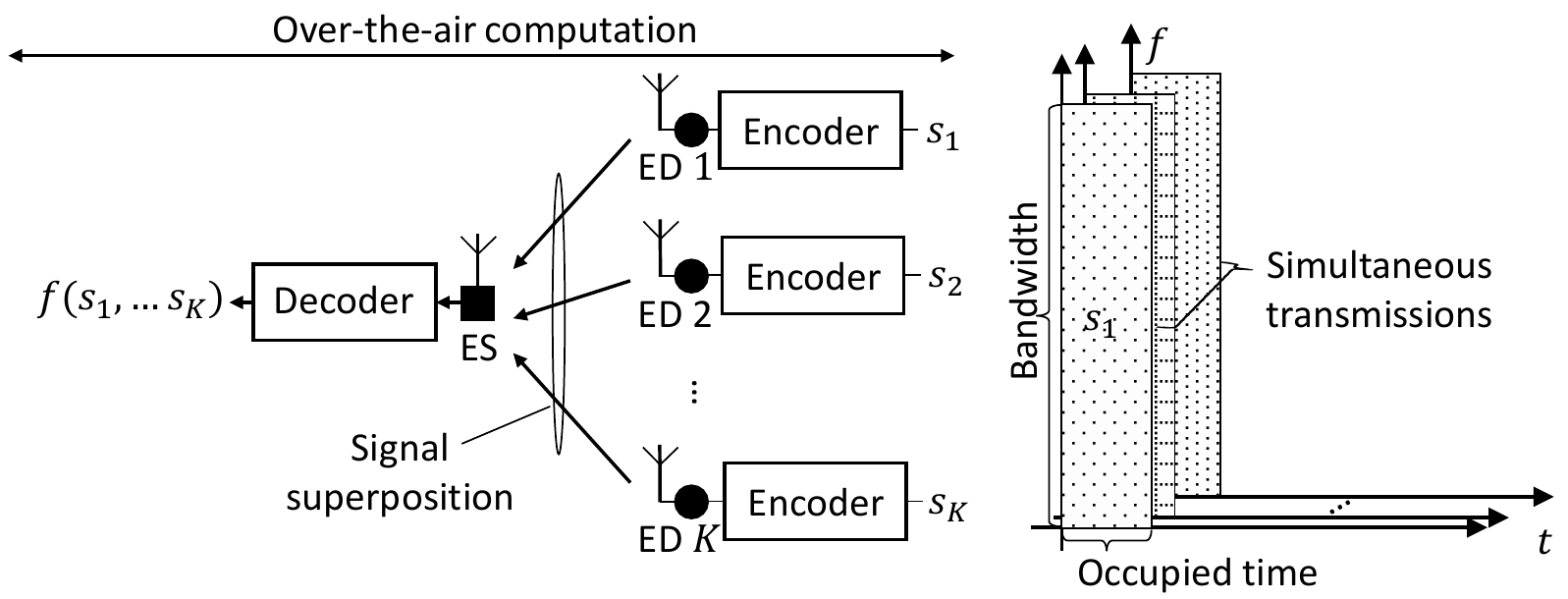}
		\label{subfig:join}}	
	\caption{Separation  of communication and computation versus OAC.}
	\label{fig:separationVsJoint}	
\end{figure}

\Ac{OAC} refers to the computation of mathematical functions by exploiting the signal superposition property of wireless multiple access channels.  The distinct feature of \ac{OAC} is that the local data at the   \acp{ED}, such as smartphones, laptops, tablets, vehicles, or sensors, are not acquired  over {\em orthogonal} channels to perform a computation task at a fusion node, e.g., an \ac{ES} at a base station or an access point.  Instead, the computation is handled by harnessing the interference via {\em simultaneous transmissions}. For example, suppose that the goal is to evaluate a function $\functionArbitrary[](\symbolVectorEle[1],\mydots,\symbolVectorEle[\numberOfEdgeDevices])$ at an \ac{ES}, where $\symbolVectorEle[\indexED]$ is the symbol at the $\indexED$th \ac{ED}. With the separation of communication and computation tasks, the function is computed at the fusion node after each symbol is received via {\color{\reviewColor}orthogonal or non-orthogonal resources (i.e., \ac{OMA} and \ac{NOMA})}, as illustrated for \ac{TDMA}  in \figurename~\ref{fig:separationVsJoint}\subref{subfig:sep}. On the other hand, with \ac{OAC}, the function is intended to be computed {\color{\reviewColor} through signal superposition in the channel as shown in \figurename~\ref{fig:separationVsJoint}\subref{subfig:join}. 
In this example, the key observation is that if the \ac{ES} is not interested in the local information but only in a function of them, \ac{OAC} paves the way for reducing resource usage, which otherwise scales with the number of \acp{ED}.} Hence, it is a fundamental and disruptive concept to the traditional way of handling computation and communication tasks independently. 

The idea of function computation over a multiple access channel was first thoroughly analyzed in Bobak's pioneering work in \cite{Nazer_2007} and the theoretical limits of computation over multiple access channels were investigated for a fixed many-to-one function. In \cite{goldenbaum2013harnessing}, Goldenbaum made the first connection between nomographic functions and \ac{OAC}. In \cite{Jeon_2014} and \cite{goldenbaum2015nomographic}, it was shown that \ac{OAC} can provide a significantly higher achievable computation rate than separating communication and computation. Given the promising information theoretical results,
\ac{OAC} has been drawing more and more attention in the literature.
Initially, it has been  applied to communication problems in the interference channel, e.g.,
physical layer network coding \cite{Katti_2007networkcoding,zhang_plnc2006}, compute-and-forward relaying strategy \cite{Nazer_2011}, and \acp{WSN} to address the issues like acceleration in gossip networks \cite{Nazer_2011gossip} and several computation tasks \cite{Mergen_2006tsp, Mergen_2007}.
With the increased interest in applications that require heavy computation, it has recently been utilized in multi-disciplinary fields such as machine learning over wireless networks \cite{zhu2021over}, wireless control systems \cite{Cai_2018}, and computing frameworks like wireless data centers \cite{Xiugang_2016}  and wireless intra-chip computations~\cite{Robert_2022}. 

The exciting applications have led to the investigation of \ac{OAC}  from various perspectives and resulted in a wide-variety of computation strategies.
This paper aims to discuss these \ac{OAC} schemes   without losing the mathematical rigor and how these methods hamd;e the challenges such as the detrimental impact of wireless channels on computation, synchronization errors, maintaining accurate and fresh \ac{CSI} at the radios, security, and hardware impairments such as power amplifier non-linearity. 


\subsection{Relation to other surveys and our contributions}

The reader can find relevant discussions on distributed inference over sensor networks in \cite{predd2006distributed}. The methods relying on compute-and-forward relaying scheme and uncoded strategies for physical layer network coding are comprehensively discussed and compared in \cite{LIEW20134,Nazer_2011survey}.
 To reduce the per-round communication latency for  the implementation of distributed learning over a wireless network, \ac{OAC} has been  used in many recent works as an enabler. We refer the readers interested in wireless systems for machine learning in general  to the excellent survey papers in \cite{park2019wireless,chen2021distributed,hellstrom2020wireless,AhmedSurvey_2022,Popovski_2021,Yiming_2022,Xiaowen_2022,Liu_2022flmetaSurvey} and the references therein.   
 In  \cite{Xiaowen_2022}, \ac{FEEL}, i.e., implementation of \ac{FL} \cite{pmlr-v54-mcmahan17a} over a wireless network, and the resource management for \ac{FEEL} are surveyed. 
 In \cite{zhu2021over}, several exciting applications of \ac{OAC} and  research directions in this area are discussed without mathematical details. 
In \cite{deniz2022_survey}, semantic communication is thoroughly surveyed and \ac{OAC} is mentioned as one of the task-oriented semantic communication paradigms. 
{\color{\reviewColor}In \cite{Guangxu_2022isacSurvey,Xiaoyang_2022surveyOACISCC}, \ac{OAC} is particularly analyzed from the perspective of integrated sensing, communication, and computation. In \cite{lee_2023visions}, over-the-air distributed computing for artificial intelligence applications is envisioned for 6G wireless  networks.}
In \cite{Altun_2021survey},  the particular interest is in the applications that enjoy signal superposition in general. Besides \ac{OAC}, the topics such as \ac{NOMA}, interference alignment, multiple antenna systems, security, and spectrum sensing are investigated. 
In \cite{Yamansavascilar_2022}, the design of aeronautical networks with computation paradigms such as edge computing and off-loading are surveyed. {\color{\reviewColor}We also acknowledge the reference \cite{Zhibin_2022oac} which discusses the \ac{OAC} from the perspective of various network architectures and provides an excellent survey on the OAC based on multiple antennas at the devices.\footnote{\color{\reviewColor}Our paper and \cite{Zhibin_2022oac} are independently developed and compensate each other from the perspective of classifications of available OAC approaches. The corresponding pre-prints were listed on arXiv.org one day apart (October 19, 2022).} }

The main focus of this study is {\color{\reviewColor} to investigate {\em how to compute a function over a wireless network reliably and efficiently}. Our priority is to form a composition that can provide a relative comparison of the state-of-the-art \ac{OAC} techniques with pros and cons,  particularly from the perspective of the physical layer of communication systems. Since a wide variety of applications can benefit from the \ac{OAC}, in this study, we focus on the computation itself, rather than a particular application.  We seek answers to three main questions:}
\begin{enumerate}
\item What functions can potentially be calculated with OAC? To answer this question, we review the nomographic functions that appear in both mathematics and communication literature.
\item What are the OAC schemes in the state-of-the-art and their trade-offs to deal with the distortion in wireless channels? To address this question, we first give a general system model  along with fundamental metrics on OAC. Under this framework, we evaluate the methods based on how they achieve computation under the fading channel and the encoding strategies.
\item What are the mechanisms that play a role in achieving a reliable OAC? To answer this question, we review the impacts of synchronization impairments, power management, and channel estimation on OAC and elaborate on security aspects and computation architectures for \ac{OAC}.
\end{enumerate}
Finally, we provide an overview of the applications of OAC in the literature and point out the potential areas that can be improved for \ac{OAC}.

{\em Organization:} The rest of the study is organized as follows. In Section~\ref{sec:whatCanBeCalculated}, we provide an overview of the fundamentals and discuss the functions that can potentially be computed via \ac{OAC}. In Section~\ref{sec:whatAreTheSchemes}, we discuss  the state-of-the-art \ac{OAC} schemes, comprehensively. 
In Section~\ref{sec:EnablingMech}, we discuss the enabling mechanisms to achieve a reliable computation. We summarize the potential applications of \ac{OAC} in various fields in Section~\ref{sec:WhatAreApps}. We finalize our discussions with various topics that need to be investigated further in Section~\ref{sec:WhatNeedsToBeImproved}.

{\em Notation:} The complex  and real numbers are denoted by $\complexNumbers$ and  $\realNumbers$, respectively. The $\numberOfEdgeDevices$-times Cartesian product of space $\aDomain$ is shown as $\aDomain^\numberOfEdgeDevices$. $\functionSpace[\aDomain]$ represents the space of every function that maps $\aDomain$ to $\realNumbers$. $\compactInterval$ denotes the unit interval $[0,1]$.
$\expectationOperator[\cdot][]$ denotes the expectation  over all random variables.
The function $\signNormal[\cdot]$ results in $1$, $-1$, or $0$ for a positive, a negative, or a zero-valued argument, respectively.  
The symbol $\circledast$ denotes linear convolution.
The function $\indicatorFunction[\cdot]$ results in $1$  if  its argument holds, otherwise it is $0$.  $\probability[\cdot]$ is the probability of an event. 
The zero-mean  multivariate  complex Gaussian distribution with the covariance matrix ${\textbf{\textrm{C}}_{\numberOfActiveSubcarriers}}$ of an $\numberOfActiveSubcarriers$-dimensional random column vector $\textbf{\textrm{x}}\in\complexNumbers^{\numberOfActiveSubcarriers}$ is denoted by
$\textbf{\textrm{x}}\sim\complexGaussian[\zeroVector[\numberOfActiveSubcarriers]][{\textbf{\textrm{C}}_{\numberOfActiveSubcarriers}}]$. $\gaussian[\meanG][\sigmaG^2]$ is the normal distribution with the mean $\meanG$ and the variance $\sigmaG^2$. The trace of a matrix is denoted by  $\traceOperator[\cdot]$. The continuous uniform distribution is denoted by $\uniformDistribution[a][b]$, where $a$ and $b$ are the minimum and the maximum values, respectively. The function $\log_2^{+}(x)$ is defined as $\max(\log_2(x),0)$. Kronecker delta is expressed as $\kroneckerDelta[i][j]$.

\section{What can be calculated with OAC?}
\label{sec:whatCanBeCalculated}
\ac{OAC} aims to compute a multivariate function  by relying on its representation that can structurally match with the underlying operation that  multiple access channel naturally performs.
In wireless communications, multiple access channels are modeled with additive property, i.e., the signal superposition. With this property,  the \ac{OAC} problem boils down to the representation of a target function with a special function, called a {\em nomographic} function, or a set of nomographic functions over multiple wireless resources. These  functions are called nomographic  because they are inline with the nomographs that solve certain equations through some graphs, i.e., analog computing. A well-known example of a nomograph is the Smith chart which assists in solving problems related to transmission lines. While the nomographs allow quick and accurate computations, the use cases of nomographs diminished historically due to the effectiveness of digital computers. Nevertheless, the fundamental theories about nomography are intricate, arguably connected to the neural networks, and pave the way for addressing the scenarios where digital computation suffers from latency, power consumption, and limited-communication bandwidth. In this section, we discuss the preliminaries on nomographic functions to reveal what can be calculated with \ac{OAC}.

\subsection{Preliminaries}
\label{subsec:problemStatement}
%
%
%

%
%

\begin{definition}[Nomographic function \cite{goldenbaum2013harnessing,Goldenbaum_2012icassp,Goldenbaum_ISWCS2011,Goldenbaum_2012CISS}]
\rm 
Let $\aCompactMetricSpace^{\numberOfEdgeDevices}$, $\numberOfEdgeDevices\ge2$, be a compact metric space. A function $\functionArbitrary[]:\aCompactMetricSpace^{\numberOfEdgeDevices}\rightarrow\realNumbers$ for which there exist functions $\preProcessingFunctionW[\indexED]\in\functionSpace[\aCompactMetricSpace]$, $\indexED\in\{1,\mydots,\numberOfEdgeDevices\}$, and  $\postProcessingFunctionW[]\in\functionSpace[\realNumbers]$ such that $\functionArbitrary[]$ can be represented as 
\begin{align}
	\functionArbitrary[](\symbolVectorEle[1],\symbolVectorEle[2],\dots, \symbolVectorEle[\numberOfEdgeDevices])=\postProcessingFunction[][{\sum_{\indexED=1}^{\numberOfEdgeDevices}\preProcessingFunction[\indexED][{\symbolVectorEle[{\indexED}]}]}]~,
	\label{eq:nomographicfcn}
\end{align}
is called nomographic function and $\spaceNomographic[\aCompactMetricSpace^{\numberOfEdgeDevices}]$ is the space of nomographic functions with the domain  $\aCompactMetricSpace^{\numberOfEdgeDevices}$. 
\label{def:nomog}
\end{definition}

The functions $\preProcessingFunctionW[\indexED]$, $\forall\indexED$, and the function $\postProcessingFunctionW[]$ are further called  pre-processing functions (or inner functions) and post-processing function (or outer function), respectively. Equation \eqref{eq:nomographicfcn} reveals why a nomographic function is relevant to \ac{OAC}: Equation \eqref{eq:nomographicfcn} can be interpreted as an evaluation of the function $\functionArbitrary[]$ in an ideal \ac{UL} channel (i.e., no noise, no multi-path channel distortion), where $\symbolVectorEle[{\indexED}]$ and $\preProcessingFunctionW[\indexED]$ are the symbols and the pre-processing functions at $\indexED$th data-generating node, respectively, the sum of the signals from $\numberOfEdgeDevices$ nodes corresponds to the superposition that naturally occurs in the channel, and $\postProcessingFunctionW[]$ is the post-processing function at the fusion center. To the best of our knowledge, this connection is first made in Goldenbaum's work in \cite{goldenbaum2013harnessing,Goldenbaum_2012icassp,Goldenbaum_ISWCS2011,Goldenbaum_2012CISS} while the non-linear function examples in the form of \eqref{eq:nomographicfcn} appear in \cite{Goldenbaum_2009wcnc,Goldenbaum2010asilomar,Goldenbaum_2013tcom} without discussing the family of nomographic functions.

It is worth noting that the compactness mentioned in Definition~\ref{def:nomog} is an important assumption, especially in the analysis of continuous functions. For example, the range of a continuous function $\functionArbitrary[](\symbolVectorEle[1],\symbolVectorEle[2],\dots, \symbolVectorEle[\numberOfEdgeDevices])$ on a compact space $\aCompactMetricSpace^{\numberOfEdgeDevices}$ is compact. Since the function is bounded, one can ensure that the limits exist, or that suprema and infima are taken by the function. If the space is not compact, it can be harder to analyze the behavior of a given function and more structural properties related to the function need to be known. From the perspective of \ac{OAC}, compactness is inherited due to practical limitations. For instance, the measure space of a sensor is typically compact because a sensor can quantify values in a finite closed interval, e.g., $0^\circ \text{C} \leq \symbolVectorEle[\indexED]\leq 100^\circ \text{C}$, $\forall\indexED$. Hence, to make general statements about entire function spaces and not only about specific examples,  the space $\aCompactMetricSpace^{\numberOfEdgeDevices}$  in Definition~\ref{def:nomog} is  considered to be compact.

 Now, let us denote the space of nomographic functions, the space of nomographic functions with the restriction of continuous pre- and post-processing functions, and the space of continuous functions  with the domain $\compactInterval^{\numberOfEdgeDevices}$ as $\spaceNomographic[\compactInterval^{\numberOfEdgeDevices}]$, $\spaceContinuousNomographic[{\compactInterval^{\numberOfEdgeDevices}}]$, and $\spaceContinuous[{\compactInterval^{\numberOfEdgeDevices}}]$, respectively.\footnote{Nomographic functions in mathematics are often investigated by defining the compact space $\aCompactMetricSpace$ as $\compactInterval$.} Sprecher and Buck  provide insights into the representation of a function $\functionArbitrary[]\in\spaceContinuous[{\compactInterval^{\numberOfEdgeDevices}}]$ as a nomographic function as follows:
\begin{theorem}[Sprecher'65 \cite{Sprecher_1965}]
	\rm 
	Every function $\functionArbitrary[]\in\spaceContinuous[{\compactInterval^{\numberOfEdgeDevices}}]$ can be represented with real, monotonic increasing pre-processing functions and possibly a discontinuous post-processing function.
\end{theorem}
\begin{theorem}[Buck'79 \cite{buck_1979}]\rm
	Every function $\functionArbitrary[]\in\functionSpace[{\compactInterval^{\numberOfEdgeDevices}}]$ is nomographic (i.e. $\spaceNomographic[\compactInterval^{\numberOfEdgeDevices}]=\functionSpace[{\compactInterval^{\numberOfEdgeDevices}}]$).
	\label{th:allFunctionsNomo}
\end{theorem}
The key idea for the proof of Theorem~\ref{th:allFunctionsNomo} is to show there exists a one-to-one mapping  from $\compactInterval^{\numberOfEdgeDevices}$ to a space $\aMetricSpace\subset\realNumbers$ in the form of $\forwardMap(\symbolVectorEle[1],\mydots,\symbolVectorEle[\numberOfEdgeDevices])={\sum_{\indexED=1}^{\numberOfEdgeDevices}\preProcessingFunction[\indexED][{\symbolVectorEle[{\indexED}]}]}$. Given the existence of such $\forwardMap$ (therefore, the pre-processing functions exist), the post-processing function can then be expressed as $\postProcessingFunctionW[](x)=\functionArbitrary[](\inverseMap(x))$, where $\inverseMap$ is the inverse function that maps $x\in\aMetricSpace$ to $(\symbolVectorEle[1],\mydots,\symbolVectorEle[\numberOfEdgeDevices])$. Without any restriction on the pre-functions and the post-processing function, such a map can be obtained by choosing $\aMetricSpace=\compactInterval$ and constructing the binary representation  of $x\in\aMetricSpace$ by uniformly interleaving the digits of the binary representations of the symbol $\symbolVectorEle[\indexED]$, $\forall\indexED$ (see \cite[p. 287]{buck_1979} and \cite[p. 2]{matthiasthesis2023}). For this specific constructive proof, $\preProcessingFunctionW[\indexED]$ relies on reading the binary representation of $\symbolVectorEle[\indexED]$ in base $2^\numberOfEdgeDevices$, which implicitly causes discontinuity in its range. The proof  also shows the existence of special nomographic functions with an interesting property:
\begin{definition}[Universality]
\rm
The pre-processing functions are {\em universal} if they are fixed and can be used to calculate every function in $\functionSpace[{\compactInterval^{\numberOfEdgeDevices}}]$.
\end{definition}
The universality is a desirable property for \ac{OAC} because the pre-processing functions do not need to be re-designed (i.e., less communication overhead)  if the target function changes over time. This property is exploited in  \cite{goldenbaum2013harnessing,Goldenbaum_2012icassp} for multi-cluster computation as discussed in Section~\ref{subsec:arch}. It is also mentioned that universality provides robustness against changes in network topology (via dropping and joining devices) in the sense that transmitting nodes do not need to adapt
their pre-processing functions.

If one desires the pre- and post-processing functions to be continuous for an arbitrary continuous function $\functionArbitrary[]$, Theorem~\ref{th:allFunctionsNomo} is unfortunately not valid:
\begin{theorem}[Buck'82 \cite{buck_1982}]\rm 
	$\spaceContinuousNomographic[{\compactInterval^{\numberOfEdgeDevices}}]$ is nowhere dense in $\spaceContinuous[{\compactInterval^{\numberOfEdgeDevices}}]$.
	\label{th:nodense}
\end{theorem}

A canonical example of Theorem~\ref{th:nodense} is geometric mean, i.e., $\functionArbitrary[](\symbolVectorEle[1],\symbolVectorEle[2],\dots, \symbolVectorEle[\numberOfEdgeDevices])=(\prod_{\indexED}\symbolVectorEle[\indexED])^{\frac{1}{\numberOfEdgeDevices}}$. This function cannot be represented as $\postProcessingFunction[][{\sum_{\indexED=1}\preProcessingFunction[\indexED][{\symbolVectorEle[{\indexED}]}]}]$
with the continuous functions $\preProcessingFunctionW[1],\mydots,\preProcessingFunctionW[\numberOfEdgeDevices],\postProcessingFunctionW[]$ on $\compactInterval$  as demonstrated for $\numberOfEdgeDevices=2$ by Arnold \cite{Arnold1957} and for an arbitrary $\numberOfEdgeDevices$ by Goldenbaum \cite{goldenbaum2013harnessing}. Theorem~\ref{th:nodense}  implies that  there exist infinite number of continuous functions in $\spaceContinuous[{\compactInterval^{\numberOfEdgeDevices}}]$ that cannot be approximated with a nomographic function in $\spaceContinuousNomographic[{\compactInterval^{\numberOfEdgeDevices}}]$ for a given arbitrary precision. Kolmogorov remarkably addresses the issue of representing a continuous function with a set of nomographic functions in $\spaceContinuousNomographic[{\compactInterval^{\numberOfEdgeDevices}}]$:
\begin{theorem}[Kolmogorov'57 \cite{kolmogorov:superposition}]
	\rm 
Every function  $\functionArbitrary[]\in\spaceContinuous[{\compactInterval^{\numberOfEdgeDevices}}]$ can be represented as the superposition of at most $2\numberOfEdgeDevices+1$ nomographic functions in $\spaceContinuousNomographic[{\compactInterval^{\numberOfEdgeDevices}}]$, i.e., 
\begin{align}
	\functionArbitrary[](\symbolVectorEle[1],\symbolVectorEle[2],\dots, \symbolVectorEle[\numberOfEdgeDevices])=\sum_{\indexKolmogorovSum=1}^{2\numberOfEdgeDevices+1}\postProcessingFunction[\indexKolmogorovSum][{\sum_{\indexED=1}^{\numberOfEdgeDevices}\preProcessingFunction[\indexED\indexKolmogorovSum][{\symbolVectorEle[{\indexED}]}]}]~,
	\label{eq:kolmogorov}
\end{align}
where the post-processing functions $\postProcessingFunctionW[\indexKolmogorovSum]$ depend on $\functionArbitrary[]$ and the functions $\preProcessingFunctionW[\indexED\indexKolmogorovSum]$ are independent of $\functionArbitrary[]$.
\label{th:kolmogorovSuper}
\end{theorem}
Geometrically, the $2\numberOfEdgeDevices+1$ inner sums in \eqref{eq:kolmogorov} ensure the existence of a continuous and bijective correspondence between $(\symbolVectorEle[1],\dots, \symbolVectorEle[\numberOfEdgeDevices])\in\compactInterval^\numberOfEdgeDevices$ and $(\postProcessingFunction[1][{\sum_{\indexED=1}\preProcessingFunction[\indexED\indexKolmogorovSum][{\symbolVectorEle[{\indexED}]}]}],\mydots,\postProcessingFunction[2\numberOfEdgeDevices+1][{\sum_{\indexED=1}\preProcessingFunction[\indexED\indexKolmogorovSum][{\symbolVectorEle[{\indexED}]}]}])\in\realNumbers^{2\numberOfEdgeDevices+1}$. Hence, the inner sums describe a homeomorphism that continuously embeds $\compactInterval^\numberOfEdgeDevices$ into $\realNumbers^{2\numberOfEdgeDevices+1}$.  In \cite{Sternfeld_1985}, Sternfeld enhances the statement of  Theorem~\ref{th:kolmogorovSuper} by showing that the $2\numberOfEdgeDevices+1$ nomographic functions in \eqref{eq:kolmogorov} cannot be reduced to represent every $\functionArbitrary[]\in\spaceContinuous[{\compactInterval^{\numberOfEdgeDevices}}]$. Hence, from the perspective of \ac{OAC}, Theorem~\ref{th:kolmogorovSuper} implies that at least $2\numberOfEdgeDevices+1$ wireless resources need to be allocated where each resource is dedicated to a nomographic function in $\spaceContinuousNomographic[{\compactInterval^{\numberOfEdgeDevices}}]$ to calculate every function in $\spaceContinuous[{\compactInterval^{\numberOfEdgeDevices}}]$. 

In mathematics, Theorem~\ref{th:kolmogorovSuper}, also known as Kolmogorov's superposition or Kolmogorov-Arnold representation theorem, is notable because it solves a more constrained (i.e., the function $\functionArbitrary[]$ needs to be continuous), but a more general form (i.e., the superposition of only one variable functions) of Hilbert's the thirteenth problem  in \cite{Hilbert_1902}. 
There are also other variants of Kolmogorov's superposition and constructive proofs that show how to obtain the pre- and post-processing functions.  For a comprehensive discussion on the variants and constructions, we refer the reader to \cite[Chapter 2]{Sprecher_1965single}. A variant that is mentioned in the \ac{OAC} literature \cite{goldenbaum2013harnessing} is  as follows:
\begin{theorem}[Braun'09 \cite{jurgen_2009}]
	\rm 
	For every function  $\functionArbitrary[]\in\spaceContinuous[{\compactInterval^{\numberOfEdgeDevices}}]$, there exist $2\numberOfEdgeDevices+1$ nomographic functions in $\spaceContinuousNomographic[{\compactInterval^{\numberOfEdgeDevices}}]$ such that
	\begin{align}
		\functionArbitrary[](\symbolVectorEle[1],\symbolVectorEle[2],\dots, \symbolVectorEle[\numberOfEdgeDevices])=\sum_{\indexKolmogorovSum=1}^{2\numberOfEdgeDevices+1}\postProcessingFunction[\indexKolmogorovSum][{\sum_{\indexED=1}^{\numberOfEdgeDevices}\coefa_{\indexED}\preProcessingFunction[][{\symbolVectorEle[{\indexED}]+(\indexKolmogorovSum-1)\coefb}]}]~,
		\label{eq:kolmogorovVar}
	\end{align}
	where the pre-processing function $\preProcessingFunctionW[]$ is a well-defined, continuous, monotone, and independent of $\functionArbitrary[]$, the coefficients $\coefa_{\indexED}$, $\forall\indexED$, and $\coefb$ are appropriate non-negative real constants. 
	\label{th:kolmogorovSuperVariant}
\end{theorem}
The key observation made in \cite{goldenbaum2013harnessing} based on Theorem~\ref{th:kolmogorovSuperVariant} is that to calculate every function in $\spaceContinuous[{\compactInterval^{\numberOfEdgeDevices}}]$ with continuous nomographic functions over $2\numberOfEdgeDevices+1$ resources, the pre-processing functions can be designed to be universal. 
Note that the superposition in \eqref{eq:kolmogorovVar} involves $2\numberOfEdgeDevices+1$ post-processing function and one single pre-processing function. In the literature, it is shown that the superposition  can also be expressed with a single pre-processing function and a single post-processing function as discussed in \cite[Theorem 1]{Sprecher_1965single}  and \cite[Theorem 2.14]{Braunthesis2009} by introducing a shift to the arguments of the post-processing functions in \eqref{eq:kolmogorovVar}. Also, Kolmogorov's superposition can be interpreted as a special feed-forward neural network and is useful to predict the complexity of neural networks (see the discussions in \cite{Girosi_1989, Kurkova_1991,SCHMIDTHIEBER2021119}).

In some cases, it may be desirable not to consume $2\numberOfEdgeDevices+1$ wireless resources to calculate a specific continuous function with $2\numberOfEdgeDevices+1$ continuous nomographic functions. In this case, one may follow one of two different directions: Manipulating the domain of the target function or constructing a nomographic function that approximates the target function. In the first approach, some part of the domain is cut out so that the nomographic function can be calculated with continuous pre- and post-processing functions. For instance, if $\aCompactMetricSpace$ is chosen as $[\asmallNumber,1]$ for $\asmallNumber\in(0,1)$, the geometric mean can be calculated with a nomographic function with  $\preProcessingFunction[\indexED][x]=\ln(x)$, $\forall\indexED$, and $\postProcessingFunction[][x]=\constante^{x/\numberOfEdgeDevices}$ on $\aCompactMetricSpace$. 
In the second approach, a nomographic approximation can be defined as follows \cite{goldenbaum2013harnessing}:
\begin{definition}[Nomographic approximation] \rm
Let $\precision>0$ be an arbitrary constant. The space of approximable nomographic functions with respect to the precision $\precision$ is defined by
\begin{align}
\spaceContinuousAprxNomographic[\compactInterval^\numberOfEdgeDevices]&\triangleq\Bigg\{\functionArbitrary[]\in\functionSpace[\compactInterval^\numberOfEdgeDevices]|\exists(\preProcessingFunctionW[1],\mydots,\preProcessingFunctionW[\numberOfEdgeDevices],\postProcessingFunctionW[])\in\spaceContinuous[\compactInterval]\times\mydots\nonumber\\&\hspace{-9mm}\mydots\spaceContinuous[\compactInterval]\times \spaceContinuous[\realNumbers]:
\norm*{\functionArbitrary[]-\postProcessingFunction[][{\sum_{\indexED=1}^{\numberOfEdgeDevices}\preProcessingFunction[\indexED][{\symbolVectorEle[{\indexED}]}]}]}_\infty
\le\precision\Bigg\}~.
\end{align}
If $\functionArbitrary[]\in\spaceContinuousAprxNomographic[\compactInterval^\numberOfEdgeDevices]$, we write	$\functionArbitrary[](\symbolVectorEle[1],\dots, \symbolVectorEle[\numberOfEdgeDevices])\approx\postProcessingFunction[][{\sum_{\indexED=1}\preProcessingFunction[\indexED][{\symbolVectorEle[{\indexED}]}]}]$.
\label{def:appxNomo}
\end{definition}

For example, under Definition~\ref{def:appxNomo}, the geometric mean on  $\compactInterval^\numberOfEdgeDevices$ is a function in $\spaceContinuousAprxNomographic[\compactInterval^\numberOfEdgeDevices]$ because it can be approximated with $\preProcessingFunction[\indexED][x]=\ln(x+1/\precisionF)$ and $\postProcessingFunction[][x]=\constante^{x/\numberOfEdgeDevices}$ for $\precisionF>0$. Nevertheless, a complete characterization of the approximate nomographic functions is still an area that requires more investigation as it is possible to define the space of approximable nomographic functions in different ways. For instance, in \cite[Eq. (5)]{Frey_tsp2021}, an approximate nomographic function is defined in a stochastic manner. For further theoretical investigations on approximate nomography, the reader is also referred to \cite{Khavinson1996BestAB,Limmer_2015, Bjelakovic_2019allerton,Frey_tsp2021}.

Another interesting function space is the class of symmetric functions elaborated in \cite{Giridhar_2005}:
\begin{definition}[Symmetric function]
	\rm 
Let $\permutation:\aCompactMetricSpace^{\numberOfEdgeDevices}\rightarrow\aCompactMetricSpace^{\numberOfEdgeDevices}$ denotes a permutation.
 A  function  $\functionArbitrary[]:\aCompactMetricSpace^{\numberOfEdgeDevices}\rightarrow\realNumbers$ that is invariant with respect to permutations  of its arguments, i.e.,
\begin{align}
	\functionArbitrary[](\symbolVectorEle[1],\dots, \symbolVectorEle[\numberOfEdgeDevices])=\functionArbitrary[](\permutation(\symbolVectorEle[1],\dots, \symbolVectorEle[\numberOfEdgeDevices])),~{\forall} \permutation
	\label{eq:nomographicfcnsym}
\end{align}
is called a symmetric function.
\end{definition}
A distinct feature of the space of symmetric function is that only the data itself is important, rather than its origin. From an application standpoint, many functions such as mean, maximum, minimum, median, and \ac{MV} that have either exact or approximate nomographic functions belong to this class. The second important feature is that the functions in this space can be calculated through the type function, i.e., frequency histogram \cite{Giridhar_2005,Giridhar_2006,Mergen_2006tsp,Mergen_2007}. Type function can be defined as multiple weighted arithmetic sums of indicator functions, i.e.,  counting the number of devices based on a certain set, which is also investigated under \ac{TBMA} in \cite{Mergen_2006tsp,Mergen_2007}.

\subsection{Common nomographic functions}
\label{subsec:commonNomo}
In \tablename~\ref{table:nomoFcns}, we list several exact and approximate nomographic functions discussed in the literature. While arithmetic mean, weighted sum, and \ac{MV} are used in distributed learning applications, modulo-2 sum often appears in physical layer network coding. The product operation is used for key generation in \cite{Altun_2022}. The maximum, minimum, and counting functions are used in \ac{WSN} (e.g. generating an alert if the temperature rises) or to calculate histogram (e.g., calculating measurement statistics with \ac{TBMA} \cite{Mergen_2006tsp,Mergen_2007}). Geometric mean, $p$-norm, and polynomial functions are often mentioned to provide nomographic function  examples that can be calculated over a wireless network. In particular, $p$-norm is used for  computing average-pooling or max-pooling over the air in \cite{Zhiyan_2023arxivSensing}.

An interesting direction is to calculate  an approximate nomographic function with a continuous and monotone post-processing function and continuous pre-processing functions for a given continuous function. In \cite{Limmer_2015}, an approximation is obtained by using a combination of a
dimensionwise function decomposition. In this approach, the target function is skewed with a bijective function such that the resulting function can be approximated well with a first-order \ac{ANOVA} decomposition. To calculate the skew function, Bernstein polynomials are used. It is worth noting that  Bernstein polynomials can be utilized to constructively prove the Weierstrass approximation theorem that states every continuous function can be approximated with an arbitrary precision over any finite interval by a polynomial of a sufficient order. 

Another interesting direction is to calculate the target function by expressing it as a solution to an optimization problem and solving the problem through  iterations that can be expressed with some elementary nomographic functions. For example, as discussed in Section~\ref{subsub:byzantine}, the geometric median can be calculated through iterations over-the-air by using the Weiszfeld algorithm in \cite{huang2021byzantineresilient}. In \cite{abari_2016oac}, \cite{chen_2018jiot}, and \cite{Agrawal_2019spawc}, by using the binary representations of the parameters, several non-linear functions, e.g., maximum or minimum, are proposed to be calculated through the communication rounds. For instance, to calculate the maximum of the parameters, in the first round, the ES inquires to the EDs with  bit $1$ in the most significant bit position of the binary representation of the parameter. If there is any response to the inquiry, the ES detects that the most significant bit of the maximum of the parameters is $1$; otherwise, it is $0$. In the second round,  if the most significant bit is detected as $1$, the ES inquires to the EDs with  bit $1$ in both the most and the second significant bits. Otherwise, the ES inquires to  all EDs with bit $1$ in the second significant bit position. From the responses to the second inquiry, the ES determines the second significant bit. The procedure continues until the least significant bit is detected. The same procedure can be used for computing minimum function by using the reciprocal of the parameters. The reader is also referred to \cite{Giridhar_2006} for successive partitions to compute functions. 

\begin{table}
	\centering
\caption{Example nomographic functions.}	
\resizebox{\textwidth*2/4-0.1in}{!}{
	\begin{tabular}{l|c|c|c|c}
		Description  & $\functionArbitrary[](\symbolVectorEle[1],\symbolVectorEle[2],\dots, \symbolVectorEle[\numberOfEdgeDevices])$ & $\preProcessingFunctionW[\indexED](x)$ & $\postProcessingFunctionW[](x)$   \\
		\hline\hline \begin{tabular}{@{}l@{}}Arithmetic mean\end{tabular} & $\displaystyle\frac{1}{\numberOfEdgeDevices}\sum_{\indexED=1}^\numberOfEdgeDevices\symbolVectorEle[\indexED]$ & $x$ & $\frac{x}{\numberOfEdgeDevices}$  \\
		\hline\hline \begin{tabular}{@{}l@{}}Weighted sum\end{tabular} & $\displaystyle\sum_{\indexED=1}^\numberOfEdgeDevices w_\indexED \symbolVectorEle[\indexED]$, $w_\indexED \in\realNumbers$ & $ w_\indexED x$ & $x$  \\		
		\hline \begin{tabular}{@{}l@{}}Polynomial function \end{tabular}& $\displaystyle\sum_{\indexED=1}^{\numberOfEdgeDevices} \coefPoly[\indexED] \symbolVectorEle[\indexED]^{\indexED-1}$, $\coefPoly[\indexED]\in\realNumbers$ & $ \coefPoly[\indexED-1] x^{\indexED-1}$ & $x$  \\	
		\hline \begin{tabular}{@{}l@{}}Majority vote \end{tabular}& $\displaystyle\signNormal[{\sum_{\indexED=1}^{\numberOfEdgeDevices}\signNormal[{\symbolVectorEle[\indexED]}]}]$ & $\signNormal[x]$ & $\signNormal[x]$\\		
		\hline \begin{tabular}{@{}l@{}}Counting number of EDs \\ with the class $\mathcal{C}$ \end{tabular}& $\displaystyle{\sum_{\indexED=1}^{\numberOfEdgeDevices}\indicatorFunction[{\symbolVectorEle[\indexED]}\in\mathcal{C}]}$ & $x$ & $\indicatorFunction[x\in\mathcal{C}]$\\		
		\hline \begin{tabular}{@{}l@{}}$\pForNorm$-norm\end{tabular} & $\displaystyle\left({\sum_{\indexED=1}^\numberOfEdgeDevices|\symbolVectorEle[\indexED]|^{\pForNorm}}\right)^{1/\pForNorm}$ & $|x|^{\pForNorm}$ & $x^\frac{1}{\pForNorm}$ \\		
		\hline \begin{tabular}{@{}l@{}}Modulo-2 sum\end{tabular} & $\displaystyle\symbolVectorEle[1]\oplus\symbolVectorEle[2]\mydots\oplus\symbolVectorEle[\numberOfEdgeDevices]$, $\symbolVectorEle[\indexED]\in\integers_2$ & ${x}$ & $x\mod2$\\			
		\hline \begin{tabular}{@{}l@{}} Approximation of \\ the product	\end{tabular} & $\displaystyle{\prod_{\indexED}\symbolVectorEle[\indexED]},\symbolVectorEle[\indexED]\ge0$ & $\ln\left(x+\frac{1}{\precisionF}\right)$ & $\constante^{x}$ \\		
		\hline \begin{tabular}{@{}l@{}} Approximation of \\ the geometric mean	\end{tabular} & $\displaystyle\left({\prod_{\indexED}\symbolVectorEle[\indexED]}\right)^{1/\numberOfEdgeDevices},\symbolVectorEle[\indexED]\ge0$ & $\ln\left(x+\frac{1}{\precisionF}\right)$ & $\constante^{\frac{x}{\numberOfEdgeDevices}}$ \\
		\hline \begin{tabular}{@{}l@{}} Approximation of the \\  cosine of the product	\end{tabular}  & $\displaystyle\cos\left({\prod_{\indexED}\symbolVectorEle[\indexED]}\right),\symbolVectorEle[\indexED]\ge0$ & $\ln\left(x+\frac{1}{\precisionF}\right)$ & $\cos(\constante^{x})$ \\		
		\hline\begin{tabular}{@{}l@{}} Approximation of \\  the maximum	\end{tabular} & $\displaystyle\max_{\indexED}\{\symbolVectorEle[\indexED]\},\symbolVectorEle[\indexED]\ge0$ & $x^{\precisionF}$ & $x^{\frac{1}{\precisionF}}$ \\
		\hline \begin{tabular}{@{}l@{}} Approximation of \\  the minimum\end{tabular} & $\displaystyle\min_{\indexED}\{\symbolVectorEle[\indexED]\},\symbolVectorEle[\indexED]\ge0$ & ${x^{-\precisionF}}$ & $x^{-\frac{1}{\precisionF}}$\\
		\hline
	\end{tabular}
}
\label{table:nomoFcns}
\end{table}

\section{What are the OAC schemes?}
\label{sec:whatAreTheSchemes}
\begin{figure*}[t]
	\centering
	{\includegraphics[width =5.5in]{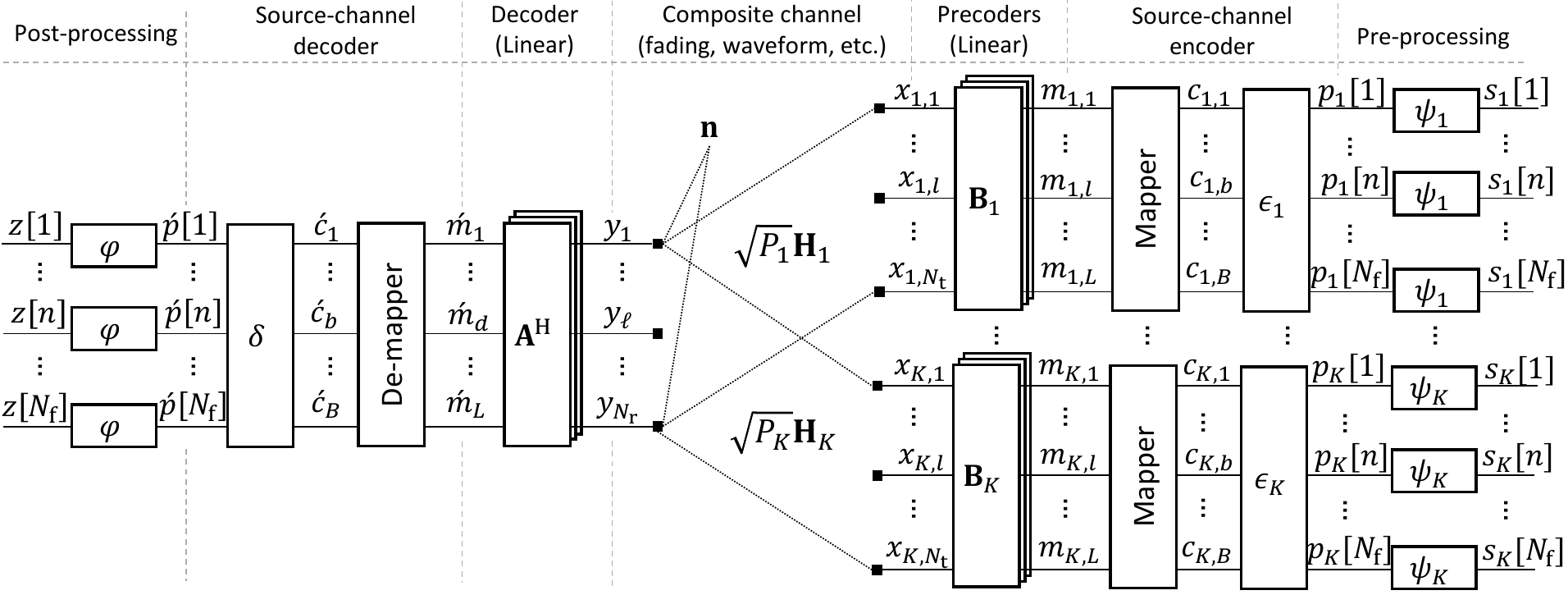}
	} 
	\caption{A generalized model for an \ac{OAC} method. The domains of the transmitted vectors $\transmittedVector[\indexED],\forall\indexED$, and the received vector $\superposedVector$ can be time, frequency, space, etc, depending on the \ac{OAC} method.} 
	\label{fig:BlockDiagram}
\end{figure*}
An \ac{OAC} scheme aims to realize \eqref{eq:nomographicfcn} (or \eqref{eq:kolmogorov})  over a wireless \ac{MAC} with a fidelity criterion.  For a rigorous classification and a generalization of the \ac{OAC} schemes in the state-of-the-art, consider an \ac{OAC} scheme that targets to calculate the nomographic function $	\postProcessedVectorEle[\indexSampleTime]=\functionArbitrary[](\symbolVector[\indexSampleTime])\triangleq	\functionArbitrary[](\symbolVectorEle[1][\indexSampleTime],\dots, \symbolVectorEle[\numberOfEdgeDevices][\indexSampleTime])$ for the symbol vector $\symbolVector[\indexSampleTime]=[\symbolVectorEle[1][\indexSampleTime],\dots, \symbolVectorEle[\numberOfEdgeDevices][\indexSampleTime]]^{\rm T}$, $\forall\indexSampleTime\in\{1,\mydots,\numberOfFunctions\}$, $\numberOfFunctions\ge1$. 
Let $\symbolVectorED[\indexED]=[\symbolVectorEle[\indexED][1],\mydots,\symbolVectorEle[\indexED][\numberOfFunctions]]^{\rm T}$ and $\preProcessedVector[\indexED]=[\preProcessedVectorEle[\indexED][1],\mydots,\preProcessedVectorEle[\indexED][\numberOfFunctions]]^{\rm T}$ denote the symbol vector  and the pre-processed symbol vector   at the $\indexED$th \ac{ED} for  $\preProcessedVectorEle[\indexED][\indexSampleTime]=\postProcessingFunction[\indexED][{\symbolVectorEle[\indexED][\indexSampleTime]}]\in\realNumbers$, $\forall\indexSampleTime$. The $\indexED$th \ac{ED} calculates the encoded vector $\encodedVector[\indexED]\in\complexNumbers^{\blockLength}$ as 
\begin{align}
	\encodedVector[\indexED] = [\encodedVectorEle[1],\mydots,\encodedVectorEle[\blockLength]]^{\rm T} =\encoder[\indexED]\left\{\preProcessedVector[\indexED]\right\}~,
\end{align}
where $\encoder[\indexED]:\realNumbers^{\numberOfFunctions}\rightarrow\complexNumbers^{\blockLength}$ is the encoder (e.g., source encoder, channel encoder, constellation mapping, or a combination of these operations) and $\blockLength$ is the number of modulation symbols in a complex-valued codeword. 

Let $\mapperResource$ be a resource mapper that maps $\blockLength$ modulation symbols to the $\blockLength$ available resources. Now,  consider $\numberOfModulationSymbols$ modulation symbols among $\blockLength$ symbols, denoted by $\modulationVector[\indexED]=[\modulationVectorEle[\indexED,1],\mydots,\modulationVectorEle[\indexED,\numberOfModulationSymbols]]^{\rm T}$, that are processed with a linear precoder $\precoder[\indexED]\in\complexNumbers^{\numberOfDimensionAtTX\times\numberOfModulationSymbols}$ as
\begin{align}
\transmittedVector[\indexED] = [\transmittedVectorEle[\indexED,1],\mydots,\transmittedVectorEle[\indexED,\numberOfDimensionAtTX]]^{\rm T} =\precoder[\indexED]\modulationVector[\indexED]~,
\label{eq:transmittedSymbols}
\end{align}
where $\transmittedVector[\indexED]\in\complexNumbers^{\numberOfDimensionAtTX}$ is the transmitted symbols from the $\indexED$th \ac{ED} over $\numberOfDimensionAtTX$ dimensions. Hence, each \ac{ED} applies $\numberOfTransforms\triangleq\ceil{\blockLength/\numberOfModulationSymbols}$ linear precoders to the modulation symbols in total.

The received vector at the \ac{ES}, denoted by $\superposedVector\in\complexNumbers^{\numberOfDimensionAtRX}$,  can be written as
\begin{align}
	\superposedVector = [\superposedVectorEle[1],\mydots,\superposedVectorEle[\numberOfDimensionAtRX]]^{\rm T} =& \sum_{\indexED=1}^{\numberOfEdgeDevices} \sqrt{\powerED[\indexED]}\channelMatrix[\indexED]\transmittedVector[\indexED] +\noiseVector \nonumber\\=&  \sum_{\indexED=1}^{\numberOfEdgeDevices}\sqrt{\powerED[\indexED]}\channelMatrix[\indexED]\precoder[\indexED]\modulationVector[\indexED]+\noiseVector~,
	\label{eq:superPosOAC}
\end{align}
where $\channelMatrix[\indexED]\in\complexNumbers^{\numberOfDimensionAtRX\times\numberOfDimensionAtTX}$ is the  channel matrix between the $\indexED$th \ac{ED} and the \ac{ES} with the assumptions of  $\expectationOperator[{{\channelMatrix[\indexED]\channelMatrix[\indexED]^{\rm H}}}][]=\numberOfDimensionAtTX\identityMatrix[\numberOfDimensionAtRX]$ {\color{\reviewColor} and each element of $\channelMatrix[\indexED]$ is modeled as a zero-mean symmetric complex Gaussian random variable, unless otherwise stated},  $\noiseVector=[\noiseVectorEle[1],\mydots,\noiseVectorEle[\numberOfDimensionAtRX]]^{\rm T}\in\complexNumbers^{\numberOfDimensionAtRX}
\sim\complexGaussian[\zeroVector[\numberOfDimensionAtRX]][{\noiseVariance\identityMatrix[\numberOfDimensionAtRX]}]
$ is the zero-mean symmetric noise vector with the variance $\noiseVariance$,  $\numberOfDimensionAtRX$ is the number of available dimensions at the \ac{ES}, and $\powerED[\indexED]\in\realNumbers$ denotes the received signal  power for the $\indexED$th \ac{ED}, which is a function of the large-scale channel model, power control, waveform, \ac{PA} non-linearity, and \ac{ACLR} requirements (see Section~\ref{subsec:powerControl} for further discussions). The receiver at the \ac{ES} processes the superposed vector  $\superposedVector$ with a linear decoder $\aggregator^{\rm H}\in\complexNumbers^{\numberOfModulationSymbols\times\numberOfDimensionAtRX}$  (e.g., to overcome the impact of the channel on the superposition along with $\precoder[\indexED]$, $\forall\indexED$) as
\begin{align}
	\modulationVectorSuperposed = [\modulationVectorSuperposedEle[1],\mydots,\modulationVectorSuperposedEle[\numberOfModulationSymbols]]^{\rm T} &= \aggregator^{\rm H}\superposedVector \nonumber\\ &= \sum_{\indexED=1}^{\numberOfEdgeDevices}\sqrt{\powerED[\indexED]}\aggregator^{\rm H}\channelMatrix[\indexED]\precoder[\indexED]\modulationVector[\indexED]+\aggregator^{\rm H}\noiseVector~.
	\label{eq:modSuperPosed}
\end{align}

By using $\numberOfTransforms$ outputs of the linear decoders, the resource de-mapper $\demapperResource$ first constructs the superposed codeword $\demappedVector=[\demappedVectorEle[1],\mydots,\demappedVectorEle[\blockLength]]^{\rm T}\in\complexNumbers^{\blockLength}$. Afterward, the receiver calculates an estimate of the superposed pre-processed symbols as
\begin{align}
	\decodedVector= [\decodedVectorEle[1],\mydots,\decodedVectorEle[\numberOfFunctions]]^{\rm T} = \decoder\left\{\demappedVector\right\}~,
\end{align}
where $\decoder:\complexNumbers^{\blockLength}\rightarrow\realNumbers^{\numberOfFunctions}$ is the decoder (e.g., constellation de-mapper, channel decoder, source decoder, or a combination of these operations) at the \ac{ES}. 
Finally, the target functions are evaluated with the post-processing function as
$\postProcessedVector =[	\postProcessedVectorEle[{1}], \mydots, \postProcessedVectorEle[{\numberOfFunctions}]]^{\rm T}
\in\realNumbers^{\numberOfFunctions}$ for $\postProcessedVectorEle[\indexSampleTime]=\postProcessingFunction[][{\decodedVectorEle[{\indexSampleTime}]}]\in\realNumbers$, $\forall\indexSampleTime\in\{1,\mydots,\numberOfFunctions\}$.

In \figurename~\ref{fig:BlockDiagram}, we provide a block diagram based on aforementioned transmitter and receiver operations for OAC. It is worth noting that we do not specify the domains of the transmitted vectors $\transmittedVector[\indexED]$, $\forall\indexED$, and the received vector $\superposedVector$. Without loss of generality, the domain can be time, frequency, or space, depending on the scheme.  In addition, if the target function is based on Kolmogorov's superposition, $2\numberOfEdgeDevices+1$ nomographic functions in  \eqref{eq:kolmogorov} can be computed over orthogonal resources (see \cite{Goldenbaum_2012CISS} for an example) with a general \ac{OAC} scheme. 

{\color{\reviewColor}\subsection{Metrics}
\label{sec:whatAreTheMetrics}
In this subsection, we discuss widely used metrics  for assessing \ac{OAC} schemes.}
\subsubsection{Error definitions}
Let $\functionArbitraryEstimate[](\symbolVector[\indexSampleTime])$ be an estimate of  $\functionArbitrary[](\symbolVector[\indexSampleTime])$ for $\symbolVector[\indexSampleTime]\triangleq[\symbolVectorEle[1][\indexSampleTime],\dots, \symbolVectorEle[\numberOfEdgeDevices][\indexSampleTime]]^{\rm T}$, for $\indexSampleTime\in\{1,2,\mydots,\numberOfFunctions\}$. The \ac{FEE} can be expressed as
\begin{align} \estimationError[{\functionArbitraryEstimate[](\symbolVector[\indexSampleTime])}]\triangleq\functionArbitraryEstimate[](\symbolVector[\indexSampleTime])-\functionArbitrary[](\symbolVector[\indexSampleTime])~.
\end{align} 
In  \cite{Goldenbaum_2013tcom} and \cite{Jeon_2018}, for $\functionArbitrary[](\symbolVector[\indexSampleTime])\in[\fmin,\fmax]$, $\forall\indexSampleTime$, a normalized \ac{FEE} is defined by
\begin{align}
	\functionEstimationError[{\functionArbitraryEstimate[](\symbolVector[\indexSampleTime])}]\triangleq\frac{\estimationError[{\functionArbitraryEstimate[](\symbolVector[\indexSampleTime])}]}{\fmax-\fmin}~.
	\label{eq:fcnEstError}
\end{align}

\subsubsection{Average error}
The classical \ac{MSE} \cite{Chen_2018,liu2020over}, normalized \ac{MSE}, and Bayesian \ac{MSE} for computation can be expressed as
\begin{align}
	\meanSquareError[{\functionArbitraryEstimate[](\symbolVector[\indexSampleTime])}]\triangleq\expectationOperator[\norm{\estimationError[{\functionArbitraryEstimate[](\symbolVector[\indexSampleTime])}]}^2_2][]~, \nonumber
\end{align}
and
\begin{align}
	\normalizedMeanSquareError[{\functionArbitraryEstimate[](\symbolVector[\indexSampleTime])}]\triangleq\frac{\expectationOperator[\norm{\estimationError[{\functionArbitraryEstimate[](\symbolVector[\indexSampleTime])}]}^2_2][]}{\norm{{\functionArbitrary[](\symbolVector[\indexSampleTime])}}^2_2}~, \nonumber
\end{align}
and
\begin{align}
	\bayesianMeanSquareError\triangleq\expectationOperator[\meanSquareError[{\functionArbitraryEstimate[](\symbolVector[\indexSampleTime])}]][], \nonumber
\end{align}
respectively. In \cite{chen_2018jiot}, the \ac{MSFE} is defined by
\begin{align}
	\meanFunctionSquareError\triangleq\frac{\expectationOperator[\norm{\estimationError[{\functionArbitraryEstimate[](\symbolVector[\indexSampleTime])}]}^2_2][]}{\expectationOperator[\norm{{\functionArbitrary[](\symbolVector[\indexSampleTime])}}^2_2][]},
\end{align}
where the expectation is over both $\symbolVector[\indexSampleTime]$, $\forall\indexSampleTime$, and the channel.

\subsubsection{Outage probability}
{\color{\reviewColor} The outage probability provides a statistical view of the computation error as compared with  \ac{MSE}.}
By using \eqref{eq:fcnEstError}, it can be defined as the probability that the normalized \ac{FEE} is larger than or equal to $\outageThreshold$ \cite{Goldenbaum_2013tcom}\cite{Jeon_2018}, i.e., 
\begin{align}
	\outageProbability[\error]\triangleq\probability[{|\functionEstimationError[{\functionArbitraryEstimate[](\symbolVector[\indexSampleTime])}]|\ge\outageThreshold}]~.
\end{align}
Based on \cite{goldenbaum2015nomographic}, for a given $\error$, the block outage probability may also be defined by
\begin{align}
	\errorProbabilityCont[\error] \triangleq \probability[\bigcup_{\indexSampleTime=1}^{\numberOfFunctions}{\sup|\functionArbitraryEstimate[](\symbolVector[\indexSampleTime])-\functionArbitrary[](\symbolVector[\indexSampleTime])|>\error}]~.
\end{align}

\subsubsection{Computation error rate}
{\color{\reviewColor}
\Ac{CER} is analogous to the bit error rate in communication systems. It can be used for assessing the computation when the arguments of the functions are discrete values.}
Let $\symbolVectorEle[\indexED][\indexSampleTime]$ be a discrete random variable, $\forall\indexED,\indexSampleTime$. The \ac{CER} can be defined as
\begin{align}
	\errorProbability \triangleq \probability[{{}\functionArbitraryEstimate[](\symbolVector[\indexSampleTime])\neq\functionArbitrary[](\symbolVector[\indexSampleTime])}]~.
	\label{eq:cer}
\end{align}
If $\numberOfFunctions$ functions are taken into account for the error-rate calculation, as done in \cite{Jeon_2014} and \cite{Jeon_2016}, the \ac{BCER} can be expressed as\footnote{In \cite{Jeon_2014} and \cite{Jeon_2016}, \eqref{eq:bcer} is defined as probability of error. To differentiate \eqref{eq:bcer} from \eqref{eq:cer}, we use a different terminology for \eqref{eq:bcer} in this study.}
\begin{align}
	\errorProbabilityblock \triangleq \probability[\bigcup_{\indexSampleTime=1}^{\numberOfFunctions}{{}\functionArbitraryEstimate[](\symbolVector[\indexSampleTime])\neq\functionArbitrary[](\symbolVector[\indexSampleTime])}]~.
	\label{eq:bcer}
\end{align}

\subsubsection{Computation rate and throughput}
\label{subsec:computationRate}
{\color{\reviewColor} Computation rate $\computationRate$ and computation throughput $\computationThroughput$ can be defined as the number of functions  computed per channel use \cite{Jeon_2014,goldenbaum2015nomographic,Jeon_2016} and the number of functions computed per second, respectively. They can be respectively expressed as 
\begin{align}
		\computationRate =\frac{\numberOfFunctions}{\numberOfChannelUse}~\text{[functions/dimension]}~,
\end{align}
and
\begin{align}
	\computationThroughput = \frac{\numberOfFunctions}{\duration} ~\text{[functions/second]}~,
\end{align}
where $\numberOfChannelUse=2\numberOfTransforms\numberOfDimensionAtRX$ and  $\duration$ are the number of real dimensions and the interval used for computing $\numberOfFunctions$ functions.\footnote{The dimensions can be in time, frequency, and/or space, depending on the resource utilization.} It is worth noting that the number of real dimensions for a single-input-single-output scenario  is approximately equal to $2\bandwidth\duration$ for given bandwidth $\bandwidth$ and time interval $\duration$ \cite[Section 4.6]{Proakis2007}. Hence, for $\numberOfSpatialStreams$ spatial streams, the computation throughput can be 	approximately calculated as $	\computationThroughput \approx  2\bandwidth\numberOfSpatialStreams\times\computationRate$
}

Based on \cite{Jeon_2014}  and \cite{Jeon_2016}, the achievable computation rate can be defined as follows:
\begin{definition}[Achievable computation rate \cite{Jeon_2014,Jeon_2016}] \rm Let $\symbolVectorEle[\indexED][\indexSampleTime]$ be  a discrete random variable, $\forall\indexED,\indexSampleTime$.
	The rate $\computationRate$ is said to be achievable if there exists a sequence of length-$\blockLength$ block codes such that
	$
	\lim_{\blockLength\rightarrow\infty}\errorProbabilityblock= 0
	$.
	\label{def:achievableRateSecond}
\end{definition}

In \cite{Nazer_2011}, the computation of modulo-$\aprime$ sum function, i.e., $	\functionArbitrary[](\symbolVector[\indexSampleTime])=\bigoplus_{\indexED=1}^{\numberOfEdgeDevices} \coefInField[\indexED]\symbolVectorEle[\indexED][\indexSampleTime]
$, $\forall\indexSampleTime$, for $\symbolVectorEle[\indexED][\indexSampleTime],\coefInField[\indexED]\in\finiteField[\aprime]$, is investigated  over Gaussian channel, i.e.,
$
\demappedVectorEle[\indexEncodedVector]=\sum_{\indexED=1}^{\numberOfEdgeDevices}\channelAtSubcarrier[\indexED]\encodedVectorEle[\indexED,\indexEncodedVector]+\noiseVectorEle[\indexEncodedVector]
$ (see \figurename~\ref{fig:BlockDiagram}). For $\noiseVectorEle[\indexEncodedVector]\in\realNumbers$, in  \cite[Theorem~2]{Nazer_2011}, Nazer and Gastpar show that achievable computation rate is
\begin{align}
	\computationRate<\computationRateAchievableNoArg=\frac{\frac{1}{2} \log_2^{+}\left(\left(\norm{\integerVector}^2_2-\frac{\powerED[](\channelVector[]^{\rm T}\integerVector)^2}{\noiseVariance+\powerED[]\norm{\integerVector}^2_2}\right)^{-1}\right)}{\log_2{(\aprime)}}~,
	\label{eq:candfrate}
\end{align}
where $\integerVector=[\integerVectorEle[1],\mydots,\integerVectorEle[\indexED]]\in\integers^{\numberOfEdgeDevices}$ for $\coefInField[\indexED]=\mappingFromFieldtoInteger({\integerVectorEle[\indexED]\mod\aprime)})$, i.e., $\mappingFromFieldtoInteger$ is a map from $\{0,1,\mydots,\aprime-1\}$ to the corresponding elements in $\finiteField[\aprime]$, and $\powerED[]$ is the average transmit  power, $\channelVector[]=[\channelAtSubcarrier[1],\mydots,\channelAtSubcarrier[\numberOfEdgeDevices]]\in\realNumbers^{\numberOfEdgeDevices}$.
By observing the fact that, for a sufficiently large prime $q$, computing the modulo-$q$ sum of $\numberOfEdgeDevices$ elements for $\symbolVectorEle[\indexED][\indexSampleTime]<\aprime$ is equivalent to the arithmetic sum for $q>(\aprime-1)\numberOfEdgeDevices$ and using \cite[Theorem 1]{Nazer_2007} and \eqref{eq:candfrate} for $\integerVectorEle[\indexED]=1$, $\forall\indexED$, Jeon, Chien-Yi, and Gastpar in \cite{Jeon_2014} show that the achievable computation rate can be expressed as
\begin{align}
	\computationRate<\computationRateAchievableNoArg=\frac{\frac{1}{2} \log_2^{+}\left(\frac{1}{\numberOfEdgeDevices}+\min_{\forall\indexED}\frac{|\channelAtSubcarrier[\indexED]|^2\powerED[]}{\noiseVariance}\right)}{\entropy[{\functionArbitrary[](\symbolVector[\indexSampleTime])}]}~,
	\label{eq:jeonBound}
\end{align}
where $\entropy[\cdot]$ is the entropy function and $\symbolVectorEle[\indexED][\indexSampleTime]$ is a discrete random variable, $\forall\indexED,\indexSampleTime$. 

In \cite{goldenbaum2015nomographic}, the achievable computation rate is defined as follows:
\begin{definition}[Achievable computation rate \cite{goldenbaum2015nomographic}] \rm
	The parameter $\computationRateAchievable$ is said to be an achievable computation rate for $\functionArbitrary[]$ and $\error$, if there exists a scheme that fulfills $
	\errorProbabilityCont[\error]<\errorProbabilityConst
	$ for every rate $\computationRate<\computationRateAchievable$ and every $\errorProbabilityConst>0$.
	\label{def:achievableRateFirst}
\end{definition}
Based on Definition~\ref{def:achievableRateFirst} and the properties of nested lattice codes as done in \cite{Nazer_2007} and \cite{Nazer_2011}, for a real-valued \ac{AWGN} channel and arithmetic sum, in \cite[Theorem 5 with Equation (54)]{goldenbaum2015nomographic}, Goldenbaum shows that
\begin{align}
	\computationRate<\computationRateAchievable=\frac{\frac{1}{2} \log_2^{+}\left(\frac{\powerED[]}{\noiseVariance}\right)}{\log_2(2^{\numberOfDigitsBits(\functionArbitrary[],\error)}-1)+\log_2{\numberOfEdgeDevices}}~,
	\label{eq:goldBound}
\end{align}
where $\numberOfDigitsBits(\functionArbitrary[],\error)$ is the number of bits defined in Section~\ref{parag:genNomoFunc} for given function $\functionArbitrary[]$ and the amount of maximum distortion $\error$.

\begin{figure}[t]
	\centering
	{\includegraphics[width =3.5in]{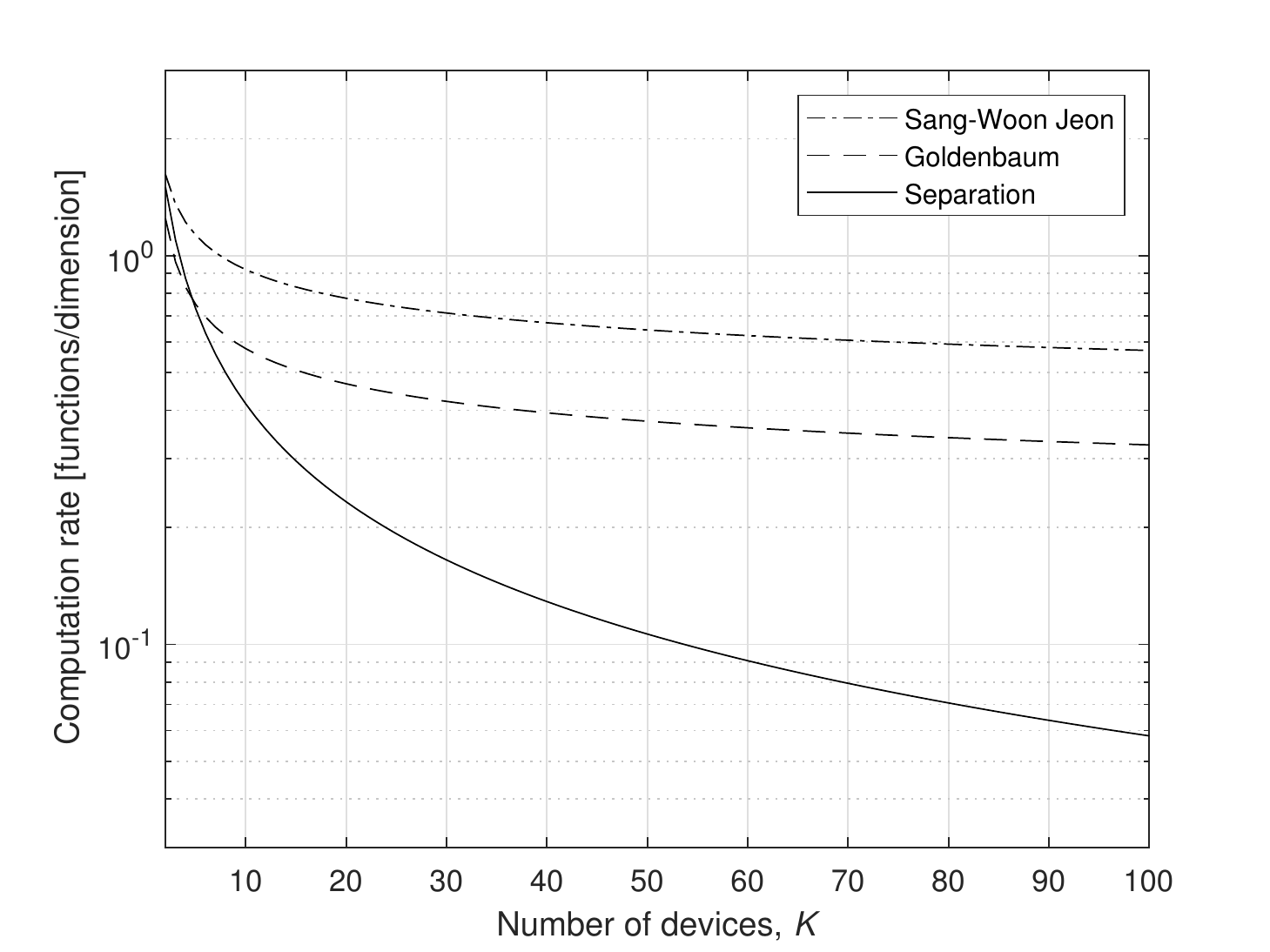}
	} 
	\caption{Achievable computation rate for the arithmetic sum of $K$ Bernoulli random variables in AWGN channel. The achievable computation rate with the separation of communication and computation can be dramatically lower than the ones with OAC.} 
	\label{fig:separationOrNot}
\end{figure}
In \figurename~\ref{fig:separationOrNot}, we demonstrate the achievable computation rates for a given number of \acp{ED} based on the example given in \cite{Jeon_2014} for arithmetic sum. Suppose the symbols $\symbolVectorEle[\indexED][\indexSampleTime]$, $\forall\indexED,\indexSampleTime$, are \ac{iid} Bernoulli random variables with the probability $1/2$ and the \ac{SNR} is 15 dB. From \cite{Chang_1979},  ${\entropy[{\functionArbitrary[](\symbolVector[\indexSampleTime])}]}$ can be calculated as ${\entropy[{\functionArbitrary[](\symbolVector[\indexSampleTime])}]}=\numberOfEdgeDevices-2^{-\numberOfEdgeDevices}\sum_{\indexED=1}^{\numberOfEdgeDevices}\binom{\numberOfEdgeDevices}{\indexED}\log_2\binom{\numberOfEdgeDevices}{\indexED}\le\frac{1}{2}\log_2{(\pi\constante\numberOfEdgeDevices/2)}$.
Also, $\numberOfDigitsBits(\functionArbitrary[],\error)=1$ is sufficient for the representation of the symbols.
 The achievable computation rate for the separation  of communication and computation can be calculated as $\frac{1}{2\numberOfEdgeDevices}\log_2(1+{\numberOfEdgeDevices\powerED[]}/{\noiseVariance})$ \cite{Jeon_2014},\cite[Equation 6.13]{tseBook}. 
As can be seen from \figurename~\ref{fig:separationOrNot}, 
the achievable computation rate with the separation of communication and computation can be dramatically lower than the ones with joint communication and computation.
{\color{\reviewColor}For example, if there are $\numberOfEdgeDevices=100$ EDs, OAC promises approximately 10 times faster reliable computations than the one with separation.}  For this example, Goldenbaum's result is more conservative than Jeon's rate as Goldenbaum's expression implicitly assumes equiprobable final outcomes. {\color{\reviewColor}In \cite[p. 4]{matthiasthesis2023}, for the same scenario,
it is argued that the EDs can send significantly less information to the ES if the objective is to compute the sum function. This is because the entropy of the computed function is much smaller than  the amount of information that needs to be transferred to the ES with the separation, i.e., ${\entropy[{\functionArbitrary[](\symbolVector[\indexSampleTime])}]}\le\sum_{\indexED=1}^{\numberOfEdgeDevices}\entropy[{\symbolVectorEle[\indexED][\indexSampleTime]}]=\numberOfEdgeDevices$.}

In \cite{Jeon_2016} and \cite{Fangzhou_twc2019}, to improve the computation rate, a construction of the target function via local functions is proposed. In \cite{Jeon_2016}, the authors assume that the channel coefficients are \ac{iid} random variables and they select a subset of EDs with the largest channel gains to gradually calculate the target function over fast-fading channels. It is shown that a non-vanishing computation rate is achievable even if the number of \ac{ED} in the network increases. In a recent work \cite{Naifu_2022}, nested lattice codes are utilized along with stochastic quantization for the \ac{AWGN} channel. A theoretical comparison between uncoded \ac{OAC},  coded \ac{OAC}, and separation  of communication and computation is provided in terms of distortion under the \ac{AWGN} channel. In \cite{Chen_2019cpmac}, the computation rates for both
homogeneous and heterogeneous
networks under large-scale fading are studied. The reader is also referred to \cite{Nazer_2011survey} for the computation rates for uncoded and coded physical layer network coding strategies.



\subsection{Classification criteria: Availability of CSI}
One of the challenges for calculating  \eqref{eq:nomographicfcn} with an \ac{OAC} method  arises due to  the fact that the impact of multi-path distortion on the transmitted symbols from the \acp{ED} (i.e., $\channelMatrix[\indexED]\transmittedVector[\indexED]$) occurs before the superposition, as expressed in \eqref{eq:superPosOAC}. Hence, an  estimator that estimates the output of the function at the \ac{ES} can be unrealized, i.e., it may not be written solely as a function of the received symbols. For example, consider a scenario where $\channelMatrix[\indexED]$ is \ac{iid} and follows Rayleigh fading and the power control ensures that the average received signal powers are perfectly aligned, e.g., $\powerED[\indexED]=1$, $\forall\indexED$. Since the superposed symbol at the \ac{ES} in this case cannot be expressed as
\begin{align}
\superposedVector =\sum_{\indexED=1}^{\numberOfEdgeDevices} \channelMatrix[\indexED]\transmittedVector[\indexED] +\noiseVector = \channelMatrix[] \sum_{\indexED=1}^{\numberOfEdgeDevices} \transmittedVector[\indexED]+\noiseVector~,
\end{align}
where  $\channelMatrix[]\in\complexNumbers^{\numberOfDimensionAtRX\times\numberOfDimensionAtTX}$ is a matrix {\em independent} of the vector $\transmittedVector[\indexED]$, $\forall\indexED$ (i.e., {\em there is no uniform  fading matrix $\channelMatrix[]$} \cite{Chen_2018}), an equalizer that relies on the availability of $\channelMatrix[]$ to obtain $\sum_{\indexED=1}^{\numberOfEdgeDevices} \transmittedVector[\indexED]$ cannot be realized. Given this observation, in this subsection, we classify the methods in the state-of-the-art based on how they deal with the fading due to the wireless channel through the precoders $\precoder[\indexED]$, $\forall\indexED$, and the decoder $\aggregator$. We group them based on the availability of \ac{CSIT} and \ac{CSIR}. For simplifying the classification, we assume  $\powerED[\indexED]=1$, $\forall\indexED$. We also provide corresponding equations for different strategies in \tablename~\ref{table:pre-equalization} {\color{\reviewColor}when the real part of the superposed symbol is used for computation.}

\setlength\tabcolsep{1.5pt}
\begin{table*}[t]
	\color{\reviewColor}
	\centering
	\caption{Computation under fading channel
	}
	\resizebox{\textwidth}{!}{
	\begin{tabular}{l|c|c|c|c|c|c|c|c|c|c|c|c|c}
		Category & Method & $\numberOfDimensionAtTX$  & $\numberOfDimensionAtRX$ & $\numberOfFunctions$& \begin{tabular}{@{}c@{}} Structure of \\ channel $\channelMatrix[\indexED]$ \end{tabular}  & \begin{tabular}{@{}c@{}}Linear \\ precoder $\precoder[\indexED]$ \end{tabular}  & \begin{tabular}{@{}c@{}} Linear \\ decoder $\aggregator$\end{tabular} & \begin{tabular}{@{}c@{}} Encoder \\ $\encoder[\indexED]\{\preProcessedVector[\indexED]\}$ \end{tabular}  & \begin{tabular}{@{}c@{}} Decoder \\ $\decoder\{\demappedVector\}$ \end{tabular}  & \begin{tabular}{@{}c@{}}Rate $\computationRate$ \\($\numberOfTransforms=1$) \end{tabular}  & \begin{tabular}{@{}c@{}} $\preProcessingFunctionW[\indexED](x)$ \\ \& $\postProcessingFunctionW[](x)$ \end{tabular} & Calculated function & Target function \\ \hline\hline
		
		\multirow{5}[25]{*}{\begin{tabular}{@{}l@{}}CSIT: \cmark \\  CSIR: \xmark \end{tabular}}   
			& \begin{tabular}{@{}c@{}}Zero  \\    forcing\end{tabular} & 1 & 1 & 1 & $\channelAtSubcarrier[\indexED]$ & $\sqrt{\normalizationCoef[\indexED]}\frac{\channelAtSubcarrier[\indexED]^*}{|\channelAtSubcarrier[\indexED]|^2}$ & 1 & $\preProcessedVectorEle[\indexED]$ & $\demappedVectorEle[]$ & $\frac{1}{2}$ & x   & $\displaystyle \Re\left\{\sum_{\indexED=1}^{\numberOfEdgeDevices}\sqrt{\normalizationCoef[\indexED]}\symbolVectorEle[\indexED]+\noiseVectorEle[]\right\}$ & $\displaystyle {\sum_{\indexED=1}^{\numberOfEdgeDevices}\symbolVectorEle[\indexED]}$ \\\cline{2-14}
		
		 	&\begin{tabular}{@{}c@{}}Truncated  \\  channel \\ inversion\end{tabular}  & 1 & 1 & 1 & $\channelAtSubcarrier[\indexED]$  &  $\sqrt{\normalizationCoef[\indexED]}\frac{\channelAtSubcarrier[\indexED]^*}{|\channelAtSubcarrier[\indexED]|^2}\indicatorFunction[{|\channelAtSubcarrier[\indexED]|^2>\thresholdTCI}]$ & 1& $\preProcessedVectorEle[\indexED]$ & $\demappedVectorEle[]$ & $\frac{1}{2}$& x  &   $\displaystyle
			\Re\left\{{\sum_{\indexED=1}^{\numberOfEdgeDevices}\sqrt{\normalizationCoef[\indexED]}\indicatorFunction[{|\channelAtSubcarrier[\indexED]|^2>\thresholdTCI}]\symbolVectorEle[\indexED]}+\noiseVectorEle[]\right\}$ & $\displaystyle
			{\sum_{\indexED=1}^{\numberOfEdgeDevices}\symbolVectorEle[\indexED]}$\\\cline{2-14}

			& \begin{tabular}{@{}c@{}}Phase  \\   correction\end{tabular}  & 1 & 1 & 1 & $\channelAtSubcarrier[\indexED]$  &  $\frac{\channelAtSubcarrier[\indexED]^*}{|\channelAtSubcarrier[\indexED]|}$  & 1 & $\preProcessedVectorEle[\indexED]$ & $\demappedVectorEle[]$ & $\frac{1}{2}$  & x   & $\displaystyle
			\Re\left\{\sum_{\indexED=1}^{\numberOfEdgeDevices}|\channelAtSubcarrier[\indexED]|\symbolVectorEle[\indexED]+\noiseVectorEle[]\right\}$ & $\displaystyle
			{\sum_{\indexED=1}^{\numberOfEdgeDevices}\symbolVectorEle[\indexED]}$\\\cline{2-14}
			
			& \begin{tabular}{@{}c@{}}Maximum \\ratio  \\   transmission\end{tabular}  & $\sequenceLength$ & 1 & 1 & $[\channelAtSubcarrier[\indexED,1],\mydots,\channelAtSubcarrier[\indexED,\sequenceLength]]$  &  $\sqrt{\normalizationCoef[\indexED]}\channelVector[\indexED]^{\rm H}$ & 1 & $\preProcessedVectorEle[\indexED]$ & $\demappedVectorEle[]$ & $\frac{1}{2}$ & x  &   $\displaystyle \Re\left\{\sum_{\indexED=1}^{\numberOfEdgeDevices}\sqrt{\normalizationCoef[\indexED]}\norm{\channelVector[\indexED]}^2_2\symbolVectorEle[\indexED]+\noiseVectorEle[]\right\}$ & $\displaystyle
			{\sum_{\indexED=1}^{\numberOfEdgeDevices}\symbolVectorEle[\indexED]}$ \\\cline{2-14}
		
			& \begin{tabular}{@{}c@{}}Amplitude \\ correction\\ + energy \\ estimation\end{tabular}  &  $\sequenceLength$ & $\sequenceLength$ & $1$ & $\diagOperation[{[\channelAtSubcarrier[\indexED,1],\mydots,\channelAtSubcarrier[\indexED,\sequenceLength]]}] $  &  $ 	\diagOperation[{[\frac{\sqrt{\normalizationCoef[\indexED]}}{|\channelAtSubcarrier[\indexED,1]|} ,\mydots, \frac{\sqrt{\normalizationCoef[\indexED]}}{|\channelAtSubcarrier[\indexED,\sequenceLength]|}]}]$  & $\frac{1}{\sequenceLength}\superposedVector$   &	\begin{tabular}{c} $\sqrt{g(\preProcessedVectorEle[\indexED])}\times$  \\   $ 	{\begin{bmatrix}
						{\constante^{\constantj\theta_1}} ,\mydots, {\constante^{\constantj\theta_\sequenceLength}}
					\end{bmatrix}^{\rm T}}$ \\$\affineEnc$ is affine \end{tabular} & \begin{tabular}{c}$\affineDec(\demappedVectorEle[])$ \\$\affineDec$ is affine \end{tabular} &$\frac{1}{2\sequenceLength}$ & x &  \begin{tabular}{c} $\displaystyle \affineDec\left(\sum_{\indexED=1}^{\numberOfEdgeDevices}{\normalizationCoef[\indexED]}\affineEnc(\symbolVectorEle[\indexED])+\text{noise}+\text{inter.}\right)$ \end{tabular}   & $\displaystyle {\sum_{\indexED=1}^{\numberOfEdgeDevices}\symbolVectorEle[\indexED]}$  \\ \hline

		\multirow{2}[7]{*}{\begin{tabular}{@{}l@{}}CSIT: \cmark \\  CSIR: \cmark \end{tabular}}   
			& \begin{tabular}{@{}c@{}}ZF  \\ coordination\end{tabular}  &  $\sequenceLength$ & $\sequenceLengthOther$ & $\numberOfModulationSymbols$ & 		$\begin{bmatrix}
				{\channelAtSubcarrier[\indexED,1,1] ,\mydots,\channelAtSubcarrier[\indexED,\sequenceLength,1]} \\
				\vdots\\
				{\channelAtSubcarrier[\indexED,1,\sequenceLengthOther] ,\mydots,\channelAtSubcarrier[\indexED,\sequenceLength,\sequenceLengthOther]} 
			\end{bmatrix}$	  &  \eqref{eq:zfInversion}  & \eqref{eq:ZFproblem}  & $\preProcessedVector[\indexED]$ & $\demappedVector$ & $\frac{\numberOfModulationSymbols}{2\sequenceLengthOther}$ & x &   $\displaystyle \Re\left\{\sum_{\indexED=1}^{\numberOfEdgeDevices}\symbolVectorED[\indexED]+(\aggregator^*)^{\rm H}\noiseVector\right\}$ & $\displaystyle {\sum_{\indexED=1}^{\numberOfEdgeDevices}\symbolVectorED[\indexED]}$ \\\cline{2-14}
			
			& \begin{tabular}{@{}c@{}}MMSE  \\ coordination\end{tabular}  &  $\sequenceLength$ & $\sequenceLengthOther$ & $\numberOfModulationSymbols$ & 		
		$\begin{bmatrix}
			{\channelAtSubcarrier[\indexED,1,1] ,\mydots,\channelAtSubcarrier[\indexED,\sequenceLength,1]} \\
			\vdots\\
			{\channelAtSubcarrier[\indexED,1,\sequenceLengthOther] ,\mydots,\channelAtSubcarrier[\indexED,\sequenceLength,\sequenceLengthOther]} 
		\end{bmatrix}$					
			  &  \eqref{eq:MMSEproblem} & \eqref{eq:MMSEproblem}  & $\preProcessedVector[\indexED]$ & $\demappedVector$ & $\frac{\numberOfModulationSymbols}{2\sequenceLengthOther}$ & x &   $\displaystyle \Re\left\{\sum_{\indexED=1}^{\numberOfEdgeDevices}(\aggregator^*)^{\rm H}\channelMatrix[\indexED]\precoder[\indexED]^*\symbolVectorED[\indexED]+(\aggregator^*)^{\rm H}\noiseVector\right\}$ & $\displaystyle {\sum_{\indexED=1}^{\numberOfEdgeDevices}\symbolVectorED[\indexED]}$ \\\hline				
	
		\begin{tabular}{@{}l@{}}CSIT: \xmark \\  CSIR: \cmark \end{tabular} 
			& \begin{tabular}{@{}c@{}}Channel  \\  hardening \\ via \\ aggregated \\ CSI\end{tabular}  & 1 & $\sequenceLengthOther$ & 1 & $[\channelAtSubcarrier[\indexED,1],\mydots,\channelAtSubcarrier[\indexED,\sequenceLengthOther]]^{\rm T}$  &  1 & $\displaystyle\frac{1}{\sequenceLengthOther}\sum_{\indexED=1}^{\numberOfEdgeDevices}\channelVector[\indexED]$ & $\preProcessedVectorEle[\indexED]$ & $\demappedVectorEle[]$ & $\frac{1}{2\sequenceLengthOther}$ & x  &  $\displaystyle \Re\left\{\sum_{\indexED=1}^{\numberOfEdgeDevices}\frac{\norm{\channelVector[\indexED]}^2_2}{\sequenceLengthOther}\symbolVectorEle[\indexED]+\frac{1}{\sequenceLengthOther}\sum_{\indexED=1}^{\numberOfEdgeDevices}\channelVector[\indexED]^{\rm H}\noiseVector+\text{inter.}\right\}$ & $\displaystyle	{\sum_{\indexED=1}^{\numberOfEdgeDevices}\symbolVectorEle[\indexED]}$\\		\hline

		\multirow{1}[13]{*}{\begin{tabular}{@{}l@{}}CSIT: \xmark \\  CSIR: \xmark \end{tabular}} 
			& \begin{tabular}{@{}c@{}}Channel  \\  hardening \\ + energy \\ estimation\end{tabular}  &  $\sequenceLength$ & $\sequenceLength\sequenceLengthOther$ & 1 & 
		$\begin{bmatrix}
			\diagOperation[{[\channelAtSubcarrier[\indexED,1,1] ,\mydots,\channelAtSubcarrier[\indexED,\sequenceLength,1]]}] \\
			\vdots\\
			\diagOperation[{[\channelAtSubcarrier[\indexED,1,\sequenceLengthOther] ,\mydots,\channelAtSubcarrier[\indexED,\sequenceLength,\sequenceLengthOther]]}] 
		\end{bmatrix}$
		   &  $\identityMatrix[N]$   & $\frac{1}{\sequenceLength\sequenceLengthOther}\superposedVector$  & 	\begin{tabular}{c} $\sqrt{g(\preProcessedVectorEle[\indexED])}\times$  \\   $ 	{\begin{bmatrix}
				{\constante^{\constantj\theta_1}} ,\mydots, {\constante^{\constantj\theta_\sequenceLength}}
			\end{bmatrix}^{\rm T}}$ \\$\affineEnc$ is affine \end{tabular} & \begin{tabular}{c}$\affineDec(\demappedVectorEle[])$ \\$\affineDec$ is affine \end{tabular} &	$\frac{1}{2\sequenceLength\sequenceLengthOther}$  & x &  \begin{tabular}{c}$\displaystyle \affineDec\left(\sum_{\indexED=1}^{\numberOfEdgeDevices} \frac{\traceOperator[{\channelMatrix[\indexED]^{\rm H}\channelMatrix[\indexED]}]}{\sequenceLength\sequenceLengthOther}\affineEnc(\symbolVectorEle[\indexED])+\text{noise}+\text{inter.}\right)$ \end{tabular}   & $\displaystyle	{\sum_{\indexED=1}^{\numberOfEdgeDevices}\symbolVectorEle[\indexED]}$\\\cline{2-14}

			&\begin{tabular}{@{}c@{}}Orthogonal\\ signaling \end{tabular}  &  $2$ & $2$ & $1$ &		$\diagOperation[{[\channelAtSubcarrier[\indexED]^{'} ~\channelAtSubcarrier[\indexED]^{''}]}]$			&  $\identityMatrix[2]$ & $\identityMatrix[2]$ & $\begin{bmatrix}
\indicatorFunction[{\preProcessedVectorEle[\indexED]=1}] \\ \indicatorFunction[{\preProcessedVectorEle[\indexED]=-1}]
			\end{bmatrix}$ & \begin{tabular}{@{}c@{}}Energy\\ comparator \end{tabular} & $\frac{1}{4}$ & $\signNormal[x]$  &   $\displaystyle \signNormal[{\abs*{\sum_{\substack{\indexED=1\\\symbolVectorEle[\indexED]>0}}^{\numberOfEdgeDevices}\channelAtSubcarrier[\indexED]^{'}+\noiseVectorEle[1]}^2-{\abs*{\sum_{\substack{\indexED=1\\\symbolVectorEle[\indexED]<0}}^{\numberOfEdgeDevices}\channelAtSubcarrier[\indexED]^{''}+\noiseVectorEle[2]}^2}}] $ & $\displaystyle\signNormal[{\sum_{\indexED=1}^{\numberOfEdgeDevices}\signNormal[{s_{\indexED}}]}]$ \\ \hline

		\end{tabular}
	}
	\label{table:pre-equalization}
\end{table*}
\subsubsection{CSIT: Available, CSIR: Not Available}
In this category, we assume that all \acp{ED} have their {\em own} \ac{CSI} (i.e., the $\indexED$th \ac{ED} knows $\channelMatrix[\indexED]$, but not the set $\{\channelMatrix[\indexED']|\indexED'\neq\indexED\}$) and the \ac{CSI} is not available to the \ac{ES}, i.e., no \ac{CSIR}. Hence, the \acp{ED} cannot make coordinated transmissions. However, each \ac{ED} can pre-distort its own transmitted symbols under an average transmit power constraint or an instantaneous transmit power constraint by designing the precoders, i.e., $\precoder[\indexED], \forall\indexED$, such that the superposed symbol at the receiver is a good approximation to the desired value.  In this category, the  equalization used in traditional communication is shifted from the receiver to the transmitter. {\color{\reviewColor}The computation rate of these schemes can be as high as $\computationRate=1$ function/dimension if both imaginary and real parts of the received symbols are exploited.}


\paragraph{Channel inversion}
\label{parag:TCI}
\Ac{TCI} is a pre-equalization technique considered in many studies in the literature, e.g., \cite{Guangxu_2020, Amiri_2020, Guangxu_2021, Amiri2021Convergence,park_2022tcom,Houssem_arxiv2022,Liang_2022,Zezhong_2022,Krouka_2022, Wenzhi_2022, Wei_2022, Luteng_2022gc,Mitsiou_2022arxiv,gafni2023federated}.
In this approach, to reverse the effect of the wireless channel on the transmitted symbols, the symbol is multiplied with the inverse of the  channel coefficient {\em if} the absolute square of the  corresponding channel coefficient is larger than a pre-set threshold $\thresholdTCI$. Hence, \Ac{TCI} is inline with \ac{ZF} \cite{Min_uav2022,razavikiaGC_2022,sato_2022gc}  (or pseudo-inverse of the channel matrix in general for $\numberOfDimensionAtRX\le\numberOfDimensionAtTX$). Three potential issues can arise with \ac{TCI}: 1) Due to the inversion,  the transmit power can increase instantaneously. As a result, it can cause instantaneous spectral growth for a non-linear hardware. 2) Due to the truncation, some of the \acp{ED} may not participate in the computation, e.g., $\symbolVectorEle[\indexED]=0$ for a non-participating ED. The implication of non-participating \acp{ED} on the performance depends on the application. 3) A power normalization  that takes the average transmit power into account needs to be applied to the symbols, which is a function of the channel statistics. In the literature, the power control factor $\normalizationCoef[\indexED]$ for \ac{TCI} is often designed based on the average transmit power  after the truncation for Rayleigh fading channel  (See \cite[Eq. 13]{Guangxu_2020} and \cite[Eq. 13]{Guangxu_2021} and the corresponding encoding in \tablename~\ref{table:pre-equalization}), i.e.,
\begin{align}
	\normalizationCoef[\indexED]=\frac{1}{{\exponentialIntegral[\thresholdTCI]}}~,
\end{align}
where $\exponentialIntegral[\cdot]$ is the exponential integral defined as $\exponentialIntegral[\thresholdTCI]\triangleq\int_{\thresholdTCI}^{\infty}\frac{1}{x}\exp(-x)\text{d}x$.
 Note that \ac{TCI} is also mentioned in Goldenbaum's early work in \cite[Remark 4]{Goldenbaum_2013tcom}, but not analyzed. In \cite{Mital_2022}, \ac{TCI} is investigated in a scenario where the same
symbol is transmitted over different subcarriers to achieve a diversity gain.

\paragraph{Phase correction}
\Ac{PC} is a pre-equalization strategy to achieve a coherent superposition by rotating the phase of the transmitted symbols based on the amount of the phase distortion due to the channel \cite{Mergen_2007,sery2020analog,Seif_2020isit,Zhong_2022uav,sato_2022gc, CarlosGC_2022}. The main advantage of this approach is that the power normalization due to the pre-equalization is not needed. On the other hand,  \ac{PC} does not correct the amplitude of the superposed symbols.
For some applications, this is not an issue. For example, in \cite{Wei_2021}, it is shown that amplitude alignment is not necessarily the best strategy for \ac{FEEL}.  In \cite{sery2020analog}, it is emphasized that correcting the phase with an error less than $\pi/2$ can be sufficient to yield a constructive aggregation at the receiver.
The \ac{PC} is also utilized with \ac{BAA} in \cite{paul2021accelerated} and \ac{OBDA} in  \cite{Ruichen_2020} for \ac{FEEL}. In \cite{Jihoon_2022Platooning}, \ac{PC} is  applied only for the real part of the symbols, which corresponds to a sign multiplication.

\paragraph{Amplitude correction and energy estimation}
\label{parag:ac}
In \cite{Goldenbaum_2013tcom,Goldenbaum2010asilomar,Goldenbaum_2009wcnc,Goldenbaum_2010ICASSP}, Goldenbaum and Sta\'nczak propose to calculate the energy of a sequence of superposed symbols along with a channel inversion to compute a general nomographic function. In this approach, the output of the pre-processing function is processed further with a function $\affineEnc$ that results in a non-negative value and afterward the {\em square root} of the resulting value is multiplied with  
a sequence of length $\sequenceLength$ as $\sqrt{g(\preProcessedVectorEle[\indexED])}\times[
		{\constante^{\constantj\theta_1}} ,\mydots, {\constante^{\constantj\theta_\sequenceLength}}]$.  At the receiver, the energy of the received sequence is calculated to achieve the \ac{OAC} and the superposed symbol is processed with another affine function $\affineDec$ to reverse the impact of $\affineEnc$ on the superposed symbols (see \tablename~\ref{table:pre-equalization} for the final expression and Section~\ref{parag:affine}). Three interesting observations  were made:  1) Only amplitude correction is needed as the receiver calculates the energy of the symbol. This implies that the \acp{ED}  need only modulus \ac{CSI} \cite[Proposition 1]{Goldenbaum_wcl2014}. 2) The sequences for the proposed scheme should be designed to harness interference as a common goal, instead of eliminating the interference as in a traditional code division multiple access systems \cite{Goldenbaum_2009wcnc}. The need for a sequence set that should satisfy the property of mutually-orthogonal complementary codes or Z-complementary code set in general (e.g., \cite[Eq. 2]{Wu_2021}). In  \cite{Goldenbaum_2013tcom}, unimodular sequences with random phases are adopted to reduce the interference. 3) As the \ac{ES} calculates the energy of the received sequence, the proposed scheme is not sensitive to time and phase synchronization errors (see \cite[Figure 2]{Goldenbaum_2013tcom} for an illustration). By following Golderbaum's work \cite{Goldenbaum_2013tcom}, in \cite{Jeon_2018}, 
it is assumed that the range space of each pre-processing function is a compact subset of non-negative reals. At the \acp{ED}, amplitude correction with truncation is applied to the output of the pre-processing function. To ensure both peak and average transmit power constraints, a sequential symbol power adaptation strategy that uses only the observed channel coefficients  is proposed. At the \ac{ES}, the energy of the superposed sequence is calculated to compute the nomographic function. Note that truncation is also used in \cite{Kortke_2014} to realize Goldenbaum and Sta\'nczak's scheme  in practice.

\paragraph{Maximum-ratio transmission without CSIR}
With \ac{MRT} for traditional  communications, a symbol is transmitted on the strongest eigenmode of the channel matrix and the received signals are combined using \ac{MRC}, which requires both \ac{CSIT} and \ac{CSIR} \cite{Lo_1999}.
In \cite{Liqun_2021}, \ac{MRT} without \ac{CSIR} is proposed   for a scenario with single-antenna \ac{ES} and \acp{ED}. In this approach,
 the symbols are multiplied with the conjugate of the channel coefficients. Hence, the transmit power can be more effectively utilized as compared to that of  \acp{PC} if the symbols at the \ac{ED} observe channel coefficients with different magnitudes. However, the norm-square of the channel appears on the calculated function (see \tablename~\ref{table:pre-equalization}). 
The power control factor for this approach can be designed based on average transmit power (i.e., $\normalizationCoef[\indexED]=1/\numberOfDimensionAtTX$) or instantaneous transmit power (i.e., $\normalizationCoef[\indexED]=1/\norm{\channelVector[\indexED]}_2^2$ for $\numberOfDimensionAtTX>1$) \cite{Jingon_2018}.


\paragraph{What can go wrong?} 
\label{parag:whatCanGoWrongOne}
The methods in this category requires  {\em accurate} \acp{CSIT} at the \acp{ED}. However, as discussed in \cite{Goldenbaum_2013tcom}, this can impose stringent requirements on the underlying mechanisms such as  time and phase synchronizations, channel estimation, and channel prediction, which can be challenging to satisfy without a clock synchronization and/or under the non-stationary  channel conditions in mobile wireless networks \cite{Careem_2020,Haejoon_2021}. A sample-level precise synchronization (i.e., not just within the \ac{CP} duration of an \ac{OFDM} symbol) is needed for the methods that are sensitive to the phase distortion as the \ac{CSI} is a function of the synchronization point at the \ac{ES} and the time-of-arrivals of the transmitted signals from the \acp{ED}. The second challenge is that 
it is not trivial to extend these methods to the cases with $\numberOfDimensionAtRX>\numberOfDimensionAtTX$ (e.g., an \ac{ES} with multiple antennas or a multi-cell computation) \cite{Sami_2022}. This is due to the fact that the precoder cannot achieve a channel inversion for a random channel matrix without interference  (i.e., $\channelMatrix[\indexED]\precoder[\indexED]\neq\identityMatrix[\numberOfDimensionAtRX]$  for  $\numberOfDimensionAtRX>\numberOfDimensionAtTX$). One potential solution to this issue is coordination among the \acp{ED} (e.g., through some orchestration by an \ac{ES} or multiple \acp{ES}) as done in uniform forcing, which unfortunately requires \ac{CSIR}.

\subsubsection{CSIT: Available, CSIR: Available}
In this category, we assume that the \ac{CSI} is available at both \ac{ES} and \acp{ED}. Hence, it is the most flexible framework for optimizing the \ac{OAC} performance as the precoders, i.e., $\precoder[\indexED], \forall\indexED$, and the decoder, i.e., $\aggregator$, can be designed jointly.

{\color{\reviewColor}
\paragraph{ZF and MMSE coordinations}

For \ac{ZF} coordination \cite{Chen_2018,chen_2018jiot,Guangxu_2019iotj,Yang_2020, Sedaghat_2022coordination}, the noise at the \ac{ES} is ignored. In this case, if $\aggregator^{\rm H}\channelMatrix[\indexED]\channelMatrix[\indexED]^{\rm H}\aggregator$ is an invertible matrix, $\forall\indexED$,  for a given  $\aggregator\triangleq\aUnitaryMatrix/\sqrt{\normalizationCoef[]}$ for any $\aUnitaryMatrix\in\complexNumbers^{\numberOfModulationSymbols\times\numberOfDimensionAtRX}$ such that $\traceOperator[{\aUnitaryMatrix^{\rm H}\aUnitaryMatrix}]=\numberOfModulationSymbols$, the precoder $\precoder[\indexED]$, $\forall\indexED$, can be obtained as 
\begin{align}
\precoder[\indexED]=(\aggregator^{\rm H}\channelMatrix[\indexED])^{\dagger}=\sqrt{\normalizationCoef[]}(\aUnitaryMatrix^{\rm H}\channelMatrix[\indexED])^{\rm H}(\aUnitaryMatrix^{\rm H}\channelMatrix[\indexED]\channelMatrix[\indexED]^{\rm H}\aUnitaryMatrix)^{-1}~,
\label{eq:zfInversion}
\end{align}
where  $\max_\indexED\traceOperator[{\precoder[\indexED]\precoder[\indexED]^{H}}]\le\aFixedPowerValue$ must hold true for a given maximum transmit power $\aFixedPowerValue$. By evaluating the condition further, the power control factor $\normalizationCoef[]$ can be obtained as
\begin{align}
\normalizationCoef[]=\frac{\aFixedPowerValue}{\min_{\indexED}\traceOperator[({\aUnitaryMatrix^{\rm H}\channelMatrix[\indexED]\channelMatrix[\indexED]^{\rm H}\aUnitaryMatrix})^{-1}]}.
\label{eq:powerNormFactor}
\end{align}
Hence, the corresponding \ac{MSE} of the superposed modulation symbols can be expressed as
\begin{align}
	\meanSquareErrorOfMod[{\normalizationCoef[],\aUnitaryMatrix}]&\triangleq\expectationOperator[\norm*{\modulationVectorSuperposed-\sum_{\indexED=1}^{\numberOfEdgeDevices}\modulationVector[\indexED]}_2^2][] =\noiseVariance\frac{\traceOperator[{\aUnitaryMatrix^{\rm H}\aUnitaryMatrix}]}{\normalizationCoef[]}~.
	\label{eq:zfcoorerror}
\end{align}
By substituting $\normalizationCoef[]$ into \eqref{eq:zfcoorerror}, the optimization problem for \ac{ZF} coordination can be written by 
\begin{align}
	\aUnitaryMatrix^*=\arg\max_{\substack{\aUnitaryMatrix} }\min_{\indexED}\quad&\traceOperator[({\aUnitaryMatrix^{\rm H}\channelMatrix[\indexED]\channelMatrix[\indexED]^{\rm H}\aUnitaryMatrix})^{-1}]\label{eq:ZFproblem}\\
\textrm{s.t.} \quad& \traceOperator[{\aUnitaryMatrix^{\rm H}\aUnitaryMatrix}]=\numberOfModulationSymbols~.
\nonumber
\end{align}
Since $\aggregator^{\rm H}\channelMatrix[\indexED]\channelMatrix[\indexED]^{\rm H}\aggregator$ is assumed to be an invertible matrix, $\forall\indexED$, the maximum number of computable functions is $\numberOfModulationSymbols\le\min\{\sequenceLength,\sequenceLengthOther\}$.
The computation rate can be calculated as $\numberOfModulationSymbols/\sequenceLengthOther$ if both real and imaginary parts of the symbols are used for computation. If ${\precoder[\indexED]}$ and $\aggregator$ represent the precoder and decoder for a multi-antenna system, respectively,  $\numberOfSpatialStreams=\numberOfModulationSymbols$ spatial streams via multiple antennas can be utilized to compute $\numberOfModulationSymbols$ functions in parallel, i.e., a higher computation throughput (see the definition in Section~\ref{subsec:computationRate}).
 Also, note that the  design problem for \ac{OAC} is the dual of the beamforming
optimization for the \ac{DL} multi-casting \cite{Chen_2018,Guangxu_2019iotj}.

In \cite{Chen_2018,Yang_2020, Sedaghat_2022coordination}, ZF coordination is investigated for a scenario with single-antenna \acp{ED} and a multiple-antenna \ac{ES}. In \cite{Chen_2018}, an algorithmic approach based on semidefinite relaxation is considered to solve \eqref{eq:ZFproblem}. In \cite{Yang_2020} and \cite{Sedaghat_2022coordination}, \ac{ZF}-coordination is studied based on device subset selection. In \cite[Theorem 1]{Sedaghat_2022coordination}, a closed-form solution is provided when only a single ED transmits with the maximum power under certain channel conditions. The case  where both \acp{ED} and \ac{ES} have multiple antennas is studied in \cite{chen_2018jiot} and \cite{Guangxu_2019iotj}. In \cite{chen_2018jiot}, \ac{ZF} coordination is investigated for $\aUnitaryMatrix^*=\identityMatrix[\numberOfModulationSymbols]$ and $\numberOfDimensionAtTX\ge\numberOfDimensionAtRX=\numberOfModulationSymbols$. An approximate solution for    $\aUnitaryMatrix^{\rm H}\aUnitaryMatrix=\identityMatrix[\numberOfModulationSymbols]$ is provided in \cite[Eq. 14]{Guangxu_2019iotj}.

With \ac{MMSE} coordination \cite{Sedaghat_2022coordination,cao2020Optimized,liu2020over,Fan_2019tvt,Shusen_2023}, the main goal is to minimize the \ac{MSE} of the superposed modulation symbol vector $\modulationVectorSuperposed$, where the MSE can be written as a function of $\aggregator$ and $\precoder[\indexED], \forall\indexED$, by
\begin{align}
	\meanSquareErrorOfMod[{\aggregator,\{\precoder[\indexED]\}}]&\triangleq\expectationOperator[\norm*{\modulationVectorSuperposed-\sum_{\indexED=1}^{\numberOfEdgeDevices}\modulationVector[\indexED]}_2^2][] \nonumber \\&= \expectationOperator[\norm*{\sum_{\indexED=1}^{\numberOfEdgeDevices}(\aggregator^{\rm H}\channelMatrix[\indexED]\precoder[\indexED]-\identityMatrix[\numberOfModulationSymbols])\modulationVector[\indexED]+\aggregator^{\rm H}\noiseVector}_2^2][]\nonumber\\
	&=\sum_{\indexED=1}^{\numberOfEdgeDevices}\traceOperator[({\aggregator^{\rm H}\channelMatrix[\indexED]\precoder[\indexED]-\identityMatrix[\numberOfModulationSymbols]})({\aggregator^{\rm H}\channelMatrix[\indexED]\precoder[\indexED]-\identityMatrix[\numberOfModulationSymbols]})^{\rm H}]
	\nonumber\\&~~~~~~~~~~~~~~~~~~~~~~~~~~~~~~~~~~~+\noiseVariance\traceOperator[{\aggregator^{\rm H}\aggregator}]~,\nonumber
\end{align}
for $\expectationOperator[{\modulationVector[i]\modulationVector[j]^{\rm H}}][]=\kroneckerDelta[i][j]\identityMatrix[\numberOfModulationSymbols]$, $\forall i,j$. 
For \ac{MMSE} coordination, the optimization problem can be expressed as
\begin{align}
	({{\aggregator^*,\{\precoder[\indexED]^*\}}})=\arg\min_{\substack{\aggregator,\{\precoder[\indexED]\}} }\quad&\meanSquareErrorOfMod[{\aggregator,\{\precoder[\indexED]\}}]\label{eq:MMSEproblem}\\
	\textrm{s.t.} \quad& \traceOperator[{\precoder[\indexED]\precoder[\indexED]^{H}}]\le\aFixedPowerValue,\forall\indexED
	\nonumber
\end{align}

The MMSE coordination differs from the ZF coordination in that it leads to a combination of maximum power transmission
and channel inversion across the EDs, and the optimal precoders and decoder are functions of the noise variance at the ES. In \cite{cao2020Optimized} and \cite{Evgenidis2023}, the authors investigate \ac{MMSE}-coordination for a scenario with single-antenna ES and EDs. \cite{cao2020Optimized}  extends the optimization problem to time-varying channels while \cite{Evgenidis2023} investigates the
effect of imperfect CSI on the computation and provides several closed-form solutions based on some approximations. In \cite{liu2020over} and \cite{Sedaghat_2022coordination}, a scenario where the EDs are equipped with a single antenna while ES has multiple antennas is considered. It is shown that the MMSE coordination is related to the device subset selection problem.
In \cite{Fan_2019tvt}, a single function, i.e., $\numberOfModulationSymbols=1$, is aim to be computed in \ac{MIMO} channel and an iterative algorithm described in \cite[Eqs. 11-13]{Fan_2019tvt} is adopted for \ac{MMSE} coordination. In \cite{Park_2023mdpi}, \ac{MMSE} coordination is considered by taking the normalization of the parameters into account for distributed sensing.}

In the literature, the optimization of the aforementioned linear precoders and decoder have been investigated for various interesting applications and scenarios.
In \cite{Wang_2022}, analog beamforming is investigated for \ac{OAC}, where the main objective is to minimize a bound on the loss function for \ac{FEEL}, instead of \ac{MSE}.
In \cite{Haoyu_2022}, the scenario is extended to  a multi-cluster network and uniform forcing is investigated under inter-cluster interference  in selective fading.
In \cite{Lee_2020}, the receive beamforming is optimized based on antenna selection.  In \cite{jiang2021joint}, the same scenario is investigated with the consideration of a relay network with multiple antennas and the beamforming vectors at the \acp{ED}, \acp{ES}, and relays are jointly optimized. 
To calculate multiple linear functions, the design of beamforming vectors at the \ac{ES} and \ac{ED}  is discussed in \cite{Huang_2015}. In \cite{Zhenyi_2022}, a general distribution optimization problem is investigated when the radios are equipped with a large number of antennas for transmission and full-duplex capability. Beamforming vectors are proposed to be optimized to support multiple concurrent computations. In \cite{Bereyhi_2022,Shi_2022}, several channel inversion techniques along with scheduling are investigated when there are multiple antennas at the ES. For a given maximum tolerable computation error \cite{Bereyhi_2022} or \ac{FEEL} performance  \cite{Shi_2022}, greedy scheduling algorithms are proposed. In \cite{NakaiGC_2022}, the authors design the precoders  at the EDs and the decoder at the ES with the considerations of both spatial correlation and heterogeneous data correlation to minimize \ac{MSE} further. In \cite{xufengGC_2022}, simultaneous federated learning with \ac{OAC} and information transmission is studied.

\paragraph{Diversity-oriented techniques}
In the literature, some methods exploit time and/or frequency diversity techniques to improve the performance of \ac{OAC} based on pre-equalization. For example, in \cite{qin2021over}, a \ac{WSN} scenario, where each sensor sends a symbol over multiple subcarriers, is considered. The pre-equalization vector at the sensors and the aggregation vector that combines the copies at the fusion center are jointly optimized such that \ac{MSE} is minimized under a power constraint. The utilized precoder and combining vectors are similar to the ones with multiple antennas \cite{Yang_2020} since the effective channel can be expressed as a diagonal \ac{MIMO} matrix. 
In \cite{Zheng_2022}, a space-time approach for multiple \acp{ED} and \acp{ES} is also investigated  to minimize the \ac{MSE} of the computation. Similarly, in  \cite{tang2021multislot}, a multi-slot \ac{OAC} is proposed for fast-fading channels and time diversity is exploited to mitigate the impact of fading channels on \ac{OAC}.

In \cite{Joung_2021}, an \ac{OAC} strategy using a \ac{STLC} \cite{Jingon_2018} is proposed to achieve a receive diversity gain. In this approach, each \ac{ED} is equipped with a single antenna while the \ac{ES} has two antennas.  Each \ac{ED} performs two \ac{STLC} symbol transmissions back-to-back as a function of the \ac{CSI}, where \ac{STLC} symbols are generated from the same sensor information. The \ac{ES} combines the received symbols at different time slots and antennas blindly. Nevertheless, the power normalization factor still needs to be known at the \acp{ED}, where the \ac{ES} calculates the factor based on the feedback transmitted from all \acp{ED} in orthogonal channels. 

\paragraph{Channel manipulation}
One way of achieving favorable propagation conditions for \ac{OAC} is to manipulate the multi-path channel itself with technologies like \ac{RIS} \cite{Saifullah_2022}. For example,  in \cite{Wenzhi_2021twc}, \ac{RIS} is utilized to boost or null the received signal power at desired locations.  In \cite{Hang_2021twc}, device selection for \ac{OAC} is studied along with \ac{RIS} with the availability of \ac{CSIT}. In \cite{Liu_2022ris_wcl}, the RSI phase shifts are exploited for over-the-air model
aggregation with the consideration of the cascaded channel
coefficients to eliminate the need for \ac{CSI} at the \acp{ED}.
In \cite{Fusheng_2022ris}, multiple \acp{RIS} are investigated for a similar scenario.
 In \cite{Zhang_2022clt} and \cite{Zheng_2022twc}, the authors consider the optimization problem of transceiver design and RIS phase selection with the consideration of imperfect CSI.  In \cite{Li_2022}, the \ac{signSGD} is exploited for \ac{OAC} along \ac{RIS}.  In \cite{Bouzinis_2022ris}, average \ac{MSE}
with respect to a target function is minimized by jointly optimizing the \ac{RIS} phase-shift vector and the transmission and reception scaling factors of \ac{ED}. In \cite{sanchez2022airnn}, \ac{RIS} is proposed to compute convolution over the air to realize a \ac{CNN} in the wireless channel.  In \cite{Zixin_2023}, the authors investigate a
joint optimization problem concerning the transmit power, denoising factor, and \ac{RIS} phase-shifts for a graph neural network.  We refer the reader  to \cite{Ni_2022}  and the references therein for further optimization frameworks on phase shift design.

\paragraph{What can go wrong?} 
The methods in this category can introduce a major computation complexity  to both \acp{ED} and \ac{ES}. They are also prone to imperfections caused by underlying enabling mechanisms, e.g.,  imperfect coordination among \acp{ED} within the coherence time, and phase, time, and frequency synchronization errors. Hence, the \ac{OAC} performance of these methods is a strong function of how much the underlying link-layer procedures in practice can make the assumptions (e.g., accurate and fresh \ac{CSI} estimates at both \ac{ES} and \ac{ED} within the coherence time) hold.

\subsubsection{CSIT: Not available, CSIR: Available}
This category is dedicated to blind \acp{ED}, i.e., \acp{ED} cannot access the \ac{CSI}, but the \ac{ES} has some knowledge about the \ac{CSI}. The methods in this category particularly rely on channel hardening techniques.  
\paragraph{Channel hardening by using the aggregated CSI}
One way of achieving channel hardening for \ac{OAC} is to use an estimate of the superposed channel, i.e., the sum of the channel gains from all the \acp{ED} (i.e., $\sum_\indexED\channelVector[\indexED]$) to derive the linear decoder $\aggregator$ at the \ac{ES} \cite{Amiria_2021,weiICC_2022,aygunICC_2022,aygunArXiv_2022,Ozan_2022}. Since the \ac{CSI} between each \ac{ED} and \ac{ES} is not needed with this strategy,  the channel estimation overhead significantly reduces at the expense of some interference due to the uncoordinated transmissions of the \acp{ED}. In \cite{Amiria_2021,weiICC_2022}, this approach is adopted based on multiple antennas at the \ac{ES}. It is assumed that the \ac{ES} has a noisy estimate of the aggregated channel from all the devices to each antenna and employs \ac{MRC}. It is shown that the variance of  interference on the superposed symbol is scaled by $(\numberOfEdgeDevices-1)/\numberOfDimensionAtRX$, where $\numberOfDimensionAtRX$ is the number of antennas at the \ac{ES}  \cite{Amiria_2021,weiICC_2022}. In \cite{Busra_2023},  the same scenario is investigated for time-varying channels and 
it is shown that the time-variation in typical wireless channels does not reduce the
\ac{FEEL} performance  and  the inter-carrier interference reduces with the increasing number of receive antennas.

\paragraph{Advanced receivers}
In \cite{ZhaoICC_2022,ZhaoICC_2022_extension}, the digital \ac{OAC} problem is interpreted as a multi-user detection problem. Considering an asynchronous multi-user \ac{OFDM} scenario, it is demonstrated that multi-user detection algorithms can be applied to the superposed signals for convolution code or \ac{LDPC}. In \cite{Lizhao_2022lora}, a similar multi-user detection and aggregation approach is considered for \ac{LoRa} networks. In \cite{Becirovic_2022}, it is proposed to  separate the transmitted
signal of each client from the superposed signals so that independent sparsification patterns can be applied at the \acp{ED} by assuming that the number of antennas is larger than the number of \acp{ED}.

\paragraph{What can go wrong?}
Channel hardening relies on the existence of a large number of \ac{DoF} at the \ac{ES}, i.e., $\numberOfDimensionAtRX$, to decrease  the variance of  interference, which can increase the cost {\color{\reviewColor}and decrease the computation rate $\computationRate$ as can be seen from Table~\ref{table:pre-equalization}}. Also, if $\numberOfDimensionAtRX$ is larger than or equal to $\numberOfEdgeDevices\numberOfDimensionAtTX$ and the complete \ac{CSIR} is available at the \ac{ES}, the separation of computation and communication tasks can be more reliable than the \ac{OAC} as the number of observations can be larger than or equal to the number of unknowns. The main limitation of advanced receivers for the detection of codewords from the superposed signals is that it is not trivial to extend the computation to a large number of \acp{ED} as the computation and storage complexity can be prohibitively high.

\subsubsection{CSIT: Not available, CSIR: Not available}
\label{subsubsec:noncoherent}
The methods in this category is the most restrictive in the sense of optimization as the \ac{CSI} is available neither at the \acp{ED} nor the \ac{ES}. Although this appears as a prohibiting factor for a reliable \ac{OAC}, the main benefits gained by not using \ac{CSI} are the robustness against time-variation of the wireless channel and synchronization errors and a major overhead reduction as compared to the methods relying on the availability of \ac{CSIT} or \ac{CSIR}.

\paragraph{Orthogonal signaling}
\label{par:orthSignaling}
In \cite{sahinCommnet_2021,sahinWCNC_2022,safiPIMRC_2022,mohammadICC_2022}, the authors consider distributed training by the \ac{MV} with \ac{signSGD} \cite{Bernstein_2018} and calculate the \ac{MV} based on orthogonal signaling at the \acp{ED} and non-coherent detection at the \ac{ES}.
Since the arguments and the output of the nomographic function in this specific application consist of discrete states, i.e., $\{-1,1\}$,  the schemes in \cite{sahinCommnet_2021,Sahin_2022MVjournal,sahinWCNC_2022,safiPIMRC_2022}, and  \cite{safiINFOCOM_2023} use \ac{FSK}  over \ac{OFDM}, \ac{PPM}, and \ac{CSK} over \ac{DFT-s-OFDM}, where the symbols are determined based on $\signNormal[{s_{\indexED}}]$. An energy comparator is used to detect the \ac{MV} at the \ac{ES}.  The authors show the efficacy of this approach for distributed learning under time synchronization errors without requiring phase synchronization (see \cite{sahinGC_2022} for demonstration). Since these schemes do not use \ac{CSIT} and \ac{CSIR}, they can also be used for multi-cell computation, where there are multiple fusion nodes \cite{mohammadICC_2022}. 
The authors in \cite{Akshay_IPSN2020} also propose to utilize orthogonal resources for negative and positive-valued measurements. As compared to \cite{Akshay_IPSN2020}, in \cite{NingningINFOCOM_2023}, multiple orthogonal resources are used for data encoding while \cite{safiINFOCOM_2023} uses them for multiple \ac{MV} computations. Note that the distributed training based on \ac{MV} with \ac{signSGD} is considered in \cite{Guangxu_2021} and \cite{Li_2022}, where the \ac{OAC} relies on \ac{TCI} discussed in Section~\ref{parag:TCI} and the use of \ac{RIS}, respectively. \ac{signSGD} along with error feedback is investigated in \cite{liu2023overtheair},
However, phase synchronization is needed for these methods as \cite{Guangxu_2021,Li_2022,liu2023overtheair} use the same resource for negative and positive gradients. 

The keying idea is generalized in \cite{sahin2022md,sahinGCbalanced_2022} to compute general nomographic functions by using balanced numerals. 
With similar motivations, i.e., robustness against synchronization errors, Valenti and Ferrent study \ac{FSK} for \ac{PLNC} in \cite{Valenti_2011,Ferrett_2018}, where the target function is the XOR of the transmitted bits from two devices. In \cite{yang2021revisiting,Zihan_2023}, it is proposed to project the gradient vector onto a set of orthogonal basis.
In \cite{Jakimovski_2011}, a positional encoding for \ac{OAC} without using \ac{CSI} is discussed. Based on the fraction of time slots occupied among a fixed number of available slots, the activated classes that encode temperature ranges are estimated. In \cite{Meiyi_2022tbma}, a codebook for \ac{TBMA} is proposed to be designed via a neural network with the considerations of channel and source statistics.

\paragraph{Channel hardening  and energy estimation}
\label{parag:ch_second}
In \cite{Goldenbaum_wcl2014}, Goldenbaum and Sta\'nczak re-evaluate their scheme presented in \cite{Goldenbaum_2013tcom}  when there is no \ac{CSIT} under a scenario where the receiver is equipped with multiple
antennas. In this approach, each \ac{ED} transmits a sequence as discussed in Section~\ref{parag:ac}. However, it does not apply any amplitude correction and the receiver at the \ac{ES} calculates the energy of the received sequence over the multiple antennas.  One of the main conclusions drawn from this approach is that the rapid changes of the channel coefficients can be beneficial to improve the convergence when \ac{CSIT} is not available for \ac{iid} small-fading. Note that a similar approach is also adopted in more recent work in \cite{Bjelakovic_2019allerton,Fabio_2021,Fabio_2022}, and \cite[Section VI]{Frey_tsp2021}. To mitigate the channel fading coefficients on \ac{OAC}, averaging over multiple antennas  are investigated in \cite{Choi_2022jsait,sahin2022md}. The impact of an uncompensated channel along with multiple transmissions is  discussed for collaborative-decision making  in \cite{CarlosGC_2022}.

\paragraph{Joint channel and parameter estimation}
 In \cite{dong2020blind}, the blind \ac{OAC} is interpreted as a joint channel and parameter estimation problem. To estimate the channel coefficients and parameters, a randomly initialized Wirtinger flow is proposed. It is demonstrated
 that the proposed approach results in small  estimation errors with sufficient samples.

\paragraph{What can go wrong?} 
The {\color{\reviewColor} error rate or MSE} with orthogonal signaling can be worse than the methods relying on pre-equalization or multi-antenna techniques as the impact of the channel on the symbol is not compensated. To reduce the estimation errors, averaging can consume time, frequency, or space resources {\color{\reviewColor}(i.e., a lower computation rate)} while introducing a complex algorithm to the ES can increase the receiver complexity.

\subsection{Classification criteria: Encoding}
The classification in this category particularly concerns  how the outputs of the pre-processing functions are processed before the linear encoders take place to overcome the fading channel. We group the approaches under analog encoding and digital encoding. While analog encoding  deals with the continuous-valued symbols to realize the desired nomographic function (over an analog modulation), digital encoding uses some form of quantization, compression, and/or source-channel coding for the same goal (over a digital modulation).

\subsubsection{Analog encoding}
In the literature, a majority of the  \ac{OAC} methods exploit the available \ac{DoF} without a particular encoding as long as the reliable superposition is maintained under the fading channel  with aid of the precoders $\precoder[\indexED], \forall\indexED$, and the decoder $\aggregator$. Nevertheless, it has been shown that additional processing can still be helpful for certain goals.

\paragraph{Linear analog encoders}
A linear analog encoder for \ac{OAC} can be defined as
\begin{align}
\encoder[\indexED]\{\preProcessedVector[\indexED]\}=\projectionMatrix\preProcessedVector[\indexED]~,\forall\indexED~,
\label{eq:linAnalogEnc}
\end{align}
where $\projectionMatrix$ is a $\blockLength\times\numberOfFunctions$ matrix for $\blockLength<\numberOfFunctions$ (i.e., compression) \cite{Jha_2021}. Hence, the encoder projects the vector $\preProcessedVector[\indexED]$ into a lower dimensional space and reduces the number of resources to be utilized to compute $\numberOfFunctions$. A linear encoder is a suitable operation for \ac{OAC} because  the sum of the projected vectors is identical to the projection of the sum of  vectors.  

Linear encoders are especially applied to applications 
where the pre-processed symbol vector $\preProcessedVector[\indexED]$ is inherently sparse or can be sparsified with a tolerable distortion. For example, in \cite{amiri2020machine} and \cite{Amiri_2020},  for distributed learning, the \acp{ED} sparsify their gradient estimates before they project them into a lower dimensional space with \eqref{eq:linAnalogEnc}. These projections are directly used with an analog \ac{OAC} scheme. At the \ac{ES}, an approximate message passing algorithm is proposed to recover the superposed symbols. In \cite{Jha_2021}, with a similar motivation, $\projectionMatrix$  is constructed randomly from the rows of the rotated Walsh-Hadamard matrix. At the \ac{ES}, the projected vector is mapped to the superposed symbol  based on a general norm-minimization problem, i.e., $\decodedVector=\decoder\{\demappedVector\}=\arg\min_{\decodedVector'}\norm{\demappedVector-\projectionMatrix\decodedVector'}_2$. It is shown that linear analog encoders in the \ac{AWGN} channel can perform well at low \acp{SNR}. In \cite{Haoming_2022}, the sparsification is achieved by selecting $k$ gradients with the greatest magnitudes, where $k$ is a predetermined integer. In this study, a discrete cosine transform matrix is used for the compression. In \cite{Becirovic_2022}, a similar compression approach is considered for the model updates and the measurement matrix is constructed based on \ac{iid} Gaussian distribution.  

One of the concerns on \ac{OAC} with sparsification is that the sum of the sparse vectors may not be sparse after the aggregation, i.e., the sparsity level can deteriorate depending on the number of \acp{ED}. In \cite{Becirovic_2022}, the authors study a uniform sparsity pattern and independent sparsity patterns for \ac{FEEL} and show that a higher test accuracy can be achieved when the \acp{ED} decide their sparsity pattern, especially when the data distribution is heterogeneous.

\paragraph{Affine analog encoders}
\label{parag:affine}
 In \cite{Goldenbaum_2013tcom,Goldenbaum2010asilomar,Zheng_cn2017, Zheng_tgcn2017,Frey_tsp2021, razavikiaGC_2022,Busra_2023}, it is proposed to process the output of the pre-processing function with an affine function as
\begin{align}
\encoder[\indexED]\{\preProcessedVectorEle[\indexED][\indexSampleTime]\}= a\preProcessedVectorEle[\indexED][\indexSampleTime]+b~,\forall\indexED
\end{align} 
for $a,b\in\realNumbers$, where $a$ and $b$ are chosen such that the output of the encoder always results in a non-negative value (see \cite[Eq. (28)]{Frey_tsp2021} for an example under a power normalization constraint and discussions in Section~\ref{parag:ac}). Hence, the sign ambiguity at the receiver is eliminated. After the energy calculation of the received sequence at the receiver, the superposed symbol is processed with the inverse function $\decoder\{x\}=(x-\numberOfEdgeDevices b)/a$. Note that the encoder must be an affine function to be able to calculate the desired nomographic function \cite[Proposition 1]{Goldenbaum_2013tcom}. 

{\color{\reviewColor}As discussed in Section~\ref{parag:ac}, the encoder proposed in \cite{Goldenbaum_2013tcom} multiplies the square-root of the affine function with a sequence as
\begin{align}
	\encoder[\indexED]\{\preProcessedVectorEle[\indexED][\indexSampleTime]\}=\sqrt{a\preProcessedVectorEle[\indexED][\indexSampleTime]+b}\times[
	{\constante^{\constantj\theta_1}} ,\mydots, {\constante^{\constantj\theta_\sequenceLength}}]~,\forall\indexED.
\end{align}
The fundamental reason for the square-root operation is that the \ac{OAC} relies on the estimation of the energy of the sequence at the receiver.}

\paragraph{Non-linear encoders}
In \cite{park_2022tcom, Guangxu_2021,Shao_2021,aygunICC_2022,Becirovic_2022,Haoming_2022,Zhong_2022uav,zhong2021overtheair}, the imaginary and real parts of the  symbols are proposed to be utilized as
 \begin{align}
	\encoder[\indexED]\{\preProcessedVectorEle[\indexED][\indexSampleTime],\preProcessedVectorEle[\indexED][\indexSampleTime+1]\}= \preProcessedVectorEle[\indexED][\indexSampleTime]+\constantj\preProcessedVectorEle[\indexED][\indexSampleTime+1],
\end{align}
for all odd positive $\indexSampleTime$. Therefore, the spectrum efficiency of the transmission is improved as $\blockLength$ is equal to $\ceil{\numberOfFunctions/2}$.

In \cite{jankowski2022airnet}, Shannon-Kotelnikov mapping (see \cite{Hekland_2009,floor2022shannonkotelnikov} and the references therein) is utilized to map the neural network parameters onto continuous-values on Archimedes’ spirals for unequal error
protection and improving bandwidth efficiency. Although the proposed method is not used for OAC, introducing similar analog coding approaches to OAC is an unexplored area and may improve the reliability of the computation.

\subsubsection{Digital encoding}
In this subsection, we discuss the methods that use some form of source and channel coding for \ac{OAC}.
\paragraph{Nested lattice codes}
\label{parag:genNomoFunc}
The linearity of the nested lattice codes are first used in \cite{Nazer_2007} and \cite{Nazer_2011} for achieving computation over Gaussian channels in finite fields (see Section~\ref{subsec:computationRate} for further discussions).
In \cite{goldenbaum2015nomographic} and \cite{Goldenbaum_2013icassp}, a digital \ac{OAC} scheme along with a nested lattice coding is introduced to  increase the reliability of the computation of $\numberOfFunctions$ functions in a real-valued \ac{AWGN} channel. To calculate $\sum_{\indexED=1}\preProcessedVectorEle[\indexED][\indexSampleTime]$, $\forall\indexSampleTime$, the steps taken at the $\indexED$th \ac{ED}, $\forall\indexED$, based on \cite[Proof of Theorem 5]{goldenbaum2015nomographic}, can be listed as follows:
\begin{itemize}
\item Step 1 (Quantization): Let $\numberOfDigitsBits$ be the quantization parameter
in bits, which is specified based on the amount of tolerable quantization error. The  symbol $\preProcessedVectorEle[\indexED][\indexSampleTime]$, $\forall\indexSampleTime$, is mapped to a positive integer $\integersGoldenbaum[\indexED][\indexSampleTime]\in\{0,1,\mydots,2^\numberOfDigitsBits-1\}$  for a given  $\numberOfDigitsBits$.
For example, for $\numberOfDigitsBits=3$ bits, $\numberOfEdgeDevices=3$ \acp{ED}, and $\numberOfFunctions=4$ functions to be computed, suppose that the \acp{ED}' symbols within the range $[0,1]$ are given by
\begin{align}
	\begin{tabular}{c|c|c|c|c|}
		$\preProcessedVectorEle[\indexED][\indexSampleTime]$ & $\indexSampleTime = 1$ & $\indexSampleTime =2$ & $\indexSampleTime =3$ & $\indexSampleTime =4$\\\hline
		$\indexED=1$& 1&1&0.9&0.4\\\hline
		$\indexED=2$&1&1&0&0.7\\\hline
		$\indexED=3$&1&1&0.2&0.1\\\hline
	\end{tabular}~.
\end{align}
By uniformly dividing the range $[0,1]$ into $2^3$ parts and using a natural code, the corresponding integers can be obtained as
\begin{align}
	\begin{tabular}{c|c|c|c|c|}
		$\integersGoldenbaum[\indexED][\indexSampleTime]$ & $\indexSampleTime = 1$ & $\indexSampleTime =2$ & $\indexSampleTime =3$ & $\indexSampleTime =4$\\\hline
		$\indexED=1$&7&7&6&3\\\hline
		$\indexED=2$&7&7&0&5\\\hline
		$\indexED=3$&7&7&1&1\\\hline
	\end{tabular}~.
	\label{eq:quan}
\end{align}

\item Step 2 (Source encoder): The integers obtained from the quantization step, i.e., $\integersGoldenbaum[\indexED][\indexSampleTime]$, $\forall\indexSampleTime$, are divided into $\numberOfGroups$ sub-sequences and each sub-sequence is mapped to a message as an integer. The procedure is as follows: Each sub-sequence contains  $\numberOfParametersGoldenbaum=\numberOfFunctions/\numberOfGroups$  integers for a given $\numberOfGroups$. The $\numberOfParametersGoldenbaum$ integers in each sub-sequence is first assigned to the digits of a number, e.g., $(\integersGoldenbaum[\indexED][\numberOfParametersGoldenbaum]\mydots\integersGoldenbaum[\indexED][2]\integersGoldenbaum[\indexED][1])_\base$, in the base-$\base$ positional numeral system for $\base\triangleq\numberOfEdgeDevices(2^\numberOfDigitsBits-1)+1$. Afterward, each number in base $\base$ is converted to an integer, e.g., $\sum_{t=1}^{\numberOfParametersGoldenbaum}\integersGoldenbaum[\indexED][t]\base^{t-1}$. Note that the source encoding results in $\numberOfGroups$ messages, where the maximum value of a message is $(\base^\numberOfParametersGoldenbaum-1)/\numberOfEdgeDevices$.

The reason for choosing $\base$ as a function of number of \acp{ED} is to eliminate a potential carry digit for the superposed message across $\numberOfEdgeDevices$ \acp{ED} so that  $\sum_\indexED(\integersGoldenbaum[\indexED][\numberOfParametersGoldenbaum]\mydots\integersGoldenbaum[\indexED][2]\integersGoldenbaum[\indexED][1])_\base$  can be expressed as $(\sum_\indexED\integersGoldenbaum[\indexED][\numberOfParametersGoldenbaum]\mydots\sum_\indexED\integersGoldenbaum[\indexED][2]\sum_\indexED\integersGoldenbaum[\indexED][1])_\base$.  For example, consider the integers in \eqref{eq:quan}. Suppose that the source encoder results in $\numberOfGroups=2$ messages. Hence, each message is calculated over $\numberOfParametersGoldenbaum=\numberOfFunctions/\numberOfGroups=2$ integers by mapping them to the digits of a number in a positional numeral system. The corresponding numbers are then converted to the messages as
\begin{align}
	\begin{tabular}{c|c|c|c|c|}
		$\indexED=1$& $(77)_\base$&$(63)_\base$\\\hline
		$\indexED=2$&$(77)_\base$&$(05)_\base$\\\hline
		$\indexED=3$&$(77)_\base$&$(11)_\base$\\\hline
	\end{tabular}\xrightarrow{\substack{\text{From base}~\base\\\text{to base 10}}}\begin{tabular}{c|c|c|c|c|}
		$\indexED=1$& $161$&$135$\\\hline
		$\indexED=2$&$161$&$5$\\\hline
		$\indexED=3$&$161$&$23$\\\hline
	\end{tabular}~,
	\label{eq:convDec}
\end{align}
for $\base=\numberOfEdgeDevices(2^\numberOfDigitsBits-1)+1=22$. Hence, the sum of the messages across $\numberOfEdgeDevices=3$ \acp{ED} can be calculated without a carry digit, e.g., the sum $(77)_{22}+(77)_{22}+(77)_{22}$ is $(21,21)_{22}\equiv(483)_{10} $ for $\indexSampleTime = 1$. Note that the maximum value of the superposed message for this example is $\base^\numberOfParametersGoldenbaum-1=483$ while the maximum value of a message is $(\base^\numberOfParametersGoldenbaum-1)/\numberOfEdgeDevices=161$ as can be seen from \eqref{eq:convDec}.

\item Step 3:  (Channel encoder): A nested lattice code from $\integers_{\aprime}^{\numberOfGroups}$ to $\realNumbers^{\blockLength}$  is constructed for a prime $\aprime\ge\base^\numberOfParametersGoldenbaum$. The $\numberOfGroups$ messages calculated from the previous step are used to calculate the codeword $\encodedVector[\indexED]\in\complexNumbers^{\blockLength}$ with the corresponding generator matrix $\generatorMatrix\in \realNumbers^{\blockLength\times\numberOfGroups}$ of the lattice   under an average power constraint. For example, the codewords at  \acp{ED} for the messages in \eqref{eq:convDec} can be calculated as 
\begin{align}
	\encodedVector[\indexED=1]=\generatorMatrix\begin{bmatrix}161 \\ 135\end{bmatrix},
	\encodedVector[\indexED=2]=\generatorMatrix\begin{bmatrix}161 \\ 5\end{bmatrix},
	\encodedVector[\indexED=3]=\generatorMatrix\begin{bmatrix}161 \\ 23\end{bmatrix}~.
	\label{eq:codewords}
\end{align}

\end{itemize}
The \ac{ES} receives the sum of the codeword, i.e., $\demappedVector$. For example,  for the codewords in \eqref{eq:codewords},  $\demappedVector$ can be expressed as
\begin{align}
	\demappedVector=\sum_{\indexED=1}^{3}\encodedVector[\indexED].
	\nonumber
\end{align}
The superposed vector $\demappedVector$ at the \ac{ES} is processed as follows:
\begin{itemize}
\item Step 1 (Channel decoder):  By using an Euclidean nearest neighbor decoder,  the decoder obtains the message, which corresponds to the superposed message due to the linearity of the code. For example,  for the codewords in \eqref{eq:codewords},  $\demappedVector$ can be expressed as
\begin{align}
	\demappedVector=\generatorMatrix\begin{bmatrix}483 \\ 163\end{bmatrix}.
	\nonumber
\end{align}
and the decoder results in  $483$ and $163$ as the superposed messages if there is no error.  

\item Step 2 (Source decoder): 
Each superposed message is expressed  in base $\base$ to obtain the sum of the quantization results, e.g., $(\sum_\indexED\integersGoldenbaum[\indexED][\numberOfParametersGoldenbaum]\mydots\sum_\indexED\integersGoldenbaum[\indexED][2]\sum_\indexED\integersGoldenbaum[\indexED][1])_\base$. For example, $(483)_{10}\equiv(21,21)_{22}$ and $(163)_{10}\equiv(79)_{22}$. Hence, the source decoder returns 
$\decodedVector= [\decodedVectorEle[1],\decodedVectorEle[2],\decodedVectorEle[3],\decodedVectorEle[4]]^{\rm T} =[21,21,7,9]^{\rm T}$, i.e., the sums of the values on each column in \eqref{eq:quan}.
\end{itemize}


\paragraph{Encoding based on numeral systems}

 In \cite{Minjie_2022}, \ac{PAM} is utilized with a radix-based encoding. In this approach, the binary representation of a parameter $\preProcessedVectorEle[\indexED]$ is first partitioned into subgroups. Afterward, the decimal representation of the bits on each subgroup is mapped to a \ac{PAM} symbol to achieve a processing gain (see \cite[Fig.~4]{Minjie_2022}). This approach generalizes the encoder that  encodes $\preProcessedVectorEle[\indexED]$ into an $M$-\ac{PAM} symbol after  $\preProcessedVectorEle[\indexED]$ is quantized into $\log_2 M$ bits \cite[Eq. (13)]{Jha_2021}. Note that \ac{PAM} can be extended to square $M$-\ac{QAM} constellations, where real and imaginary part of the constellation dedicates to two symbols, as done for gradient aggregation for \ac{FEEL} in \cite{Liang_2022}.


In \cite{Xiugang_2016}, the authors propose to calculate 
$	{\sum_{\indexED=1}^{\numberOfEdgeDevices}{\preProcessedVectorEle[\indexED]}}$ where $\preProcessedVectorEle[\indexED]=\weight[\indexED]\symbolVectorEle[\indexED]$ for non-negative integers $\symbolVectorEle[\indexED],\weight[\indexED]$.  They consider the binary representation of each symbol and map each bit to a \ac{BPSK} symbol. By using the linearity of the decomposition, the weights $\weight[\indexED]$ are incorporated to the transmit power at the source nodes to calculate the weighted sum. The main observation is that the received symbol is a point in a non-standard \ac{PAM} after the superposition. The destination node exploits the discrete nature of the superposition to compute the arithmetic sum.  The authors also derive the \ac{CER} (See Section~\ref{sec:whatAreTheMetrics}). In \cite{RazavikiaDigital2023}, the authors investigate the computation problem by mapping the image of the desired function to the discrete points in the constellation after the signal superposition.

In \cite{sahin2022md,sahinGCbalanced_2022}, the authors propose to utilize the balanced number systems for \ac{OAC} to represent a negative parameter efficiently. A balanced numeral system consists of negative numerals, i.e., signed digits. Hence, it can represent a negative number without using a dedicated sign symbol. In this method, the parameters are first represented in a balanced number system. For each numeral, one of the corresponding \ac{OFDM} subcarriers are activated. At the receivers, the sum of the parameters is estimated. \ac{MSE} is also analyzed for multiple antennas at the receiver. If only a single digit is used with balanced ternary system, the encoding in \cite{sahin2022md,sahinGCbalanced_2022} corresponds to the keying methods discussed in \cite{sahinCommnet_2021,mohammadICC_2022, safiPIMRC_2022, sahinWCNC_2022}. The proposed encoding method in \cite{sahinCommnet_2021,mohammadICC_2022, safiPIMRC_2022, sahinWCNC_2022} is designed for calculating the \ac{MV} (see \tablename~\ref{table:nomoFcns}).   
Since the nomographic function for the \ac{MV} consist of discrete states,  the modulation symbols are determined based on keying methods such as \ac{FSK}, \ac{CSK}, and \ac{PPM}, and non-coherent detectors are used to obtain the \ac{MV} at the \ac{ES}.  Note that the same nomographic function is targeted to be computed in \cite{Guangxu_2021} based on \ac{TCI}. In \cite{Sahin_2022MVjournal}, it is shown that using a tri-state encoder (i.e., $\{-1,0,1\}$) that eliminates the EDs with small gradients from the MV computation) can largely address the bias due to the data heterogeneity  and imperfect power control in the learning when OAC is used for \ac{FEEL}. In \cite{safiINFOCOM_2023}, CSK is extended to $M$-ary CSK. The sign of $\log_2 M$ parameters  are mapped to a chirp index  and the authors show how to calculate \acp{MV} for this specific encoding by exploiting the binary representations of the indices.

In \cite{Choi_2022jsait}, the authors propose to encode a local gradient vector of length $L$ into a sequence in an orthogonal sequence set of size $2L$ for \ac{OAC}, where the encoding relies on stochastic quantization. In this scheme, the transmitted sequence is chosen with the probability that is proportional to the magnitude of the elements of the normalized local gradient vector (i.e., the coefficients derived from the decomposition of the normalized local gradient vector over a scaled cross polytope constructed over standard basis vectors). At the receiver, the received signal is correlated  with each sequence in the set to estimate the probabilities. By combining the results over realizations, the superposed gradient vector is obtained. It is worth noting that this scheme also separates the sign information to the orthogonal resources as in \cite{sahinCommnet_2021,mohammadICC_2022, safiPIMRC_2022, sahinWCNC_2022}, but it differs from these studies as it targets a continuous-valued computation through a probabilistic choice of the activated resource over $2L$ resources. In \cite{Michelusi_2022}, the same idea is investigated for the consensus problem.

\section{What are the enabling mechanisms for OAC?}
\label{sec:EnablingMech}
In this section, we discuss the underlying mechanisms which maintain reliable computation and elaborate on security issues.

\subsection{Synchronization}
\label{parag:synch}
{\color{\reviewColor}
	\begin{figure}
	\centering
	\subfloat[Frequency and phase offsets.]{\includegraphics[width =3.5in]{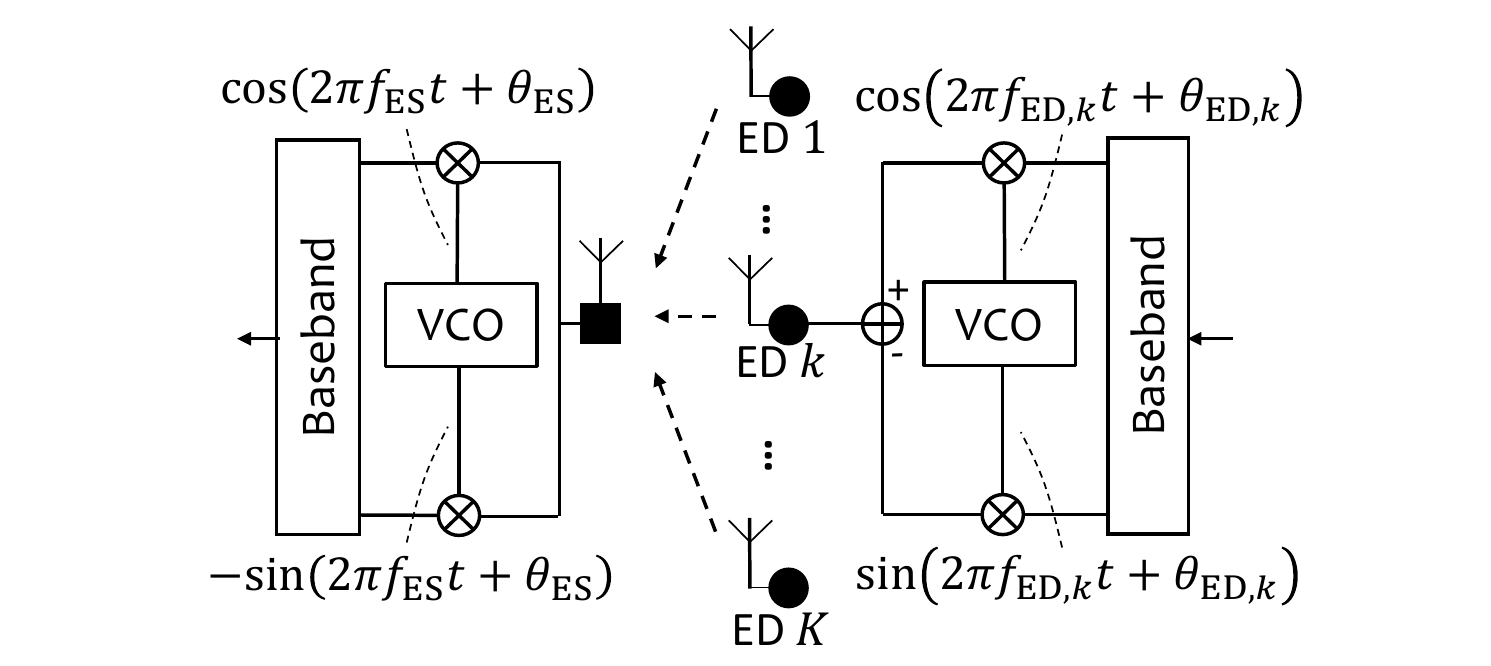}
	\label{subfig:freq}}	
	\\	
	\subfloat[Imperfect time-of-arrivals and the synchronization error at the ES relative to the ideal synchronization point $\syncPoint$.]{\includegraphics[width =3.5in]{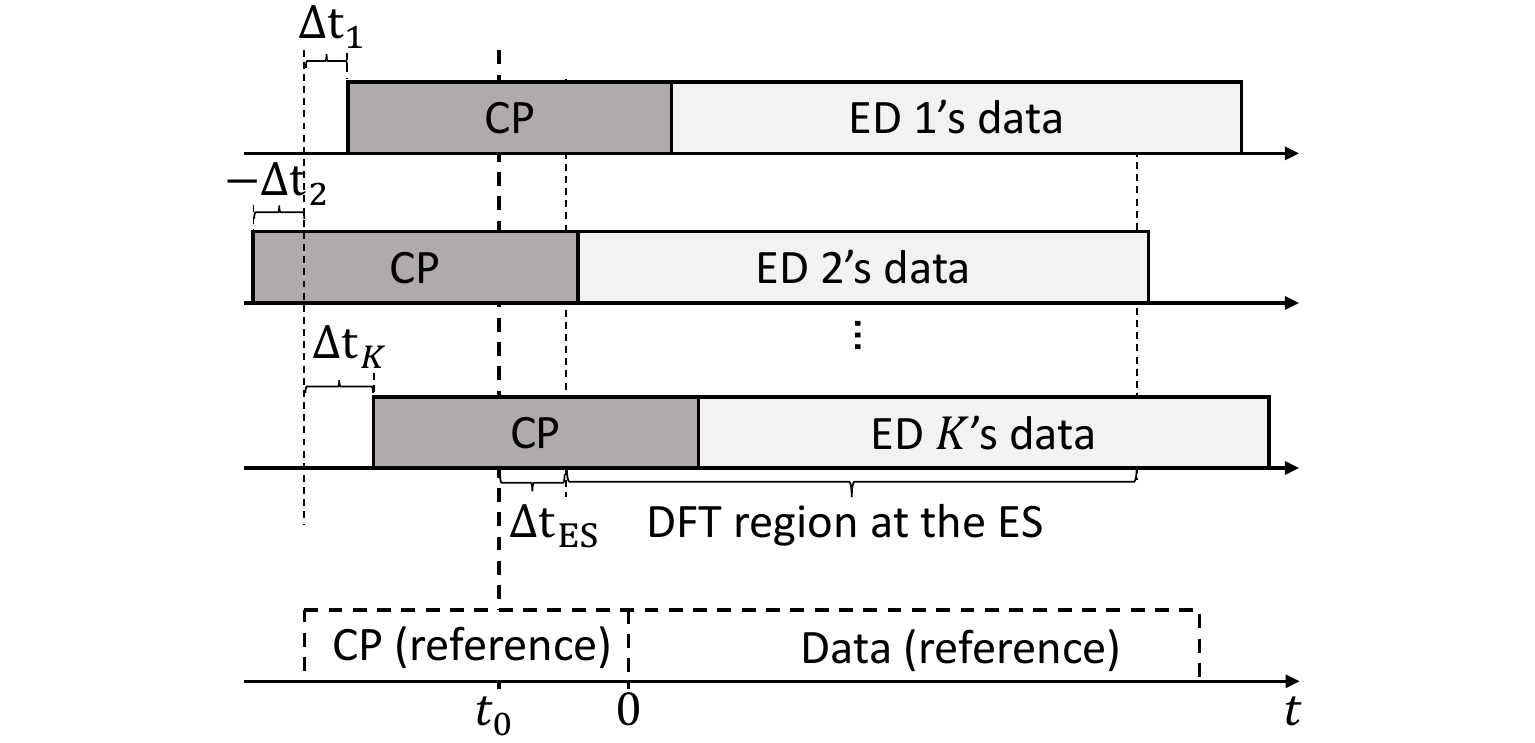}
	\label{subfig:time}}
	\caption{Synchronization imperfections in time and frequency.}	
	\label{fig:sync}	
\end{figure}
To elaborate the effect of synchronization impairments  on \ac{OAC}, let $\fcarrierES$ and $\fcarrierED[\indexED]$ denote the carrier frequencies at the \ac{ES} and the $\indexED$th \ac{ED}, respectively, where the phase of the corresponding oscillators are $\phaseES$ and $\phaseED[\indexED]$, as shown in \figurename~\ref{fig:sync}\subref{subfig:freq}. Thus, the \ac{CFO} and the \ac{PO} between $\indexED$th \ac{ED} and the \ac{ES} are  $\cfo[\indexED]=\fcarrierED[\indexED]-\fcarrierES$ and $\po[\indexED]=\phaseED[\indexED]-\phaseES$, respectively. 
Similarly, the \ac{TO} between transmitter and receiver can be described as the following: Let $\syncPoint$ denote the ideal synchronization point at  the ES. Assume that the synchronization point at the ES and the time-of-arrival instant of the $\indexED$th \ac{ED}'signal at the ES location deviate by $\so$~seconds and $\to[\indexED]$~seconds, respectively. Thus, the overall \ac{TO}  can be expressed as $\toComplete[\indexED]=\to[\indexED]-\so-\syncPoint$.


Let $\transmittedSignal[\indexED][\timeSymbol]\in\complexNumbers$ be the baseband signal  for the $\indexED$th \ac{ED} to be transmitted. Hence, the passband signal of the $\indexED$th ED from the perspective of the ES can be expressed as
\begin{align}
	\transmittedSignalPass[\indexED][\timeSymbol] = \Re\{&\transmittedSignal[\indexED][{\timeSymbol-\toComplete[\indexED]}]\constante^{\constantj2\pi(\fcarrierES+\cfo[\indexED])(\timeSymbol-\toComplete[\indexED])}\constante^{\constantj\po[\indexED]}\}~.
\end{align} 
Assume that the impulse response of the multi-path channel in the passband is given by
\begin{align}
	\channelPass[\delaySymbol]=\sum_{\pathIndex=1}^{\numberOfPaths} \gain[\indexED,\pathIndex]\deltaFunction[{\delaySymbol-\delay[\indexED,\pathIndex]}]~,
\end{align}
where $\numberOfPaths$ is the number of paths, $\gain[\indexED,\pathIndex]\in\realNumbers$ and $\delay[\indexED,\pathIndex]\in\realNumbers$ are the $\pathIndex$th path gain and path delay for the $\indexED$th \ac{ED}, respectively \cite[Chapter 2]{Tse_2005}. The received passband signal for the $\indexED$th \ac{ED} can then be expressed as
\begin{align}
	\receivedSignalPass[\indexED][\timeSymbol]=&\channelPass[\delaySymbol]\circledast\transmittedSignalPass[\indexED][\timeSymbol]=\Re\{
	\constante^{\constantj2\pi\fcarrierES\timeSymbol}\receivedSignal[\indexED][\timeSymbol]
	\}
\end{align}
where $\receivedSignal[\indexED][\timeSymbol]$ is the received complex baseband signal given by
\begin{align}
	\receivedSignal[\indexED][\timeSymbol]
	=\sum_{\pathIndex=1}^{\numberOfPaths}
	\gainComplex[\indexED,\pathIndex]
	\transmittedSignal[\indexED][{\timeSymbol-\delay[\indexED,\pathIndex]-\toComplete[\indexED]}]\constante^{\constantj2\pi\cfo[\indexED]\timeSymbol}
	\label{eq:compositeResponse}
\end{align}
for 
$\gainComplex[\indexED,\pathIndex]\triangleq \gain[\indexED,\pathIndex]\constante^{-\constantj(2\pi\fcarrierES(\delay[\indexED,\pathIndex]+\toComplete[\indexED])-\po[\indexED])}	\constante^{-\constantj2\pi\cfo[\indexED]({\delay[\indexED,\pathIndex]+\toComplete[\indexED]})}$.
Hence, based on \eqref{eq:compositeResponse}, we can infer the followings:
\begin{itemize}
	\item The timing errors at the EDs or ES not only translate the signal in time but also cause an additional phase rotation. 
	\item The \ac{CFO} causes phase error accumulation that grows over time.
	\item The CFO results in an additional phase rotation, depending on the time offset and path delays. 
\end{itemize}

Now, assume that the \ac{OAC} is based on \ac{OFDM}. For $\timeSymbol\in[0,\symbolDuration)$, we can express the baseband signal for the $\indexED$th \ac{ED} as $\transmittedSignal[\indexED][\timeSymbol]=\sum_{\indexTXsubcarrier=0}^{\numberOfActiveSubcarriers-1}\transmittedVectorEle[\indexED,\indexRXsubcarrier]\constante^{\constantj2\pi\frac{\indexTXsubcarrier}{\symbolDuration}\timeSymbol}$, where the  $\symbolDuration$ is the symbol duration, $\numberOfActiveSubcarriers$ is the number of active subcarriers, and $\transmittedVectorEle[\indexED,\indexRXsubcarrier]$ are the transmitted symbol as given in \eqref{eq:transmittedSymbols}. Assume that the channel is ideal, i.e., $\numberOfPaths=1$, $\gain[\indexED,\pathIndex]=1$, and $\delay[\indexED,\pathIndex]=0$,  $\forall\indexED$. Also, assume that the ideal synchronization point where the $\idftSize$-point \ac{DFT} is started to be applied to the received baseband signal is within the \ac{CP} duration, i.e.,  at $\syncPoint\le0$, as illustrated in \figurename~\ref{fig:sync}\subref{subfig:time}.\footnote{In practice, an OFDM receiver intentionally backs off some duration  for the DFT calculation to avoid samples from the following OFDM symbol.}
Under the synchronization impairments, 
 the received symbol on the $\indexRXsubcarrier$th subcarrier at the \ac{ES} can be expressed in \eqref{eq:syncImpairments},
 \begin{figure*}[!t]
{\color{\reviewColor}\begin{align}
\superposedVectorEle[\indexRXsubcarrier]&
= \frac{1}{\idftSize}\sum_{\indexRXsample=0}^{\idftSize-1}\sum_{\indexED=1}^{\numberOfEdgeDevices} \receivedSignal[\indexED][\timeSymbol]\bigg|_{\timeSymbol=\frac{\indexRXsample}{\idftSize}\symbolDuration}\constante^{-\constantj2\pi\frac{\indexRXsample}{\idftSize}\indexRXsubcarrier}\nonumber\\
&
= \frac{1}{\idftSize}\sum_{\indexRXsample=0}^{\idftSize-1}\sum_{\indexED=1}^{\numberOfEdgeDevices}
\constante^{-\constantj(2\pi\fcarrierES\toComplete[\indexED]-\po[\indexED])}
	\constante^{-\constantj2\pi\cfo[\indexED]{\toComplete[\indexED]}}\transmittedSignal[\indexED][{\timeSymbol-\toComplete[\indexED]}]\constante^{\constantj2\pi\cfo[\indexED]\timeSymbol}\bigg|_{\timeSymbol=\frac{\indexRXsample}{\idftSize}\symbolDuration}\constante^{-\constantj2\pi\frac{\indexRXsample}{\idftSize}\indexRXsubcarrier}\nonumber\\
&
= \frac{1}{\idftSize}\sum_{\indexRXsample=0}^{\idftSize-1}\sum_{\indexED=1}^{\numberOfEdgeDevices} \constante^{-\constantj(2\pi\fcarrierES\toComplete[\indexED]-\po[\indexED])}	\constante^{-\constantj2\pi\cfo[\indexED]{\toComplete[\indexED]}}
\sum_{\indexTXsubcarrier=0}^{\numberOfActiveSubcarriers-1}\transmittedVectorEle[\indexED,\indexRXsubcarrier]\constante^{\constantj2\pi\frac{\indexTXsubcarrier}{\symbolDuration}({\timeSymbol-\toComplete[\indexED]})}\constante^{\constantj2\pi\cfo[\indexED]\timeSymbol}\bigg|_{\timeSymbol=\frac{\indexRXsample}{\idftSize}\symbolDuration}\constante^{-\constantj2\pi\frac{\indexRXsample}{\idftSize}\indexRXsubcarrier}\nonumber\\
&
=
\sum_{\indexED=1}^{\numberOfEdgeDevices}\sum_{{\indexTXsubcarrier=0}}^{\numberOfActiveSubcarriers-1} \transmittedVectorEle[\indexED,\indexTXsubcarrier]
	\constante^{-\constantj(2\pi\fcarrierES\toComplete[\indexED]-\po[\indexED])}\constante^{-\constantj2\pi\ncfo[\indexED]\nto[\indexED]}\constante^{-\constantj2\pi\indexTXsubcarrier\nto[\indexED]}\frac{1}{\idftSize}\sum_{\indexRXsample=0}^{\idftSize-1}\constante^{\constantj2\pi\frac{\indexRXsample}{\idftSize}({\indexTXsubcarrier-\indexRXsubcarrier+\ncfo[\indexED]})}\nonumber\\
&=
\sum_{\indexED=1}^{\numberOfEdgeDevices}  \transmittedVectorEle[\indexED,\indexRXsubcarrier]
\underbrace{\constante^{-\constantj(2\pi(\ncfo[\indexED]+\indexRXsubcarrier)\nto[\indexED]+2\pi\fcarrierES\toComplete[\indexED]-\po[\indexED])}\dirichletKernel[\idftSize][{\ncfo[\indexED]}]}_{\text{distortion}}
+
\underbrace{\sum_{\indexED=1}^{\numberOfEdgeDevices} \sum_{\substack{\indexTXsubcarrier=0\\ \indexTXsubcarrier\neq\indexRXsubcarrier}}^{\numberOfActiveSubcarriers-1} \transmittedVectorEle[\indexED,\indexTXsubcarrier]
	\constante^{-\constantj(2\pi(\ncfo[\indexED]+\indexRXsubcarrier)\nto[\indexED]+2\pi\fcarrierES\toComplete[\indexED]-\po[\indexED])}\dirichletKernel[\idftSize][{\ncfo[\indexED]+\indexTXsubcarrier-\indexRXsubcarrier}]}_{\text{inter-carrier interference}} ~.
\label{eq:syncImpairments}
\end{align}}
\end{figure*}
where $\ncfo[\indexED]\triangleq\cfo[\indexED]\symbolDuration$ and $\nto[\indexED]\triangleq{\toComplete[\indexED]}/\symbolDuration$ are the normalized \ac{CFO} and \ac{TO} for the $\indexED$th \ac{ED}, respectively, and $\dirichletKernel[\kernelScalar][\kernelInput]\triangleq\frac{1}{\kernelScalar}\sum_{\indexRXsample=0}^{\kernelScalar-1}\constante^{\constantj2\pi \frac{\indexRXsample}{\kernelScalar}\kernelInput}=\frac{\constante^{\constantj\pi\kernelInput\frac{\kernelScalar-1}{\kernelScalar}}}{\kernelScalar}\frac{\sin(\pi\kernelInput)}{\sin(\pi\kernelInput/\kernelScalar)}$ is the Dirichlet sinc kernel. As can be seen from \eqref{eq:syncImpairments}, the existence of  \ac{CFO} causes inter-carrier interference while \ac{TO} due to the imperfect time-of-arrivals or the synchronization errors at the \ac{ES} results in phase rotations {\em scaled} with the subcarrier index. On the other hand, the residual \ac{PO} leads to a distortion independent from the subcarrier index. Similar observations on \ac{CFO}, \ac{TO}, and \ac{PO} are also made in \cite[Section VI]{Shao_2021}, \cite{Guo_2021}, and \cite{ZhaoICC_2022_extension}.

Since the PO, TO, and CFO affect the received signal jointly, the key OAC-related metrics such as computation rate or \ac{MSE} for the computation discussed in Section~\ref{sec:whatAreTheMetrics}  are also affected. However, how these metrics are affected under such offsets is not well-assessed in the literature. It is also worth emphasizing that the sensitivity of the computation to the residual offsets depends on the scheme and the function desired to be computed. For instance, for an OAC scheme that requires phase synchronization among the EDs, the PO, TO, and CFO need to be compensated very accurately, particularly for analog aggregation. Hence, this scheme would require a precise sample-level synchronization and the corresponding MSE would be sensitive to synchronization errors (See Section~\ref{parag:whatCanGoWrongOne}). On the other hand, if the target function is an \ac{MV} computation and the OAC scheme relies on keying approaches along with a non-coherent detection, it is demonstrated that maintaining the synchronization within the \ac{CP} range can be sufficient and the \ac{MSE} does not increase with the synchronization errors \cite{sahinGC_2022} (See also Section~\ref{par:orthSignaling}).

In practice, it is challenging to mitigate the impacts of random \acp{TO}, \acp{CFO}, and \acp{PO} on an \ac{OFDM}-based \ac{OAC}. However, in the case that the offsets change slowly, they may be mitigated through  control loops. For example, in \cite{Guo_2021}, the residual \ac{TO} and \ac{PO} are estimated by tracking the phase in the frequency domain before the \ac{OAC} takes place. A protocol that feeds back the estimates of \ac{TO} and \ac{PO} to the \acp{ED} is proposed, where the \ac{ED} can compensate for the residual errors. It is also worth noting that the residual \ac{CFO} estimation errors cause the phase error accumulation for the packets with many \ac{OFDM} symbols. To mitigate the error accumulation, extra signaling or better high-precision clocks can be utilized \cite{ZhaoICC_2022_extension}.

In the literature, the synchronization for \ac{OAC} is also investigated for waveforms different from OFDM. For instance, in \cite{Shao_2021,Shao_spawc2021,shao2021bayesian}, the analyses are performed for a \ac{SC} waveform constructed with a rectangular pulse shape.\footnote{A single carrier waveform synthesized based on a rectangular pulse shape corresponds to the \ac{DC} subcarrier of OFDM. Inclusion of a guard period or a \ac{CP} duration depends on the design.} In \cite{Shao_2021} and \cite{Shao_spawc2021}, to estimate the sum of the parameters under time-synchronization errors, a post-processing called whitened matched filtering and sampling is proposed assuming that the \ac{ES} knows the \ac{TO} for each \ac{ED}. In \cite{shao2021bayesian}, the same setup is investigated from the perspective of Bayesian approaches. In \cite{razavikiaGC_2022}, sinc kernel is considered for waveform and it is proposed to recover the summation of the parameters by solving a convex semi-definite programming without any prior information on the misalignment. Note that, in \cite{Shao_2021,Shao_spawc2021,shao2021bayesian,razavikiaGC_2022},  the channel is not frequency selective and overall phase rotation is assumed to be compensated via channel inversions at the EDs and decoupled from the \ac{TO}. 

The non-coherent \ac{OAC} approaches discussed in Section~\ref{subsubsec:noncoherent} provide robustness against  \ac{PO} as these methods do not convey the information in the phase.} In \cite{sahinWCNC_2022} and \cite{safiPIMRC_2022}, in order to mitigate the interference between the adjacent symbols due to the random \acp{TO}, a guard time  between adjacent \ac{PPM} or \ac{CSK} symbols (based on \ac{DFT-s-OFDM}) is proposed, respectively. Also, a larger energy calculation window than the corresponding bins at the transmitter is used for the energy calculations to accommodate the jitter, respectively. We also refer the reader to an excellent survey in \cite{Zimmerling_sync2020} on the protocols regarding synchronous transmissions for low-power wireless networks, which can benefit to the implementation of \ac{OAC} methods in such networks.

\subsection{Power Management}
\label{subsec:powerControl}
In this section, we investigate power management under two categories. In the first category, we discuss power management from the the perspective of the receiver, e.g., power alignment at ES via power control. In the second category, we discuss power management from the perspective of transmitters and focus on the issues related to \ac{PAPR} and \ac{OBO}.
\subsubsection{Receiver side}
For power control, a perfect amplitude alignment at the ES to ensure fairness or an accurate computation, or a policy that minimizes an application-specific metric are two main objectives one can consider. In the former case, the link with the worst channel condition may dominate the performance of computation \cite{Wei_2021}. On the other hand, the latter requires a well-defined metric to be taken into account and the power control strategy becomes a function of the application. In this case, the policy can be quite diverse. For example, in \cite{Wei_2021}, it is shown that the power alignment for gradient aggregation is not always necessary for the convergence of  \ac{FEEL}.  As another example, in \cite{Zezhong_2022}, the authors consider an online power control mechanism with the considerations of saddle regions for distributed \ac{PCA} computation. In \cite{hellstrom2021spawc} and \cite{hellstrom2021over}, for static fading channels,  channel inversion that minimizes \ac{MSE} for multiple transmissions is studied. It is shown that the optimal channel inversion coefficient is a function of the number of re-transmission. In \cite{hellstrom2021overGC_2022}, channel inversion that particularly minimizes the bias for a given number of re-transmissions is investigated for fast-fading channels (i.e., time diversity).
In \cite{Xie_2022gc}, it is shown that the optimal power allocation policy across OFDM subcarriers for channel inversion yields a proportional fairness scheme.
In another application, Byzantine attacks are taken into account for power control \cite{Fan_2022byzan}. We refer the reader  to \cite{Xiaowen_2022} and references therein for a comprehensive analysis  on various power control methods for \ac{FEEL}.


Most of the state-of-the-art power control strategies for \ac{OAC} rely on a small-scale channel model when a precoder such as \ac{TCI} is employed at the EDs (e.g., see \cite{Wei_2022dynamic, Zezhong_2022, Fan_2022byzan, Mao_2022}). On the other hand, power control is also related to large-scale models and interactions between adjacent cells. In \cite{Cao2021cooperative}, inter-cell interference in a scenario where  \ac{OAC} occurs at different cells concurrently is taken into account and the optimal policies for controlling devices’ transmit power are investigated to minimize the \ac{MSE} of computed function. In \cite{LiICC_2022}, assuming that the inter-cell interference is harmful  to the \ac{OAC},  the beamforming vectors and transmit
powers are optimized with the consideration of proportional fairness at different cells. 
In \cite{WangZhibin_2022}, inter-cell interference is investigated in both \ac{UL} and \ac{DL} directions for a channel-inversion-based \ac{OAC}. In the literature, it is also shown that inter-cell interference can actually be harnessed for \ac{OAC}, as discussed in Section~\ref{subsec:arch}.

When there is a large number of devices participating in the computation,  the dynamic range of a superposed signal at the \ac{ES} can exceed the dynamic range of the receiver \cite{Akshay_IPSN2020,Zang_2020,Shao_2021}.  Reducing the transmit powers of the \acp{ED} or designing adaptive gain control for \ac{OAC} are two potential solutions to address this issue. In \cite{Akshay_IPSN2020}, a randomization approach is proposed to spread the energy. In \cite{Zang_2020}, the \ac{MSE} minimization of \ac{OAC} under a sum-power constraint is studied, where one of the motivations is to reduce the potential interference due to the superposed signals. In \cite{Zhang_2022clt}, the sum-power constraint is introduced for energy saving.

\def\oboMin{{OBO}_{\text{min}}}
\def\oboRef{{OBO}_{\text{ref}}}
\def\powerMax{P_max}
\def\rangePowercontrol{{r}_{\text{max}}}
\subsubsection{Transmitter side} The dynamic range of the transmitted signals is directly related to the power management in a network. A transmitted signal with a large \ac{PAPR} can cause a reduced cell size due to the power back-off, a higher adjacent channel interference due to saturation, and a reduced battery life  to accommodate the instantaneous power fluctuations. In the literature, a few \ac{OAC} schemes is analyzed from the perspective of \ac{PAPR}. In \cite{sahinCommnet_2021}, the randomization symbols are used to reduce the \ac{PAPR} for \ac{FSK-MV}  computation. It is demonstrated that if the parameters are highly correlated in the frequency domain (e.g., the elements of a stochastic gradient vector at one iteration can be highly-correlated due to the over-parameterized neural networks \cite{Xue_2022}), modulating the subcarriers with  the parameters without any precaution can cause \ac{OFDM} symbols with large \acp{PAPR}. In \cite{Wang_2022,sahinWCNC_2022} and \cite{safiPIMRC_2022}, the properties of \ac{SC} waveform and chirps are exploited to mitigate \ac{PAPR} for \ac{OAC}. In \cite{safiPIMRC_2022}, a basic model is proposed to relate \ac{OBO} to the cell size under power control: Let $\Preference$ and $\oboRef$ be the average transmit signal power in Watts and the \ac{OBO} in dB, when the link distance between the ES and an ED is $\referenceDistance$ meters, respectively. Also, let $\oboMin$ be the minimum \ac{OBO} that does not violate \ac{ACLR}  or the spectrum-emission mask requirements.  Based on the simplified path loss model, the received signal for the $\indexED$th ED can be expressed as
\begin{align}
	\powerED[\indexED]=\begin{cases}
		\left(\frac{{\distanceED[\indexED]}}{\referenceDistance}\right)^{-\pathlossExponent+\powerControl}\Preference, & \referenceDistance\le\distanceED[\indexED]<\rangePowercontrol\\
		\left(\frac{{\rangePowercontrol}}{\referenceDistance}\right)^{-\pathlossExponent+\powerControl}\Preference, & \distanceED[\indexED]\geq\rangePowercontrol
	\end{cases}~,
\end{align}
where $\distanceED[\indexED]$ is the link distance between $\indexED$th ED and ES, $\pathlossExponent$ is the path loss exponent, $\powerControl\in [0,\pathlossExponent]$ is a coefficient that determines how much path loss is compensated via power control, and $\rangePowercontrol$ is the maximum link distance beyond which the \ac{ED} is unable to increase the average transmit power and it can be expressed as
\begin{align}
	\rangePowercontrol=\referenceDistance\times 10^{\frac{\oboRef-\oboMin}{10\powerControl}}~.
	\label{eq:rangeofpowercontrol}
\end{align} 
The parameter $\rangePowercontrol$ determines the region in which the average received signal powers of the EDs located in this area can be aligned at the ES location. Based on \eqref{eq:rangeofpowercontrol}, a smaller $\oboMin$ results in a larger $\rangePowercontrol$. However, to decrease $\oboMin$, a more linear \ac{PA} or an OAC scheme with low instantaneous power fluctuations is needed. In \cite{safiPIMRC_2022}, based on the Rapp model, $\oboMin$ values are obtained for \ac{OAC} with \ac{TCI} and \ac{CSK}-based \ac{OAC} and the trade-off between computation rate and cell size is emphasized. It is shown that the reduced-cell size due to the power back-off can deteriorate the performance of the computation due to the weak signals from the EDs with large link distances.

\def\subfigSize{3.5in}
\subsection{Architecture}
\label{subsec:arch}
\begin{figure}
	\centering
	\subfloat[Single cell OAC.]{\includegraphics[width =\subfigSize]{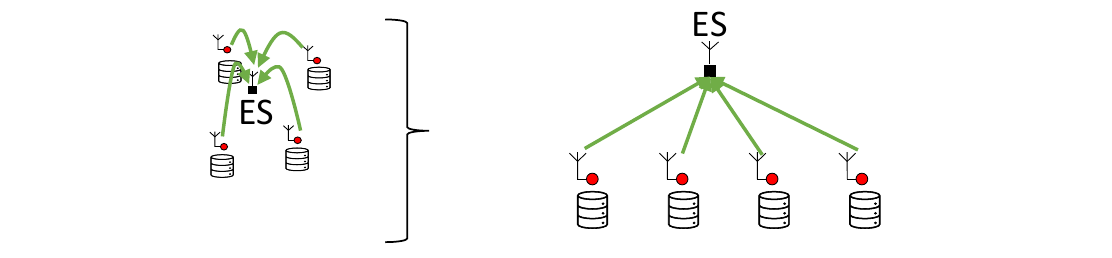}
		\label{subfig:archA}}
	\\	
	\subfloat[Uplink OAC in a multi-cell environment where the interference due to the  adjacent cells are harmful to the computation at the cells.]{\includegraphics[width =\subfigSize]{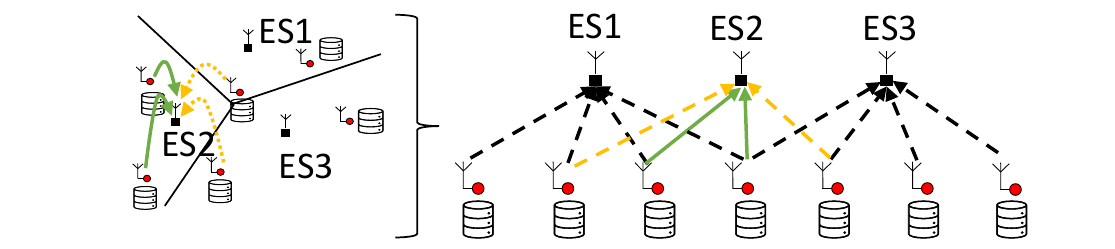}
	\label{subfig:archB}}
	\\
	\subfloat[Uplink OAC in a multi-cell environment where the interference due to the  adjacent cells are useful to the computation at the cells.]{\includegraphics[width =\subfigSize]{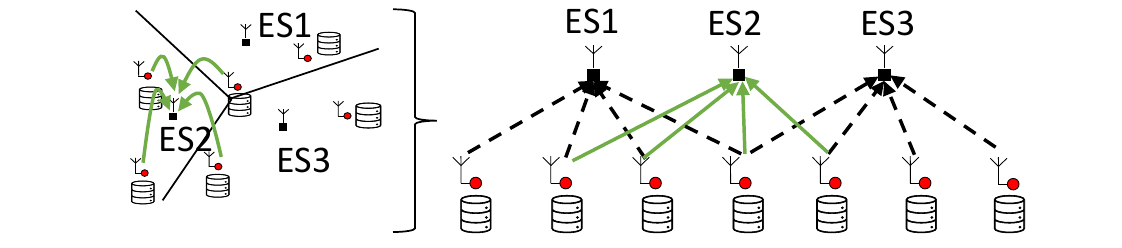}
	\label{subfig:archC}}
	\\
	\subfloat[Downlink OAC for improved OAC at the cell-edge EDs.]{\includegraphics[width =\subfigSize]{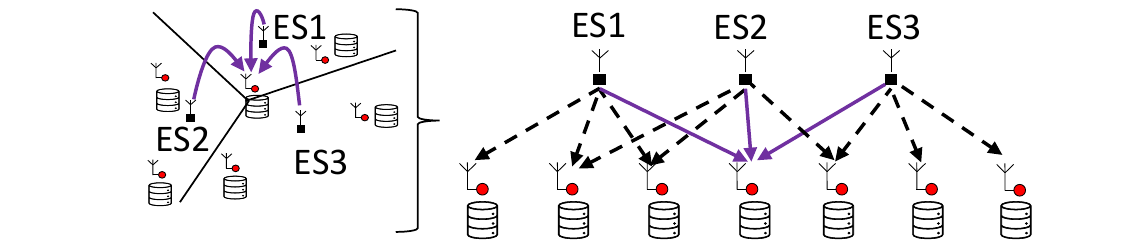}
		\label{subfig:archD}}
	\\
	\subfloat[Two-tier hierarchical OAC by exploiting non-terrestrial networks.]{\includegraphics[width =\subfigSize]{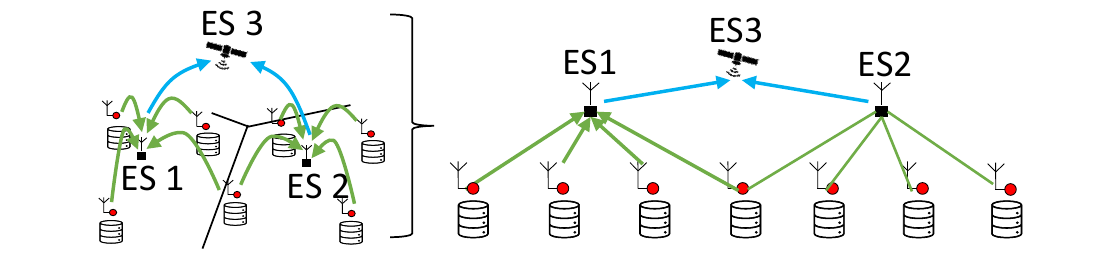}
		\label{subfig:archE}}
	\\
	\subfloat[OAC in an ad-hoc network with  half-duplex and full-duplex transcievers, where the network topology varies in time.]{\includegraphics[width =\subfigSize]{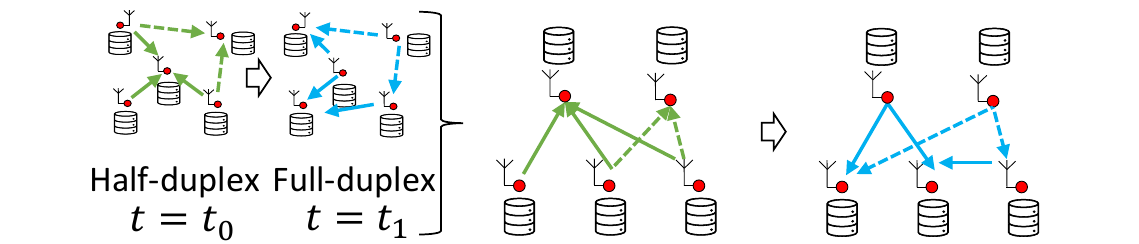}
		\label{subfig:archF}}	
	\caption{Several architectures for OAC.}
	\label{fig:arch}	
\end{figure}

In the literature, a majority of the papers on \ac{OAC} consider either a single-cell scenario or a network where there is no cooperation among the cells. On the other hand, harnessing the interference among the cells with some coordination can also result in a more global computation while addressing large-scale fading of wireless channels. 
In \figurename~\ref{fig:arch}, we illustrate several architectures for OAC. As compared to the single-cell scenario in \figurename~\ref{fig:arch}\subref{subfig:archA}, the interference due to the adjacent cells in a multi-cell scenario may be considered as harmful to the \ac{OAC} as in \figurename~\ref{fig:arch}\subref{subfig:archB} or harnessed for the computation as shown in \figurename~\ref{fig:arch}\subref{subfig:archC}. Similar to \ac{UL} OAC, \ac{OAC} can also be realized in \ac{DL} as depicted in \figurename~\ref{fig:arch}\subref{subfig:archD}. In \figurename~\ref{fig:arch}\subref{subfig:archE}, we illustrate a two-tier hierarchical computation, where the intermediate \acp{ES} can be base stations while the final fusion node can be a satellite in a non-terrestrial network (see \cite{WangGC_2022} for an example with a low-earth orbit
satellite without hierarchical computation). Finally, \figurename~\ref{fig:arch}\subref{subfig:archF} illustrates  \ac{OAC} in an ad-hoc network, e.g., to achieve consensus on the output of a function among the EDs, where the network topology can change over time  and the \acp{ED} may perform \ac{OAC} while transmitting to the other nodes with the consideration of full-duplex transceivers.

In \cite{aygunICC_2022,aygunArXiv_2022}, hierarchical \ac{FEEL} is investigated over multiple clusters to address the adverse effects of path loss on aggregation. In this approach, \ac{OAC} is employed in each cluster, where the intermediate servers are connected to a global server to aggregate their results. In \cite{Vahapoglu_2022}, hierarchical \ac{FEEL} under  data heterogeneity
is addressed via a dynamic weighting approach. The scheme used in this paper is based on \ac{TCI}. In \cite{aygunICC_2022,aygunArXiv_2022,Vahapoglu_2022}, the impact of the inter-cluster interference is not taken into account. In \cite{Wei_2022dynamic}, the inter-cluster interference is also exploited with \ac{OAC} for hierarchical \ac{FEEL}. In \cite{Sohini_2022clustered}, it is proposed to use multiple clusters where the \acp{ES} in clusters seek  a consensus for \ac{FEEL} update with \ac{OAC}, in addition to the \ac{OAC} within the clusters among the \acp{ED}.
 In \cite{Jeon_2016}, to improve the robustness against signal attenuation for long distances, a multi-hop network is considered. In this strategy, computation occurs locally in the sensors through multiple hops till final aggregation occurs in a fusion center by exploiting the fact that a target function can be decomposed into locally computable functions. 

In \cite{goldenbaum2013harnessing}, the universality property, based on Theorem~\ref{th:kolmogorovSuperVariant}, is exploited to compute multiple functions in multiple sensor clusters, where each cluster is assigned to compute one of the target functions.
In \cite{mohammadICC_2022}, interference in a multi-cell network is exploited for computation. In this approach, \ac{OAC} is used in both \ac{UL} and \ac{DL} by using \ac{FSK-MV} calculation. It is shown that such an approach can improve the minimization of independent objective functions and leads to better learning performance in a large area. In \cite{Abarghouyi_2022arxivhier}, similarly, multiple \acp{ES} are considered.  \ac{OAC} with \ac{TCI} is considered and the analysis is performed under  harmful interference due to the downlink signals for \acp{ES}. In \cite{WangZhibin_2022}, a cooperative multi-cell optimization framework is introduced to improve the average learning performance as compared to non-cooperative baseline schemes. In \cite{Zhenyi_2022}, a topology where all devices are connected with each other is investigated. The \ac{OAC} relies on the existence of a large number of antennas and full-duplex capability. In \cite{Yemini_2022}, authors consider a topology where  the EDs continue aggregation based on neighboring EDs when the EDs to ES connection is intermittently available. Although the OAC is mentioned in this study, the topology is not analyzed along with OAC. In \cite{Ozfatura_2020}, similarly, a full-duplex capability is assumed and various network topologies are investigated.  In \cite{Lee_2022gnn}, \ac{OAC} is applied to graph neural networks where each node in the graph locally processes the information. The \ac{OAC} based on \ac{PC} is investigated to improve communication efficiency and privacy preservation. While \cite{Yifan_2022mpa} uses a graph neural network with \ac{OAC} to address the power allocation problem in ad-hoc networks, \cite{zhanAirGNN} uses the channel coefficients between the nodes as part of the graph convolution operator. In \cite{Fabio_2021,Fabio_2022} a consensus protocol is proposed to achieve max-consensus in a clustered network. \ac{OAC} is utilized across the clusters with the consideration of full-duplex \cite{Fabio_2021} and half-duplex communications \cite{Fabio_2022}. For general time-varying network topology, we refer the reader to \cite{Yang_2022adhoc} and the reference therein.

\subsection{Channel Estimation}
To achieve an accurate aggregation over fading channels, accurate and fresh \ac{CSI} may need to be available at the \acp{ED} and/or the \ac{ES}, depending on the \ac{OAC} method. While an inaccurate estimate of \ac{CSI} can cause an incoherent aggregation, the aging of the \ac{CSI} estimate due to the residual \ac{CFO} or mobility can result in a larger overhead and limit the number of functions to be computed in one single packet.

For the \ac{OAC} methods relying on precoding techniques, each \ac{ED} needs its own \ac{UL} \ac{CSI} which can be acquired through a \ac{DL} signal with a time-domain duplexing system or transmitted back to the \ac{ED} based on the estimates at the \ac{ES}. While the former  requires a calibration procedure, the latter causes an overhead that is scaled with the number of \acp{ED} and increased latency. The inaccuracy due to  either of these methods may be modeled based on some error on the ground-truth \ac{CSI}. Under this model, in \cite{Yilong_2022}, the authors derive an optimal  transceiver design that minimizes \ac{MSE} with the consideration of multiple antennas at the \ac{ES} with the consideration of the imperfect \ac{CSI}. In \cite{Mao_2022}, a strategy that takes the imperfect \ac{CSI} at the EDs into account to determine the number of local update steps is proposed. In \cite{Qiao_2022arxiv}, the imperfect \ac{CSI} is taken into account for joint device selection and transceiver design, where the main goal is to maximize the number of participating \acp{ED} under an \ac{MSE} constraint. In \cite{Zhang_2022clt}, the imperfect \ac{CSI} is evaluated along with \ac{RIS}. In \cite{Evgenidis2023}, an adaptive pilot
re-transmission policy that offers a trade-off between wireless resources and gain in the computation accuracy  is proposed.

Some \ac{OAC} methods use sum-channel estimate, rather than the \ac{UL} \ac{CSI} for each link. For instance, in \cite{Fan_2019tvt}, it is proposed to use a procedure that  optimizes the beamforming vectors at the \acp{ED} and \ac{ES} iteratively by exploiting the sum-channel \ac{CSI} acquisition. In this method, the \ac{ES} first transmits a set of reference symbols and broadcasts its current beamforming  vector. After each \ac{ED} estimates the \ac{DL} channel and design its own beamforming vector, all the \acp{ED} transmits a set of common pilot symbols concurrently so that the \ac{ES} can estimate the sum channel. The key observation is that  the \ac{ES} can update its beamforming vector based on the sum channel. Common pilots are also employed in \cite{weiICC_2022,Xizixiang_2022rMIMO} to estimate the sum channel for the methods relying on channel hardening or random orthogonalization over a large number of multiple antennas. In \cite{Xizixiang_2022rMIMO}, the sum channel is also used as a precoder in the \ac{DL}. Instead of using pilots in \ac{DL}, an echo protocol where the ES broadcasts the received symbols for the sum-channel estimation back to the EDs is proposed. In the \ac{UL}, the EDs design their channel inversion coefficients based on the received symbols in the \ac{DL}.


\subsection{Security}
\ac{OAC} relies on the superposition of the transmitted signals from the \acp{ED}. As discussed in \cite{hellstrom2020wireless}, this fact has both positive and negative consequences as far as  security is concerned.
On one hand, the superposition in \ac{OAC} promotes user privacy as the transmitted signals cannot be directly observed. On the other hand, it opens up potential adversaries  to harm the computation, particularly through Byzantine attacks. This is because Byzantine attacks are launched by the nodes that are already part of the network. For example, for \ac{FEEL}, if an \ac{ED}'s local data  are deliberately labeled incorrect (i.e., one of the data poisoning attacks) or the sign of the gradients are flipped (i.e., one of the model poisoning attacks) by an adversary or due to the failure of a node, the whole learning process can be unreliable. This problem has been studied in the \ac{FL} literature by identifying the Byzantine nodes or detecting anomalies in the local signals (see \cite{Rui_2022ByzantineFEDMI} and \cite{Fan_2022byzan} and the references therein for further discussions on various attacks and defense mechanisms). However, if the aggregation is handled through an \ac{OAC} scheme for the same scenario, well-known defense strategies that rely on the observation of local information cannot be utilized directly as the \ac{ES} only observes the superposed signal, not the signals transmitted from different \acp{ED}.

\subsubsection{Byzantine Attacks}
\label{subsub:byzantine}

One way of achieving resiliency against Byzantine attacks relies on the geometric median, rather than the arithmetic mean, which can be expressed as
\begin{align}
    \optimumGeometricMean = \arg\min_{\variableForGeoemtricMean} \sum_{\indexED=1}^{\numberOfEdgeDevices}\weightDistance[\indexED]\norm{\variableForGeoemtricMean-\modelParametersAtIteration[\indexCommunicationRound][\indexED]}_{2}~,
    \label{eq:geoMedian}
\end{align}
where $\weightDistance[\indexED]$ is the weight factor for the $\indexED$th parameter. 
It is well known that \eqref{eq:geoMedian} can be solved with the Weiszfeld algorithm. In \cite{huang2021byzantineresilient} and \cite{Houssem_arxiv2022}, a modified Weiszfeld algorithm is proposed for achieving a smoothed geometric median aggregation against Byzantine attacks with \ac{OAC}. To calculate the geometric median in \eqref{eq:geoMedian},  the following iteration algorithm is proposed to realize with \ac{OAC}:
 \begin{align}
     \variableForGeoemtricMean^{(\indexCommunicationRound+1)} = \frac{\sum_{\indexED=1}^{\numberOfEdgeDevices}\medianDistance[\indexED]\modelParametersAtIteration[\indexCommunicationRound][\indexED]}{\sum_{\indexED=1}^{\numberOfEdgeDevices}\medianDistance[\indexED]}~,
     \label{eq:Weiszfeld}
 \end{align}
 where $\medianDistance[\indexED]$ is defined as
 \begin{align}
     \medianDistance[\indexED] = \frac{\weightDistance[\indexED]}{\norm{\max(\smoothFactor,\variableForGeoemtricMean^{(\indexCommunicationRound)}-\modelParametersAtIteration[\indexCommunicationRound][\indexED])}_2}~,
     \label{eq:betak}
 \end{align}
 and $\smoothFactor$ is a  smoothing factor to prevent the denominator in \eqref{eq:Weiszfeld} from yielding an unrealizable $\variableForGeoemtricMean^{(\indexCommunicationRound+1)}$ value. 
The proposed scheme considers \ac{FEEL}  based on model aggregation, where the investigated \ac{OAC} scheme is \ac{BAA} with \ac{TCI} (see Section~\ref{parag:TCI}). 
In this approach, the $\indexED$th \ac{ED} transmits the scaled model parameters $\medianDistance[\indexED]\modelParametersAtIteration[\indexCommunicationRound][\indexED]$ and the scalar $\medianDistance[\indexED]$. The \ac{ES} calculates \eqref{eq:Weiszfeld} with \ac{OAC} by using the estimates of the numerator and denominator parts and broadcast the vector $\variableForGeoemtricMean^{(\indexCommunicationRound+1)}$. This procedure continues until a certain convergence is achieved. 
The proposed scheme has two main disadvantages: First, it can cause an additional delay as the algorithm should run exclusively for each communication round of \ac{FEEL}. Second, it assumes that Byzantine users follow the proposed algorithm, which may not be the case in practice.

In \cite{park_2022tcom}, it is assumed that the \ac{ES} has a reference data set and uses its own gradient as a
reference vector to provide robustness against Byzantine \ac{ED}. In this approach, the network divides the \acp{ED} into multiple groups in orthogonal resources and compares the distances between its own gradient and the received estimate of each group with \ac{OAC} scheme that relies on analog aggregation and \ac{TCI}. In this study, \ac{OAC} based on \ac{BAA} with channel inversion is used to reduce per-round communication latency for each group, rather than directly addressing Byzantine attacks as done in \cite{huang2021byzantineresilient} and \cite{Houssem_arxiv2022}.

In \cite{Fan_2022byzan}, the \acp{ED} transmit not only their standardized gradients (i.e., the variance and the mean of the local gradients are always set to $1$ and $0$, respectively), but also the variance that is used for the standardization. It is proposed to calculate  the global gradient with a channel-inversion-based \ac{OAC} while transmitting the variance information through orthogonal channels. Assuming that Byzantine attackers follow the standardization  to avoid exposing themselves during the standardization stage, they would send the true mean and variance of their local gradients. Under this scenario, it is shown that \ac{TCI}-based \ac{OAC} for \ac{SGD} has limited defensive capability against Byzantine attacks as the \ac{TCI} aligns the amplitude levels at the \ac{ES}. The best-effort approach, i.e., using maximum power, is proposed against Byzantine attacks. The main shortcoming of the proposed approach is that a Byzantine attacker can still transmit non-standardized gradients while transmitting valid variance information.

\subsubsection{Privacy}
Differential
privacy is a well-established metric that measures the privacy of local data sets with respect to disclosed aggregate statistics \cite{Dwork_2014}. A typical approach is to randomize the
disclosed statistics by adding random noise, which causes a
trade-off between accuracy and privacy. For \ac{OAC}, random perturbations are added to the local model parameters or gradients before transmission for \ac{FEEL} \cite{Seif_2020isit,Zhang_2022,Liao_2022privacy}.
In \cite{Seif_2020isit}, it is shown that the privacy leakage per user scales as $\mathcal{O}(1/\sqrt{\numberOfEdgeDevices})$, compared to the orthogonal schemes. In \cite{seif_2021jsacPrivacy}, privacy is investigated with the consideration of random client participation and power misalignment. In \cite[Lemma 3]{Dongzhu_2021}, it is shown that such an approach guarantees differential privacy and can be
obtained without affecting the learning
performance as long as the privacy constraint level is below a
threshold. The authors also emphasize that the channel inversion for \ac{OAC} under fading is beneficial for privacy. In \cite{Krouka_2022}, the privacy concern is addressed by  incorporating
channel perturbations into the optimization problem and introducing a framework that does not explicitly transmit the Hessian or the gradient to the ES. In \cite{Lee_2022gnn}, privacy-preserving signaling and privacy-guaranteed training algorithm along with \ac{OAC} are investigated when a neural network is distributed across multiple nodes based on a graph. In \cite{Jiamo_2022tvt}, the trade-off
between data privacy and  training accuracy via
power control optimization is investigated for  channel-inversion-based \ac{OAC}. In \cite{Yan_2022}, differential
privacy is evaluated when the devices with better channel conditions are scheduled for \ac{OAC} during the training period of \ac{FEEL}. In \cite{Jiayi_2022adcpri}, it is proposed to use low-resolution \acp{ADC} at the \ac{ES} and \ac{DAC} at the \acp{ED} along with \ac{OAC} for promoting  privacy further.
\subsubsection{Eavesdropping}
Eavesdropping is one of the potential issues of \ac{OAC} as an eavesdropper can overhear the computation in the wireless channel. To address this issue, in \cite{Hyoungsuk_2011}, data confidentiality is investigated for \ac{TBMA}. The key idea in this work is that the sensors that have weaker channel gains can be utilized to confuse an eavesdropper by exploiting the independence between the desired and eavesdropping channels.
A similar idea where  a group of \acp{ED} with weaker channel conditions are selected as jammers is utilized for \ac{FEEL} in \cite{Yan_2022arxiv}.
 In \cite{Frey_2021}, the pre-processed symbols transmitted from the \acp{ED}, i.e., $\{\preProcessedVectorEle[\indexED][\indexSampleTime],\forall\indexED\}$, are intentionally distorted with jamming symbols such that the distortion can cause a substantial \ac{MSE} degradation  at the eavesdropper as compared to the one at the legitimate receiver, i.e., \ac{ES}.
The basic assumption exploited in this work is that the \ac{ES} has either precise knowledge
of the jamming symbols while the
eavesdropper only has knowledge about the distribution of the jamming signal.
Hence, under this assumption, the \ac{ES} can cancel the jamming symbols or be affected by less interference as compared to the eavesdropper.
The proposed scheme is investigated for computing an arithmetic average over an \ac{AWGN} channel. In \cite{Changjie_2022}, it is proposed to use a full-duplex transceiver  at the \ac{ES} so that a jamming noise is transmitted to degrade the eavesdropper’s links. In \cite{Luis_2022}, \ac{ES}  calculates an artificial noise vector such that it is projected into the null space of the channel vector after the simultaneous transmissions. The trade-off between the computation at the ES
and the security against the eavesdropper is emphasized.

\subsubsection{Jamming}
In \cite{zhu2019broadbandExtended}, it is proposed  to use a common spreading code assigned by the \ac{ES} to facilitate
protected model aggregation. For this scenario, it is assumed that the adversarial user does not know the spreading code. Hence, the interference due to the adversary is suppressed in the despreading/decoding process at the \ac{ES}.

\def\figureSizeApp{2.2in}
\section{What are the applications of OAC?}
\label{sec:WhatAreApps}

\begin{figure*}
	\centering
	\subfloat[Localization, e.g.,  voting-based distributed localization.]{\includegraphics[width =\figureSizeApp]{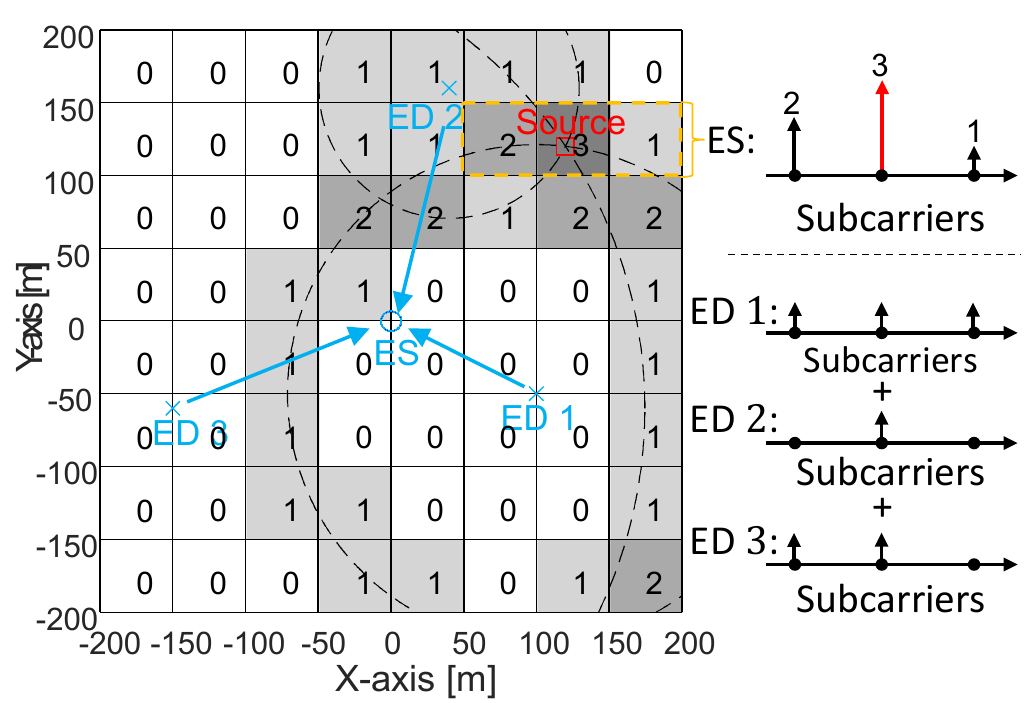}
		\label{subfig:dl}}~~
	\subfloat[Distributed optimization, e.g., federated edge learning.]{\includegraphics[width =\figureSizeApp]{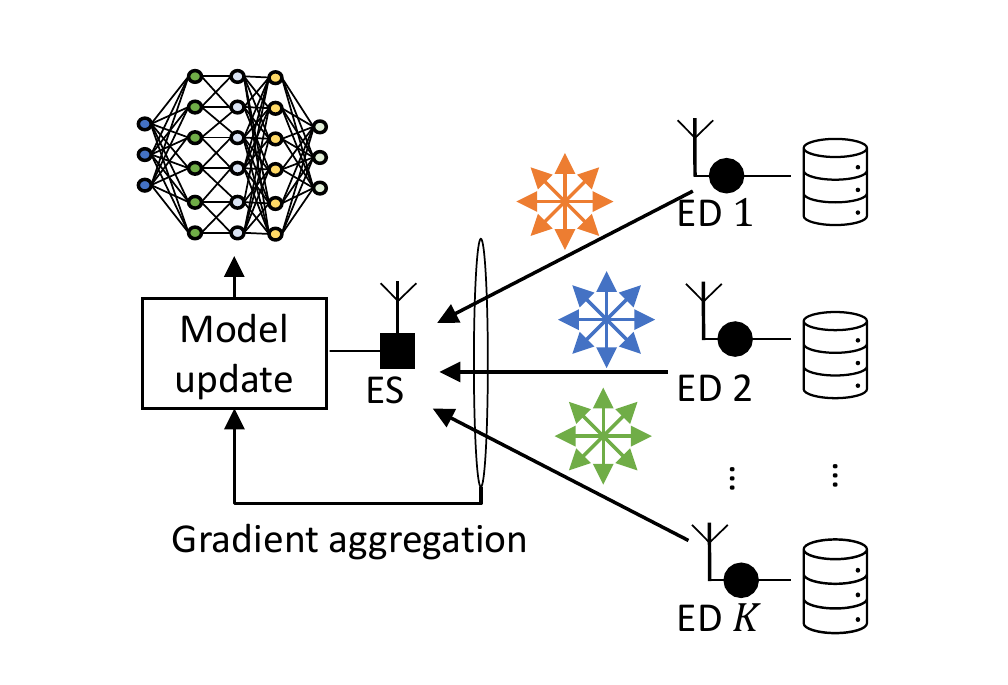}
	\label{subfig:fl}}~~
	\subfloat[Distributed optimization, e.g., split learning.]{\includegraphics[width =\figureSizeApp]{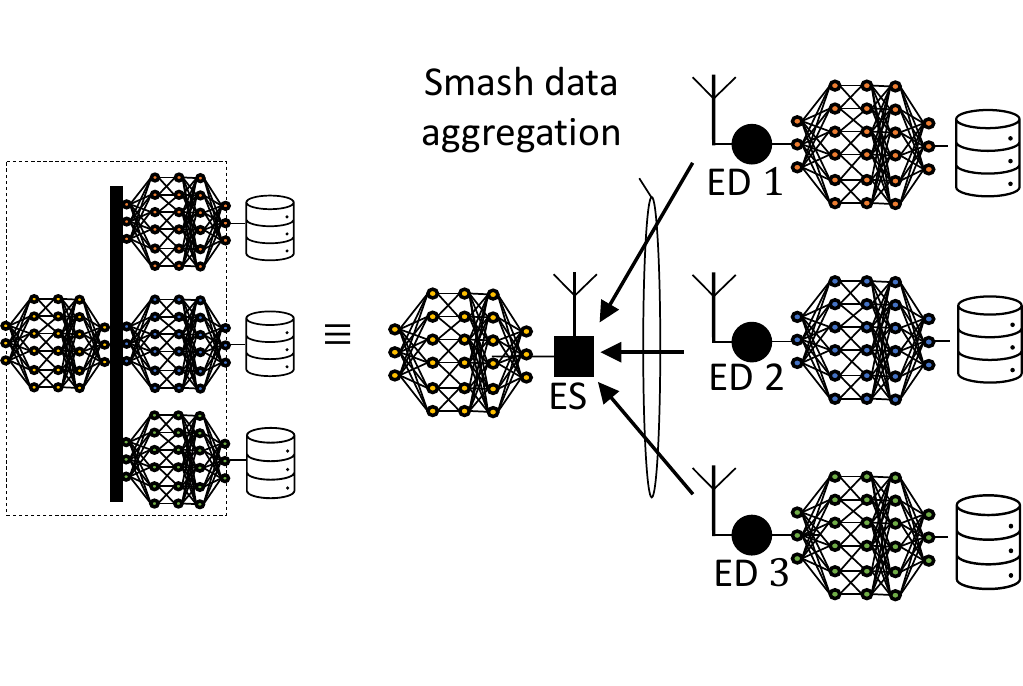}
	\label{subfig:sl}}\\
	\subfloat[Wireless communications, e.g., reliable communication in interference channel with compute-and-forward relaying strategy.]{\includegraphics[width =\figureSizeApp]{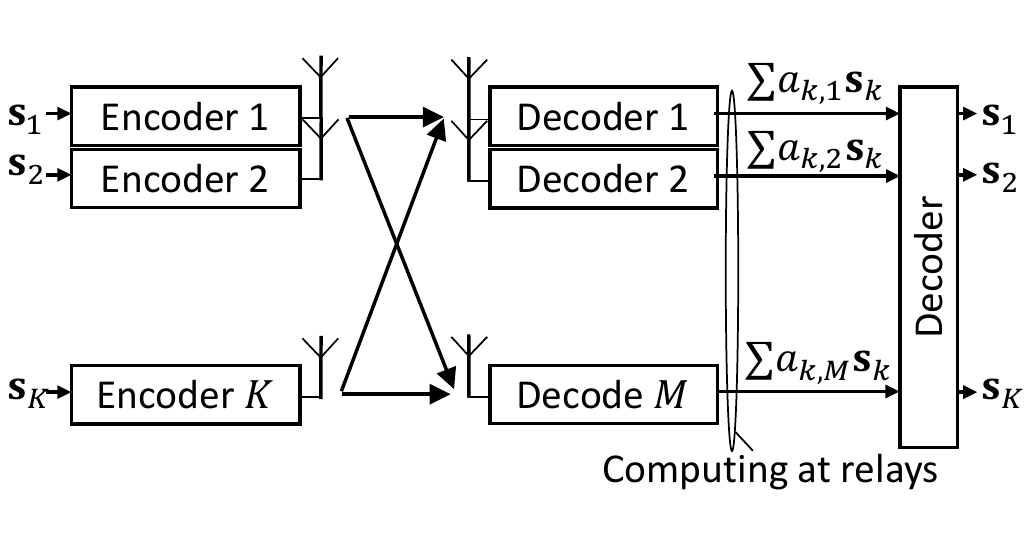}
	\label{subfig:cf}}~~
\subfloat[Wireless data centers with OAC, e.g., weighted sum computation at work nodes.]{\includegraphics[width =\figureSizeApp]{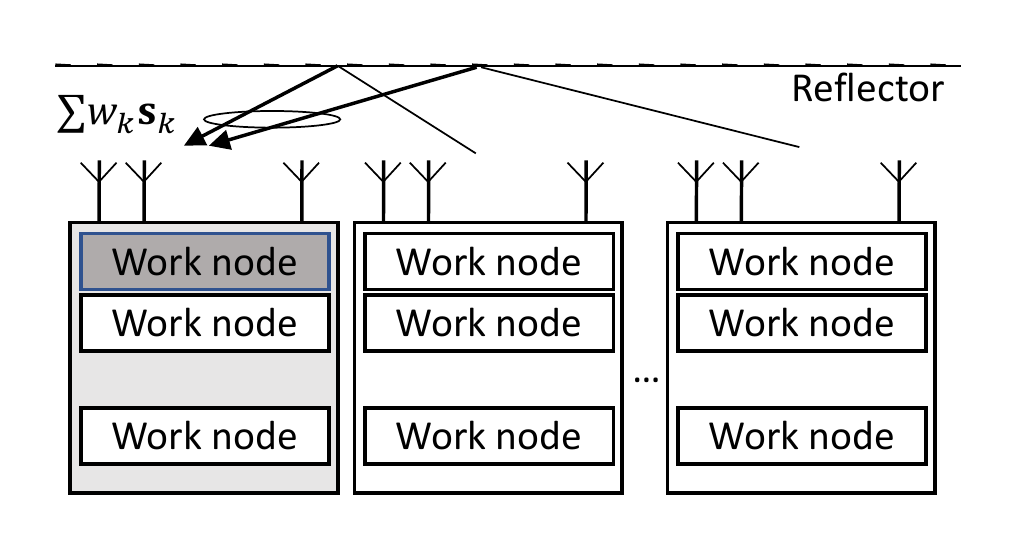}
	\label{subfig:dc}}~~
\subfloat[Wireless intra-chip computation, e.g., similarity search.]{\includegraphics[width =\figureSizeApp]{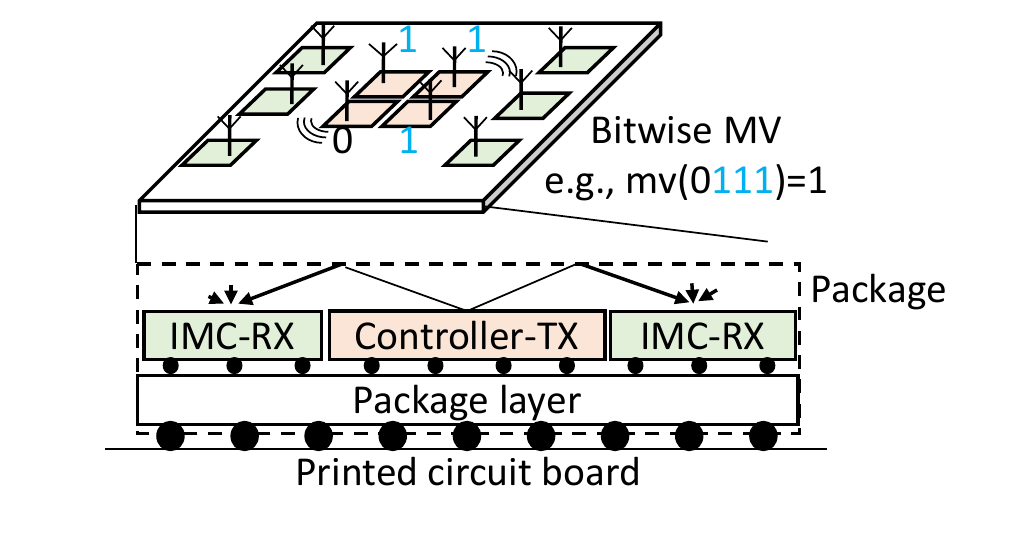}
	\label{subfig:ic}}\\
\subfloat[Wireless control systems, e.g., dynamic plants.]{\includegraphics[width =\figureSizeApp]{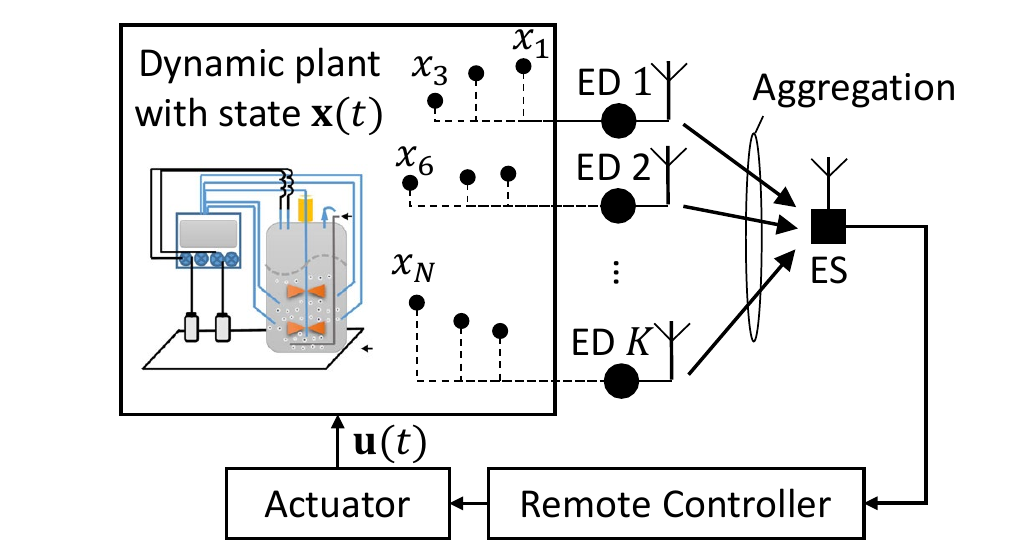}
	\label{subfig:dp}}~~
\subfloat[Wireless sensor networks, e.g., e.g., environment monitoring]{\includegraphics[width =\figureSizeApp]{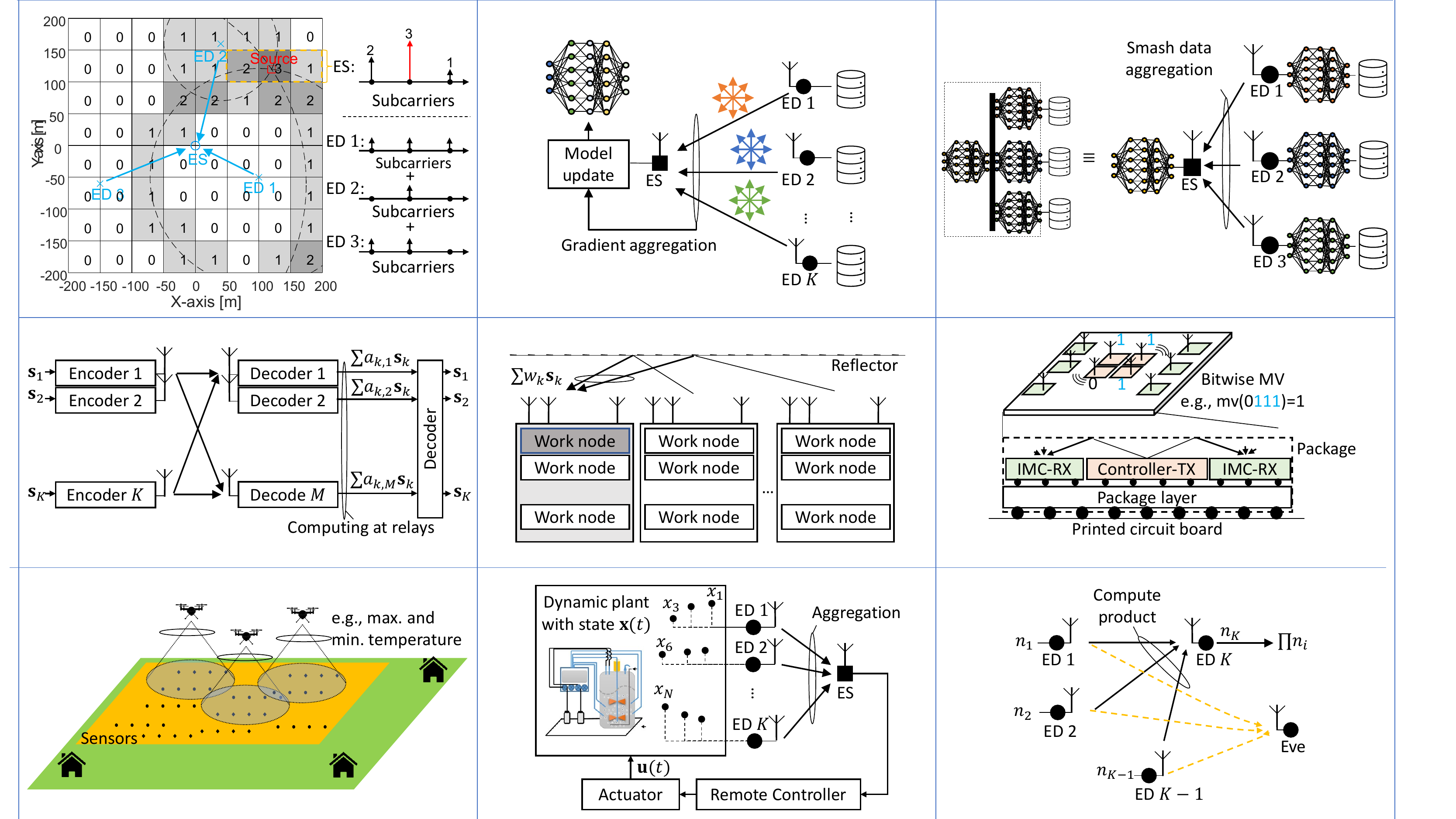}
	\label{subfig:farms}}
\subfloat[Physical layer security, e.g., key generation]{\includegraphics[width =\figureSizeApp]{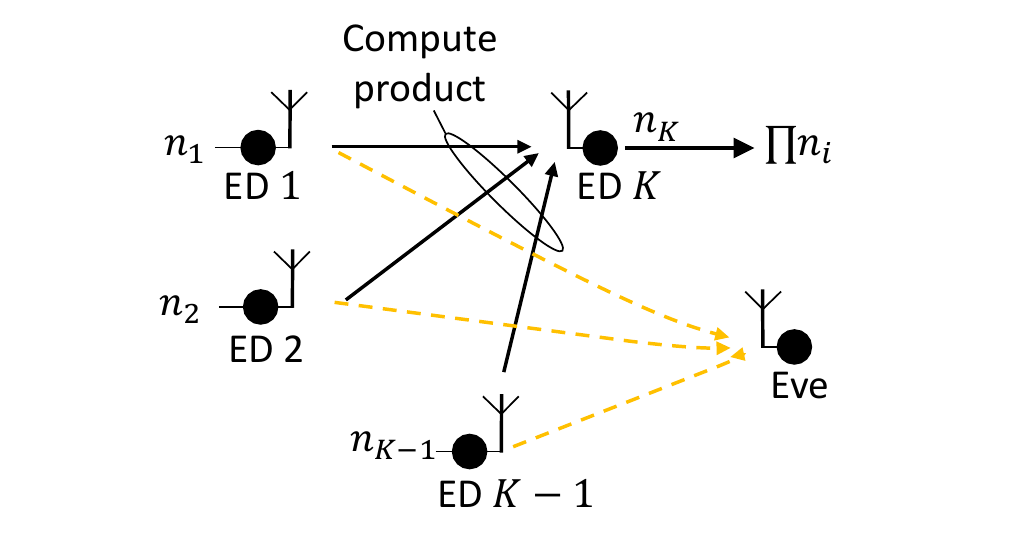}
	\label{subfig:key}}
	\caption{Applications of OAC.}
	\label{fig:apps}	
\end{figure*}

In this section, we discuss several applications of \ac{OAC} in various fields, as illustrated in \figurename~\ref{fig:apps}. We also discuss the state-of-the-art demonstrations of \ac{OAC} for certain applications.

\subsection{Distributed localization}

Consider a scenario where many sensors are deployed in an area to identify the location of a radio source emitting a signal with a known transmit power via the \ac{RSSI}. \ac{OAC} can provide a localization solution based on voting over OFDM as follows \cite{safiINFOCOM_2023}: 
\begin{itemize}
\item Step 0:  The area is divided into a grid and all sensors know their positions and the grid structure.
\item Step 1: Each sensor estimates the link distance between its location and source based on \ac{RSSI}. 
\item Step 2: Each sensor  marks the squares that intersect with the circle with the radius of the estimated link distance. 
\item Step 3: Each sensor activates the corresponding OFDM subcarriers that represent the marked squares (i.e., vote).
\item Step 4: All the sensors transmit simultaneously.
\item Step 5: The ES determines the location of the source by detecting the subcarrier that has the largest magnitude (i.e., \ac{MV}).
\end{itemize}
The procedure above is an extension of a voting-based localization \cite{liu_2005attack} with the consideration of \ac{OAC}. With \ac{OAC}, instead of using orthogonal channels to acquire the sensor information,  the \ac{ES} receives the superposed signal and obtains the locations of the radio without any extra computation. An example is illustrated in \figurename~\ref{fig:apps}\subref{subfig:dl}.

\subsection{Wireless control systems}
In control theory, a dynamic plant refers to a state-space model where the current states evolve in time. It can be modeled as a set of first-order linear difference equations, e.g., $\textbf{x}(\timeSymbol+1)=\textbf{A}\textbf{x}(\timeSymbol)+\textbf{B}\textbf{u}(\timeSymbol)+\textbf{w}(\timeSymbol)$, where $\textbf{A}$ and $\textbf{B}$ are real-valued matrices, $\textbf{u}(\timeSymbol)$ is plant control action, $\textbf{x}(\timeSymbol)$ is the plant state, and $\textbf{w}(\timeSymbol)$ is the plant noise \cite{Cai_2018}. A simple example of a dynamic plant is a pendulum, where the vector $\textbf{x}(\timeSymbol)$ represents the angular position and speed of the pendulum \cite{Minjie_2021iotj}. The vector $\textbf{x}(\timeSymbol)$ may involve a large number of spatially-distributed state variables. For example, for a chemical plant, the  state variables may be the temperature, humidity, and pressure, while they  may be atmospheric pressure, thrust, drag, speed, and acceleration for an aircraft. In \cite{Cai_2018}, a scenario where the stability of a dynamic plant is monitored by distributed sensors is considered. The main goal is to stabilize a potentially unstable plant over  limited wireless resources by acquiring the current state as quick as possible. 
By incorporating the multiple-access channel into the expression of the corresponding state-space equation of the
dynamic plant, \ac{OAC} is exploited to support a large number of sensors. We also refer reader to \cite{Minjie_2021iotj,Minjie_2021,Minjie_2022} for further details to dynamic plants, where the main goal is recover the state vector $\textbf{x}(\timeSymbol)$ by using \ac{OAC}. In \cite{Park_2021}, \ac{OAC} is applied to a general control system. The proposed scheme is concerned about the transmit power of sensors to minimize the effect of the wireless channel to the control system  under a transmit power constraint. {\color{\reviewColor}In \cite{Jihoon_2022Platooning}, distributed consensus via \ac{OAC} is applied to vehicle platooning control. \ac{OAC} is utilized to calculate the average positions of the vehicles, needed for the calculation of the accelerator at each vehicular to stabilize the platoon. For \ac{OAC}, the authors propose to multiply the parameter (a scaled version of the position of the vehicle) with the sign of the real part of the channel at the transmitters so that a coherent addition is obtained at the receiver. Only real part of the symbols are used for computation. In \cite{zhanAirGNN}, an implementation of a graph neural network is investigated with OAC for multi-robot flocking.}

\subsection{Wireless sensor networks}
In \cite{Goldenbaum_2009wcnc}, \ac{OAC} is used for estimating the portion of inactive sensors in a network. For this application, all active sensors transmit the symbol $1$ and the receiver estimates the number of active sensors from the received signal. If the \ac{ES} knows the number of sensors in the network, it immediately estimates what portion of sensors are inactive.
An example of \ac{OAC} for computing the arithmetic mean of temperatures measured by 250 sensors for
environment monitoring can be found in \cite{Goldenbaum_2009wcnc}. In \cite{abari_2016oac}, several other applications of \ac{OAC} like counting the number of sensors whose readings satisfy a certain threshold, a variance of the measured temperatures, or the best linear fit to the observed measurements, i.e., regression, are given in the area of \acp{WSN}. With the motivation of environment sensing and radio map construction, product-of-experts-based Gaussian process regression over a distributed network is investigated in \cite{sato_2022gc}, where the investigated \ac{OAC} schemes are based on \ac{ZF} precoder and \ac{PC}. {\color{\reviewColor}In \cite{Park_2023mdpi}, the authors consider a distributed sensing application, where all the sensors observe a linear combination of the data. In \cite{Min_icc2021,Min_uav2022},  the mobility of \acp{UAV} is exploited to improve the power alignment for \ac{ZF}-based \ac{OAC}. Cluster scheduling, association, and \ac{UAV} trajectory are jointly optimized with the motivation of aligning the signals within each cluster
while mitigating the inter-cluster interference. \ac{UAV}-based aggregation is also studied in \cite{Zhong_2022uav}. Similarly, \ac{UAV} trajectory optimization with the consideration of multi-slot \ac{OAC} is investigated in \cite{Xiang_ojc2022}. In \cite{Luteng_2022gc}, the authors consider the freshness of the aggregated data in addition to \ac{MSE} in a time-varying environment for OAC-based remote monitoring applications. In \cite{Akshay_IPSN2020}, \ac{OAC} is considered for  constructing the geographical heat distribution via low-power wide-area networks with the motivation of detecting forest fires.}

{\color{\reviewColor} 
\subsection{Distributed optimization over wireless networks}
One of the main motivations behind OAC is the convergence of communication and computing architectures as explicitly discussed in ITU’s report for IMT-2030 \cite{ITU-RM2516_report}. The driving force for this trend is the advances in machine learning and artificial intelligence technologies such as federated learning and split learning, and their utilization over wireless networks. The major benefit gained from OAC for a distributed optimization problem is the considerable  improvement in computation rate as compared to the traditional way of separating communication and computation tasks, particularly, when many EDs participates in computation.}

\subsubsection{Federated learning}
\ac{FL} \cite{pmlr-v54-mcmahan17a} is one of the most studied distributed learning frameworks. The task of model training  is distributed across multiple \acp{ED} and data uploading is avoided to promote the user-privacy. Instead of data samples, \acp{ED} share a large number of local stochastic gradients or local model parameters with an \ac{ES} for aggregation. For its implementation over a wireless network in general, i.e., \ac{FEEL}, we refer the reader to \cite{park2019wireless,chen2021distributed,hellstrom2020wireless,AhmedSurvey_2022,Popovski_2021,Yiming_2022,Xiaowen_2022}. 

{\color{\reviewColor}For FEEL, if the communication and computation are considered as separate tasks, for each iteration, the ES needs to acquire the local model parameters (or gradients) from $\numberOfEdgeDevices$ EDs, separately, to compute $\numberOfFunctions\sim10^6-10^8$ functions. Hence, $\numberOfFunctions\numberOfEdgeDevices$ parameters need to be transferred in the \ac{UL} and the latency grows linearly with the number of EDs for an orthogonal  multiple access scheme. On the other hand, with OAC, the cost is equal to the one with a single ED  as all the EDs transmit simultaneously to compute $\numberOfFunctions$ functions, e.g., the average of the local model parameters (or gradients). Hence, the training can be completed much faster if OAC is utilized.}

One of the crucial choices for \ac{OAC} to support \ac{FEEL}  is that the information needs to be transmitted in the \ac{UL} and \ac{DL}. This is because the \ac{UL} information can be local gradients or local parameters, while the information broadcast to the EDs in the \ac{DL} can be the updated model parameters or the aggregated gradients, leading to four different FEEL implementations. Although the gradients often have an unknown probability distribution that changes over the communication rounds \cite{Zhang_2021}, their magnitudes tend to decrease over iterations. Also, the gradients between adjacent  communication rounds across different EDs may be highly correlated, as well as the entries of a stochastic gradient vector at one iteration (see \cite{zhong2021overtheair,liang2021wyner,Xue_2022}). In addition, even if the signs of the gradients are transmitted, convergence can be achieved \cite{Bernstein_2018}. These properties are exploited in the UL in several \ac{OAC} papers, e.g., \cite{Sahin_2022MVjournal,mohammadICC_2022, Guangxu_2021, sahin2022md,sahinGCbalanced_2022, Zhang_2021, zhong2021overtheair,Xue_2022}. In DL, the broadcasting updated model parameters can ensure that the \acp{ED} calculate the gradients based on the same model parameters. On the other hand, for multi-cell \ac{OAC}, broadcasting aggregated gradients in the DL is shown to be useful for aggregation for EDs located at the cell edge  \cite{mohammadICC_2022} while promoting the personalization of the model parameters.

{\color{\reviewColor}It is also worth noting that FEEL with OAC inherits the well-known problems in FL literature such as convergence under data and device heterogeneity,  stragglers, data privacy, and various security issues. Hence, these application-specific challenges often need to be re-evaluated for a given OAC scheme. For example, it is  challenging to deal with Byzantine attacks when OAC is used for FEEL since the ES does not directly observe the gradients or the model parameters. Also, training can lead to  biased learning due to the imbalanced received signal powers \cite{sahin2022md,Sahin_2022MVjournal}}

\subsubsection{Split learning}
In \cite{gupta2018}, the authors investigate the idea of splitting a neural network over the \acp{ED} and  \ac{ES} so that the \acp{ED} can conserve the privacy of their local data sets while the computational burden is decreased on the \ac{ED} side. Under a simple configuration, the \acp{ED} train the network up to a specific layer, called {\em cut layer}, and send the output of the cut layer, i.e., {\em smashed} data, with the labels to the \ac{ES}. The \ac{ES} completes the rest of forward step starting from a layer that concatenates and aggregates the EDs' smash data, i.e., {\em aggregation} layer.
Afterward, the ES starts the back-propagation of the gradients from the last layer to the first layer of the \ac{ES}'s neural network, and sends the gradients with respect to the smash data to the \acp{ED}. The forward and back-propagation continue until the network converges. The main advantage of \ac{SL} over \ac{FL} is that the \acp{ED} have fewer layers as compared with the ones in the \ac{FL} \cite{Abhishek_2019,vepakomma2019reducing,thapa2020splitfed,vepakomma2018}. Note that the model splitting also appears as {\em vertical} \ac{FL} in the literature \cite{Wei_vfl2022}, where a neural network is divided and distributed across the network. User scheduling under fading channels for vertical \ac{FL} is discussed in \cite{Zhang_vfl2022} without using \ac{OAC}. The reader is also referred to the technical reports in \cite{tr22876,tr22874} for potential applications of \ac{SL}.

\Ac{SL} is not heavily investigated in the state-of-the-art  from the perspective of \ac{OAC}. In \cite{Krouka_2021}, the aggregation layer of \ac{SL} is proposed to be realized with \ac{BAA} along with channel inversion. In this approach, to accommodate \ac{OAC}, the weighted multiplication of the aggregation layer is moved to the EDs whereas the bias addition and activation are kept at the ESs. To address the fading, the users with deep faded channels are excluded from the training. Instead of gradient or local parameter aggregation, smash data aggregation takes place for the implementation of \ac{SL} over a wireless network as shown in \figurename~\ref{fig:apps}\subref{subfig:sl}. In \cite{YangYuzhi_2022,Yuzhi_mimoSplit2022}, by exploiting channel reciprocity and considering the wireless channel as part of the neural network, i.e., a form of \ac{OAC} for implementing a fully-connected layer over the air, forward-backward propagation for \ac{SL} is investigated over  \ac{MIMO} channels. The key observation in this study is that the backward propagation can still be maintained by transmitting the gradients  if the channel reciprocity is maintained.

\subsubsection{Other distributed computation frameworks}
 In \cite{Lee_2022gnn}, a graph neural network where the devices correspond to the edges of a graph is investigated. Each node in the graph aggregates information from its neighboring nodes in the graph. In this study, \ac{OAC} is exploited to improve the computation rate while increasing privacy. {\color{\reviewColor}In \cite{Yifan_2022mpa}, a message-passing neural network is considered to address the decentralized power allocation problem in a device-to-device network and a channel-inversion-based \ac{OAC} is used. In \cite{zhanAirGNN}, the channel coefficients are embedded into the graph convolution operator. Hence, \ac{CSI} between the link becomes part of the graph neural network.  
 In \cite{Zezhong_2022}, \ac{OAC} is utilized to perform \ac{PCA} when the data is not centralized. The authors express the centralized \ac{PCA} problem as a minimization problem. By using the corresponding gradients of the objective function, it is proposed to solve the minimization problem by aggregating the gradients over the air along with \ac{TCI}. In \cite{MOLINARI20202999}, it is proposed to obtain a cooperative solution of a linear algebraic  equation by exploiting \ac{OAC}, where each agent knows only a subset of the equations. In \cite{Mitsiou_2022arxiv}, distributed primal-dual optimization is investigated along with \ac{TCI}-based \ac{OAC}, and applied to the energy management of a smart grid system. }

In the literature, it is worth noting that there are many other statistical methods,  e.g., independent component analysis, k-means, k-SVD, that can benefit from \ac{OAC}. Exploration of such algorithms with the consideration of \ac{OAC} is currently an open topic.

\subsection{Wireless data center networks}
 A \ac{DCN} manages the communications among the work nodes across data centers to  store or process the files in a parallel manner. It is often implemented through a high-bandwidth wired network. On the other hand, a wired DCN has limited flexibility, cabling complexity, and device cost, which affects the scalability of \acp{DCN}. To address those issues, in \cite{Xiugang_2016}, \ac{OAC} is proposed to compute the arithmetic mean of the symbols at $\numberOfEdgeDevices$ source nodes for a wireless \ac{DCN}.

\subsection{Wireless intra-chip computation}
In \cite{Robert_2022}, \ac{OAC} is exploited with the motivation of addressing scaling-out wired interconnects for hyperdimensional computing. In this method, the main goal is a similarity-search task via multiple \ac{IMC} cores, where the input information at each core is the bit-wise \ac{MV} of the queries from different controllers (i.e., bundling) over a wireless channel as shown in  \figurename~\ref{fig:apps}\subref{subfig:ic}. The transmitters at the controllers use \ac{BPSK} symbols to transmit 0 and 1 based on their bits. The receivers at the \ac{IMC} cores receive a slightly different version of the superposed symbols due to the multi-path channel. To ensure that the phases are aligned at the receivers, the authors propose to optimize the phase of transmitted symbols so that the error rate of the bit \ac{MV} computation at each core is minimized.

\subsection{Wireless communications}

\subsubsection{Compute-and-forward relaying scheme}
With the compute-and-forward relaying strategy \cite{Nazer_2007, Nazer_2011}, the multiple relay nodes forward the linear functions of the transmitted messages to be decoded at the destination, as illustrated in \figurename~\ref{fig:apps}\subref{subfig:cf}. In \cite{Goldenbaum_2016}, the compute-and-forward scheme is exploited to harness the collisions for massive access. It is well-known that when a collision occurs on the channel, 
non-scheduling-based channel access protocols, like the carrier-sensing multiple access with collision avoidance used in Wi-Fi, require the involved devices to access the channel again using a back-off mechanism to reduce the probability of repeated collisions.
 To address the diminished rate in this case, in \cite{Goldenbaum_2016}, collisions are exploited at the nodes and the nodes forward a linear combination of the messages to the base station along with the corresponding coefficients for decoding.

\subsubsection{Physical-layer network coding}
\Ac{PLNC} is one of the well-studied applications of \ac{OAC} in the area of wireless communications \cite{Katti_2007networkcoding,zhang_plnc2006,Sachin_2008,Valenti_2011,Ferrett_2018,Nazer_2011}. 
A canonical example of \ac{PLNC} is  communication over a two-way relay channel. In this channel, the devices want to exchange their bits over a relay.  A {\em link-layer} network-coding strategy requires three time slots to accomplish this task: In the first two time slots, the devices send their bits to the relay, sequentially. In the third slot, the relay forwards the XOR of the bits to devices. With {\em physical-layer} network coding, the same task is accomplished in two time slots by exploiting signal superposition property: The devices transmit simultaneously their signals determined based on the bits. The relay node then forwards either the superposed signal itself (i.e., analog network coding) \cite{Katti_2007networkcoding} or the signal after some detection (i.e., digital network coding)  \cite{zhang_plnc2006,Nazer_2011,Valenti_2011,Ferrett_2018} to the devices. Since each of the devices knows its own signal, it can obtain the message at the other device.

In \cite{Katti_2007networkcoding}, the authors exploit the differential encoding along with \ac{MSK} and discuss  practical issues such as synchronization for analog network coding. In \cite{zhang_plnc2006}, it is proposed to use \ac{QPSK} symbols at the devices and the relay detects the XOR of bits, resulting in a corrupted version of the XOR of the transmitted bits.
In \cite{Valenti_2011,Ferrett_2018}, \ac{FSK} is utilized with the motivation of reducing strict requirements on power control, phase synchronization, and \ac{CFO}. The impact of the channel on the detector design at the relay for binary FSK \cite{Valenti_2011} and $M$-ary FSK along with an \ac{LDPC} code are discussed rigorously. In \cite{Nazer_2011}, Bobak shows that the nested-lattice code used in compute-and-forward relaying strategy can also be utilized in two-way relay channel to improve the reliability physical layer network coding and an excellent comparison of analog and digital network coding is provided. We also refer the reader to \cite{LIEW20134} and the references therein for the variants of physical layer network coding.

\subsubsection{Overhead reduction}
In \cite{Guangxu_2019iotj}, the authors use \ac{OAC} not only for computation but also for  overhead reduction. Instead of acquiring the \ac{CSI} feedback from each ED through orthogonal multiple access, they calculate the optimum receive beamforming vector at the ES by concurrent transmissions. It is shown  that the feedback overhead reduction can be reduced $50$ times more than the one with conventional training. In \cite{Chen_2018sensing}, the authors propose to determine the power-normalization factor for zero-forcing by calculating the minimum function through the queries discussed in Section~\ref{subsec:commonNomo}, instead of estimating the channel of each ED through orthogonal channels.

\subsubsection{Cognitive radios}
In \cite{Zheng_cn2017} and \cite{Zheng_tgcn2017}, a variant of Goldenbaum's scheme discussed in Section~\ref{parag:ac} is used for a spectrum-sensing application for cognitive radios. In this application, the fusion center desires to detect if the primary user is active or absent by using many sensors. To this end, 
 the symbols that are transmitted from the sensors are either the average signal power  or the hard-detection activity results on the primary-user band. The information across the sensors is proposed to be aggregated over the air to achieve a time-efficient cooperative spectrum sensing. Instead of only using amplitude correction, a weighted sum based on the absolute square of the channel coefficients is incorporated to Goldenbaum's scheme to avoid power boost due to the inversion, which effectively corresponds to maximum ratio transmission. In \cite{Chen_2018sensing}, the \acp{DFT} of the received signals at the sensors are proposed to be combined over the air for spectrum sensing. The \ac{OAC} in this study relies on a zero-forcing precoder. In \cite{sato_2022gc}, \ac{OAC} is applied to radio map construction. For this application, Gaussian process regression is considered for a scenario where the nodes are deployed in two-dimensional space and measure the \ac{RSSI} of a transmitter. It is shown that the proposed approach can speed up the computation time approximately 733x than the one with separation of communication and computation tasks.

\subsection{Security}
In \cite{Altun_2022}, \ac{OAC} is utilized for multiplying Gaussian prime numbers to compute a secret key that can be used in any encryption process or  to generate keys. As illustrated in \figurename~\ref{fig:apps}\subref{subfig:key}, in this approach, each node is assigned  a number and all \acp{ED} calculate the products with OAC by changing the destination node sequentially. It is argued that this can also provide physical layer security as an eavesdropper cannot directly observe the multiplication. {\color{\reviewColor}In \cite{Xie_2022blockchain}, the authors propose to use \ac{OAC} for assessing the consensus in a blockchain network. With this approach, each user maps the bits in the hash to the modulation symbols in a constellation and all the users transmit simultaneously. Since the hash should be consistent among the users, the received symbol after superposition should be one of the points in the constellation. By detection, the malicious users with  inconsistent hashes are filtered out. In this study, \ac{OAC} is based on zero forcing.}

\subsection{Demonstrations}

In the state-of-the-art, early \ac{OAC} demonstrations are mainly in the areas of \acp{WSN} and \ac{PLNC}. For example, in \cite{Jakimovski_2011},  a statistical \ac{OAC} is implemented with twenty-one RFID tags to compute the percentages of the activated classes that encode various temperature ranges. A trigger signal  is used to achieve time synchronization across the RFIDs. In \cite{Kortke_2014}, Goldenbaum and Sta\'nczak's scheme \cite{Goldenbaum_2013tcom} is implemented  with three \acp{SDR} emulating eleven sensor nodes and a fusion center. The arithmetic and geometric means of the sensor readings are computed over a 5~MHz signal. The time synchronization across the sensor nodes is maintained based on a trigger signal and the proposed method is implemented in a \acl{FPGA}. A calibration procedure is also discussed  to ensure amplitude alignment at the fusion center. In \cite{Alton2017testbed}, the summation is evaluated with a testbed that involves three \acp{SDR} as transmitters and an \ac{SDR} as a receiver. The  scheme used in this setup is based on channel inversion and {\color{\reviewColor} puts limitations on the supported
dynamic range to avoid exceeding the maximum transmit power.}
In \cite{abari_2016oac},  six \acp{USRP} represent the sensor nodes and the experiment is repeated multiple times to emulate the effect of many sensors. The sum operation is implemented by using a binary representation of the parameters. For this experiment, AirShare protocol in \cite{Abari2015Airshare} is  utilized to ensure that all transmissions are coherent. In \cite{Chen_2018sensing}, a spectrum sensing example based on \ac{OAC} is given, but the details related to the protocols for synchronization are omitted. 
{\color{\reviewColor}In \cite{Lu_2013plnc}, a real-time implementation of \ac{PLNC} is demonstrated. To achieve time synchronization, it is proposed to add a sufficiently long time that compensates for the transfer time between the computer and \acp{SDR}. In this method, the same clock/oscillator is connected to the co-located radios and  the transmission time instants are set manually.
In \cite{Lizhao_plnc2017}, by extending \cite{Lu_2013plnc} into a more general framework, a time-slotted approach is proposed   to maintain time-synchronization among the radios. In this method, the radios (i.e., EDs) first align their slot boundaries by compensating the time difference between the first sample of a reference packet transmitted from the relay (or \ac{ES}) and the first received sample (e.g., noise) marked by the \ac{USRP} hardware. By exploiting the time-stamp-based transmission feature of \ac{USRP} hardware \cite[Section III.B.2]{Lizhao_plnc2017}, the radios transmit simultaneously at the  slot boundaries. In \cite{Yihua_2018tmc}, \ac{PLNC} is implemented by using temperature-compensated oscillators and implementing custom blocks for accurate alignment.


To the best of our knowledge, the demonstrations of \ac{OAC} schemes for \ac{FEEL} are limited, but get more attention in the recent literature.
In \cite{Guo_2021},  a custom two-stage protocol that mitigates the \ac{TO} and \ac{CFO} is proposed and an \ac{OFDM}-based \ac{OAC} with channel inversion is investigated. In \cite{sahinGC_2022}, a general-purpose time synchronization method that allows a set of \acp{SDR} to transmit or receive any \acl{IQ} data simultaneously while maintaining the baseband processing in the corresponding \acp{CC} is proposed. This approach relies on the detection of a synchronization waveform and passing a pre-determined number of \ac{IQ} samples to the \ac{CC} upon its detection. All \acp{SDR} wait for a pre-determined duration for \ac{CC}-based processing and transmit simultaneously the  \ac{IQ} samples in the SDR buffers.
 By implementing this synchronization method on five \acp{SDR} (i.e., Adalm Pluto) along with a control loop that mitigates \ac{TO}, \ac{CFO}, and power offset coarsely, the performance of \ac{FSK-MV} (see Section~\ref{par:orthSignaling} for details) is practically demonstrated for hand-digit recognition task based on MNIST database.  The experiment shows that the test accuracy can reach more than 95\%  for homogeneous and heterogeneous data distributions without using channel state information at the \acp{ED} or any method for phase synchronization. The experiment is conducted in a small room, where the distances between the EDs and ES are about $5$~meters, but the channel is shown to be frequency-selective. The experiment also shows that the phase rotation in the frequency domain for OFDM is a function of the time synchronization errors at the EDs and ES, which is aligned with the analysis in Section~\ref{parag:synch}.
In \cite{Liang_2022}, a \ac{TCI}-based \ac{OAC} with two USRP N210 SDRs as \acp{ED} is used for \ac{FEEL}. This setup maintains the synchronization through a cable connection between the \acp{SDR}. In \cite{ZhaoICC_2022_extension}, the aforementioned time-stamp based transmission discussed in \cite{Lizhao_plnc2017, Lu_2013plnc} is adopted to demonstrate the multi-user detection-based computation by using a higher complexity receiver.

For a consensus application, in \cite{Xie_2022blockchain}, srsRAN is considered for implementing a channel-inversion-based OAC  with seven radios. By considering seven users, the authors propose to use user
attaching procedures for time synchronization and \ac{LTE} frame structure. However, the details related to the time synchronization among the radios are not provided explicitly.
}

{
	\color{\reviewColor}
\section{Conclusion}
\label{sec:WhatNeedsToBeImproved}

 In this section, we summarize the main takeaways based on the discussions in the previous sections and highlight several research directions.

\subsection{Takeaways}
\ac{OAC} is a concept that fundamentally disrupts the traditional way of performing communication and computation as separate tasks. The main goal of OAC is to compute a multivariate  function in the wireless channel and the arguments of the function are not intended to be obtained at the ES. While function computation requires two sequential steps (i.e., communication and computation)  with \ac{NOMA} or OMA, such separation does not occur for OAC. Hence, the main benefit gained from OAC is the improvement of the computation rate, which otherwise scales down with the number of EDs participating in the computation.

With OAC, a function that structurally matches  the underlying operation that  multiple access channel naturally performs can be computed. Due to the signal superposition property of wireless channels, the functions that can be computed are in the space of nomographic functions. By manipulating pre- and post-processing functions of a nomographic function, commonly-used functions such as arithmetic mean, weighted sum, norm, \ac{MV}, and histogram can be computed. Although the space of nomographic function with continuous pre- and post-processing functions is limited, Kolmogorov's superposition shows that every continuous function can be computed via multiple nomographic functions. Also, some functions such as maximum, minimum, and median can be computed systematically over iterations or based on some approximations.

Achieving reliable OAC  under practical wireless channels is a challenging task since the multi-path channel between an ED and ES distorts the transmitted symbols  before the signal superposition. Hence, typical linear equalization methods cannot be directly utilized for OAC to compensate for the distortion. Hence, most OAC schemes in the state-of-the-art use \ac{CSIT} along with \ac{CSIR} to combat the channel. The corresponding precoders are often derived based on an \ac{MSE} criterion or an application-specific metric such as training loss for \ac{FEEL} under certain conditions, e.g., transmit power. To deal with the distortion due to the channel at the expense of more resource consumption, there exist also blind \ac{OAC} schemes that rely on channel hardening via multiple antennas or non-coherent techniques in the state-of-the-art.

From the encoding perspective, an OAC scheme can directly use continuous-valued parameters along with an analog modulation or utilize the quantized parameters with a digital modulation. In the case of analog encoding, linear or affine transformations are shown to be effective for compression if certain properties, e.g., sparsity, are present in the parameters. For digital schemes, the family of  nested-lattice codes is often considered in the literature for reliable computation since the codes in this family can be made to be linear in $\realNumbers$. 
Nevertheless, the coding for \ac{OAC} is an area that requires more research.
From the encoding perspective, heavy quantization, e.g., 1-bit quantization for \ac{MV} computation, is shown to be an effective solution for certain applications such as distributed learning and localization while being compatible with  traditional digital communication systems.

The metrics for assessing an OAC scheme differ from the traditional communication metrics such as the data rate, bit-error rate, and block-error rate since OAC aims to compute a function. Often, the performance of an OAC scheme is measured via an \ac{MSE} analysis. For a digital OAC scheme, the probability of computing a single function (or a set of functions) incorrectly can also be used as a metric since the image of the function consists of a set of discrete values. The computation rate, i.e., the number of functions calculated per real dimension, is another metric that can be used for evaluating the efficiency of an OAC scheme. One can also obtain application-specific metrics such as test accuracy and convergence rate for FEEL when it is used with an OAC scheme. Such derivative metrics are beyond the scope of our survey paper.

As a working principle, OAC relies on simultaneous receptions of EDs' signals on the same wireless resources at the ES and
shares similar enabling mechanisms for \ac{UL}-\ac{OFDMA} and multi-user \ac{MIMO}. Reliable OAC requires underlying mechanisms such as time-frequency-phase synchronization, power management, and channel estimation or feedback mechanisms to perform well. Depending on the scheme, OAC can impose very stringent requirements. For example, if the computation relies on phase synchronization among the EDs, a sample-level time-synchronization in the network must be maintained and the phase accumulation due to the residual \acp{CFO} should be addressed. On the other hand, methods that do not rely on phase synchronization are shown to provide immunity against \ac{TO} and \ac{CFO} impairments. 

Power management for OAC has two folds: transmitter side and receiver side. From the perspective of the transmitter, an OAC scheme should consider not only maximum transmit power limitation but also the \ac{ACLR} requirements and \ac{PA} efficiency. From the perspective of the receiver, the power control mechanisms need to be utilized to align the received signal powers at the ES.  While aligning the average signal power can be managed via typical closed-loop power control mechanisms such as the ones in 4G LTE and 5G NR, a perfect amplitude alignment among the EDs at ES can be challenging as it requires accurate \ac{CSIT} and/or low-latency feedback in time-varying fading channel conditions, and channel inversion  under the transmit power limitations. Network topology for \ac{OAC}  has also a major impact on the computation as it  extends the basic single-cell OAC to a higher-complexity computation framework that involves many fusion nodes. However, the implications of OAC for a large-scale multi-cell computation are largely unknown.

OAC has both negative and positive aspects in terms of security. On the positive side,  user privacy is promoted as the transmitted signals cannot be directly observed due to the superposition. On the negative side, potential adversaries can harm the computation, particularly via Byzantine attacks. Similarly, a jammer can interfere with the superposed signals or an eavesdropper can overhear the computation in the wireless channel.

In the state-of-the-art, OAC has been considered for a wide variety of applications such as localization (e.g., voting-based distributed localization), wireless control systems (e.g., dynamic plants), wireless sensor networks (e.g., environment monitoring and UAV-trajectory optimization), distribution optimization (e.g., \ac{FEEL}), wireless data center and intra-chip computation (e.g., wireless computation and similarly search), wireless communication systems (e.g., \ac{PLNC} and spectrum sensing), and security (e.g., key generation). Among these applications, distributed optimization is currently the leading use case of OAC due to the advances in machine learning and artificial intelligence and the desire to use these techniques over wireless networks. 

\subsection{Research directions}
}
The results in the state-of-art overall advocate that OAC can address latency issues by improving the computation rate. On the other hand, \ac{OAC} needs to be evaluated further along with enabling mechanisms, applications, and corresponding algorithms. To this end, three major research directions that one can pursue are summarized as follows:

{\em \color{\reviewColor}Direction 1 -  OAC schemes with the consideration of practical limitations:} 
In the literature, \ac{OAC} primarily is investigated theoretically under certain assumptions. Hence, some of the practical aspects may be omitted. To address this issue, the methods need to be evaluated  under more challenging scenarios or designed with the considerations of imperfections and practical limitations. For instance, in practice, imperfect \ac{CSI}, the residual \ac{TO}, \ac{CFO}, \ac{PO}, mobility, and \ac{PA} non-linearity may be inevitable and their impacts on the performance depend on the robustness of the \ac{OAC} scheme  and the corresponding applications to these imperfections. Another way of addressing this issue is to generate convincing results through demonstrations. A plausible demonstration needs not only the implementation of the \ac{OAC} scheme but also the design of the underlying protocols that  maintain signal superposition. Hence, the work in this area often involves developing the corresponding protocols in addition to the \ac{OAC} scheme itself. Another practical challenge in this direction is that it is often not trivial to configure multiple standard \acp{SDR} for simultaneous transmissions. Even if there are some proof-of-concept \ac{OAC} demonstrations, there is no widely-accepted multiple access channel testbed or platform to test different \ac{OAC} schemes under realistic scenarios in controllable environments. 

{\em \color{\reviewColor} Direction 2 -  Algorithms with the consideration of OAC:}  
Another area that can be improved is the algorithms in the applications. In the literature, the algorithms for many applications are not designed with the consideration of \ac{OAC}. On the other hand, the algorithm can be designed to facilitate \ac{OAC} and relax the constraints for enabling mechanisms. For example, distributed training by \ac{MV} with \ac{signSGD}, a machine learning concept, is more compatible with digital modulation as compared to the one with \ac{SGD} and results in various \ac{OAC} schemes. Similarly, the implementations of plain \ac{FL} based on stochastic gradients (i.e., FedSGD) or parameter aggregations (FedAve) are  mathematically equivalent to each other. However, the corresponding algorithms for \ac{FEEL} relying on \ac{OAC} can perform differently as the gradients and parameters have different statistical characteristics that can be exploited for  \ac{OAC}. The algorithms that facilitate multi-cell computation under various network architectures are also needed to be developed.

{\em Direction 3 - Protocols for OAC with the consideration of standards:} 
To this date, \ac{OAC} has not been used in any communication standard or {\color{\reviewColor}a commercial system}. In fact, \ac{OAC} has recently been discussed in AI/ML \acl{TIG} for IEEE 802.11 for distributed learning \cite{sahin_2022ieee}. In 3GPP meetings, use cases and {\em potential} requirements for 5G to support machine learning applications under three main categories, i.e., \ac{FEEL}, \ac{SL}, and model distribution,  are studied \cite{tr22876,tr22874}, which highlights the need for a comprehensive system optimization under communication and computation constraints. Similarly, {\color{\reviewColor}in \ac{ITU}'s report \cite{ITU-RM2516_report}, the convergence of communication and computing architecture  in \ac{IMT} systems towards 2030 is emphasized. {\color{\reviewColor}  Currently, it is not clear if  future wireless networks will utilize OAC to facilitate this convergence} or if the existing procedures can support an \ac{OAC} scheme reliably or not. 
	This is because the current wireless standard protocols in the state-of-the-art are designed by assuming that the communication is separated from the computation. Hence, further evaluations of the underlying systems and the enablers for OAC are needed.
	To achieve a standardized \ac{OAC}, the procedures for time-frequency synchronization, power control, channel estimation, calibration, re-transmissions, compression, and security aspects along with the architectures need to be re-evaluated.}  A standardized OAC can insure interoperability among devices from different manufacturers for a large body of applications.

\acresetall

\acresetall
\bibliographystyle{IEEEtran}
\bibliography{references}

\end{document}